\documentclass[12pt]{article}
\usepackage[T1]{fontenc}
\pdfoutput=1

\usepackage{draft}
\usepackage{putex} 
\usepackage[all,cmtip]{xy}
\usepackage{fancybox}
\usepackage{subcaption}
\usepackage{graphicx}
\usepackage[dvipsnames]{xcolor}
\usepackage{cite}
\usepackage{mciteplus}
\usepackage{amsmath}
\DeclareFontFamily{OT1}{pzc}{}
\DeclareFontShape{OT1}{pzc}{m}{it}{<-> s * [1.10] pzcmi7t}{}
\DeclareMathAlphabet{\mathpzc}{OT1}{pzc}{m}{it}

\def\({\left(}
\def\){\right)}

\usepackage{tikz-cd}

\usepackage{latexsym}
\usepackage{cite}
\usepackage{framed}
\definecolor{shadecolor}{rgb}{0.95,0.95,0.97}
\definecolor{refkey}{rgb}{0.5,0.5,0}
\definecolor{labelkey}{rgb}{0.5,0.5,0}
\definecolor{citekey}{rgb}{0.5,0.5,0}
\definecolor{darkgreen}{rgb}{0,0.5,0}
\definecolor{darkblue}{cmyk}{0.9,0.9,0,0}
\definecolor{darkred}{rgb}{0.6,0,0.3}
\usepackage{ifpdf}
\ifpdf
\usepackage{graphicx} 
\usepackage{epsfig}
\usepackage[setpagesize=false,pagebackref=false, linktocpage, bookmarksopen=true, colorlinks=true, linkcolor=blue,citecolor=blue,urlcolor=blue]{hyperref}
\usepackage{hyperref}
\usepackage{trimclip}

\else
\usepackage{graphicx}

\fi

\newcommand{\U}{\text{U}}

\newcommand{\R}{\mathbb{R}}

\newcommand{\AdS}{\text{AdS}}

\newcommand{\Dd}{\text{d}}
\newcommand{\stir}[2]{%
  \begin{bmatrix}#1\\#2\end{bmatrix}}

\makeatletter
\newcommand{\rightlunate}{\mathbin{\mathpalette\rightlunate@{}}}
\newcommand{\rightlunate@}[2]{%
  \clipbox{.5\width{} 0pt 0pt 0pt}{$\m@th#1\circ$}%
}
\makeatother

\makeatletter
\newcommand{\rightlunateb}{\mathbin{\mathpalette\rightlunateb@{}}}
\newcommand{\rightlunateb@}[2]{%
  \clipbox{.5\width{} 0pt 0pt 0pt}{%
    \scalebox{1.5}{$\m@th#1\circ$}%
  }%
}
\makeatother

\def\fn#1{\footnote{#1}}

\newcommand{\td}[1]{\widetilde{#1}}
\def\eqref#1{(\ref{#1})}
\def\comma{\,,}
\def\period{\,.}

\def\bm#1{\text{\boldmath $#1$}}
\def\ket#1{|#1\rangle}

\def\XXint#1#2#3{{\setbox0=\hbox{$#1{#2#3}{\int}$}
		\vcenter{\hbox{$#2#3$}}\kern-.5\wd0}}

\newcommand{\beq}{\begin{equation}}
\newcommand{\eeq}{\end{equation}}

\def\nullify#1{}

\hypersetup{
  pdftitle={
    Orientifold and Integrability:
    N=4 Super-Yang-Mills on RP4
  },
  pdfauthor={
    Joao Caetano, Shota Komatsu,
    Paolo Soresina, Anders Wallberg
  },
  pdfsubject={
    A new integrable orientifold construction
    in the AdS/CFT correspondence.
    Combining charge conjugation with the RP4
    involution preserves planar integrability
    in N=4 super Yang-Mills theory.
    The work connects gauge-theory one-point
    functions to integrable spin-chain crosscap
    overlaps and Gaudin determinants,
    provides evidence for a holographic
    probe O1-plane in AdS5 x S5,
    and derives exact finite-N and
    finite-coupling half-BPS one-point functions
    using supersymmetric localization.
    The results connect weak-coupling gauge
    theory, supergravity and pp-wave string
    theory.
  },
  pdfkeywords={
    AdS/CFT correspondence,
    integrability in gauge and string theory,
    N=4 super Yang-Mills,
    orientifolds in string theory,
    holographic duality,
    non-orientable spacetime,
    real projective four-space RP4,
    charge conjugation,
    integrable crosscap states,
    spin-chain Bethe ansatz,
    Gaudin determinant overlap formulas,
    exact supersymmetric localization,
    nonperturbative one-point functions,
    probe O1-plane,
    Wishart-Laguerre matrix model,
    BMN pp-wave string theory
  }
}
\makeatletter
\def\section{\@startsection {section}{1}{\z@}{-3.5ex plus -1ex minus 
		-.2ex}{2.3ex plus .2ex}{\large\bf}}
\makeatother
\makeatletter
\def\subsection{\@startsection {subsection}{1}{\z@}{-3.5ex plus -1ex minus 
		-.2ex}{2.3ex plus .2ex}{\normalsize\bf}}
\makeatother
\usetikzlibrary{calc,patterns,arrows.meta,decorations.markings}
\newcommand{\propDirect}{
    \vcenter{\hbox{
    \begin{tikzpicture}
        \draw[thick, red] (0,0.15) -- (1.5, 0.15);
        \draw[CornflowerBlue, thick, dashed] (0,-0.15) -- (1.5, -0.15);
    \end{tikzpicture}
    }}
    }
\newcommand{\propCross}{
\vcenter{\hbox{
\begin{tikzpicture}
        \draw[red,thick] plot [smooth, tension=0.5] 
            coordinates{(0, 0.15) (0.5, 0.15) (1, -0.15) (1.5, -0.15)};

        \draw[CornflowerBlue, dashed,thick] plot [smooth, tension=0.5] 
            coordinates{(0, -0.15) (0.5, -0.15) (1, 0.15) (1.5, 0.15)};
    \end{tikzpicture}
    }}
    }
\tikzset{
  loop/.style={
    line width=0.9pt
  },
  loopcw/.style={
    loop,
    postaction={decorate},
    decoration={
      markings,
      mark=at position 0.19 with
        {\arrow{Stealth[length=2.5mm,width=2.2mm]}},
      mark=at position 0.24 with
        {\arrow{Stealth[length=2.5mm,width=2.2mm]}},
      mark=at position 0.69 with
        {\arrow{Stealth[length=2.5mm,width=2.2mm]}},
      mark=at position 0.74 with
        {\arrow{Stealth[length=2.5mm,width=2.2mm]}}
    }
  },
  halfdown/.style={
    loop,
    postaction={decorate},
    decoration={
      markings,
      mark=at position 0.45 with
        {\arrow{Stealth[length=2.5mm,width=2.2mm]}},
      mark=at position 0.56 with
        {\arrow{Stealth[length=2.5mm,width=2.2mm]}}
    }
  },
  dot/.style={
    circle,
    fill,
    inner sep=1.7pt
  }
}

\begin{document}
	
	\preprint{}
	\institution{Imperial}{Abdus Salam Centre for Theoretical Physics, Imperial College London, SW7 2AZ London, UK}
	\institution{CERN}{Department of Theoretical Physics, CERN, 1211 Meyrin, Switzerland}
	\institution{EPFL}{Ecole Polytechnique Federal de Lausanne (EPFL), Route de la Sorge, 1015 Lausanne, Switzerland}
	
	\title{Orientifold and Integrability:\\
         $\cN=4$ Super-Yang-Mills on $\mRP^4$
                  	}

	\authors{Joao Caetano, Shota Komatsu\worksat{\Imperial} , Paolo Soresina, Anders Wallberg\worksat{\Imperial,\CERN,\EPFL}}

	\abstract{We study planar $\mathcal{N}=4$ super Yang-Mills theory (SYM) on the real projective space  $\mRP^4$ using integrability, holography and supersymmetric localization, and discuss its relation to orientifolds in string theory. We show that the integrability of $\mathcal{N}=4$ SYM is preserved when combining the $\mathbb{Z}_2$ involution defining $\mRP^4$ with the action of the charge conjugation of the SU$(N)$ gauge group. We demonstrate this by calculating one-point functions of non-BPS single-trace  operators in various subsectors at weak coupling, which can be mapped to overlaps between energy eigenstates and crosscap states in integrable spin chains. The results exhibit a parity selection rule, a hallmark of integrability, and are given in terms of a ratio of determinants. Building on these results, we conjecture a closed formula for general single-trace operators.  We further provide evidence that this construction is holographically dual to a probe orientifold O1-plane in $\AdS_5\times \text{S}^5$ by constructing the corresponding worldsheet crosscap states in the pp-wave limit. Finally, we
use supersymmetric localization to compute one-point functions of half-BPS
operators exactly at finite $g_{\rm YM}$ and $N$. The computation reduces to
two-dimensional Yang--Mills theory on $\mathbb{RP}^2$ twisted by charge
conjugation and leads to a matrix integral of a Wishart-Laguerre type. In the
planar limit, the exact result interpolates between the weak-coupling
gauge-theory result, the strong-coupling supergravity prediction for the
orientifold, and the worldsheet crosscap in the pp-wave limit.
	
	}
	\date{}

	\maketitle
	
	\tableofcontents
	\newpage
	\section{Introduction}
Orientifolds are important yet enigmatic objects in string theory. In perturbative string theory, it is defined by gauging a combination of the worldsheet parity and the $\mathbb{Z}_2$-involution of the target spacetime. From the worldsheet perspective, this corresponds to inserting a so-called crosscap state on a worldsheet CFT (see e.g.~\cite{Sagnotti:1987tw,Ishibashi:1988kg,Fioravanti:1993hf,Sagnotti:1995ga,Huiszoon:2000ge,Fuchs:2000cm,Brunner:2002em,Gato-Rivera:2002whf,Fuchs:2003id,Harada:2025uhh} for constructions of crosscap states in CFT), which makes the worldsheet topology non-orientable. From the target space perspective, the orientifold creates an ``end-of-the-world'' wall in spacetime at the location of the fixed hypersurface of the $\mathbb{Z}_2$-involution. Such hypersurfaces are called orientifold planes, or orientifolds for short. 

Orientifolds share superficial similarities with D-branes, which are more widely discussed in the literature. Both insert special states on the worldsheet (crosscap states for orientifolds and boundary states for D-branes), and they are described by hypersurfaces in spacetime. However, in detail, there are notable differences:
\begin{itemize}
	\item D-branes host localized dynamical degrees of freedom, namely open strings, at their location. On the other hand, orientifolds do not come with any dynamical degrees of freedom. This property of orientifolds is in marked contrast with yet another similar object in string theory, namely orbifolds, which do introduce new degrees of freedom (i.e., closed strings in twisted sectors).
	\item D-branes are dynamical objects in string theory while orientifolds are not. One important consequence of this fact is that the positions of orientifolds are fixed in spacetime, unlike those of D-branes.
	\item The tension of orientifolds can be negative\fn{Nevertheless, this does not necessarily lead to instability of the entire spacetime (such as a vacuum decay) since the locations of orientifolds are fixed as mentioned above.} while the tension of D-branes is always positive.\fn{The so-called ghost D-branes or anti-branes discussed in \cite{Okuda:2006fb,Vafa:2001qf} do have negative tension. However, they are not considered to be physical objects in string theory.} Correspondingly, supergravity solutions describing orientifolds can have naked singularities.
\end{itemize}

This last property is particularly peculiar. On the one hand, this peculiarity allows one to evade the famous Maldacena-Nunez no-go theorem\footnote{The Maldacena-Nunez theorem states that, in classical supergravity with matter satisfying the positive energy condition, one cannot construct a solution describing de Sitter. Another (more popular) way to evade this theorem is to make use of the non-perturbative (instanton) effects as was done in the famous KKLT construction \cite{Kachru:2003aw}.} \cite{Maldacena:2000mw} against the construction of de Sitter vacua, and there have been various proposals in the literature which attempt to construct de Sitter vacua using orientifolds. On the other hand, one would normally think that objects with negative tension (or energy) in quantum theory are unphysical and would lead to inconsistency. This does not lead to an immediate problem since the aforementioned peculiarities (negative tension and naked singularity) arise only in some approximation (perturbative string theory and supergravity), and it is commonly believed that they will be resolved in the fully non-perturbative treatment. In fact, in some situations, non-perturbative resolutions of orientifold singularities are well-understood cf.~\cite{Sen:1996vd,Sen:1997pr,Seiberg:1996nz,Seiberg:1996bs}. However, in most cases --- especially the ones relevant for string compactification and the construction of de Sitter --- such non-perturbative descriptions are not fully understood, and it is important to develop tools to explore the fate of orientifolds at the non-perturbative level.

In this paper, we propose a setup that allows us to explore non-perturbative aspects of orientifolds using the powerful techniques of holography. Specifically, we consider $\mathcal{N}=4$ super Yang-Mills (SYM) in four dimensions, which is a prototypical example of the AdS/CFT correspondence, and place it on a non-orientable manifold $\mRP^4$. The $\mRP^4$ spacetime can be obtained from the Euclidean flat space $\mR^4$ by identifying points related by the (conformal) inversion. As shown in \cite{Caetano:2022mus}, by combining the inversion with a $\mathbb{Z}_2$-subgroup of the $R$-symmetry, one can preserve 16 super(conformal) charges and make it half-BPS. As compared to the analysis in \cite{Caetano:2022mus}, a novelty in this paper is that we combine the inversion also with the charge conjugation of the SU$(N)$ gauge group. This simple modification of the setup leads to several conceptual and technical differences, summarized below:
\begin{itemize}
	\item Without charge conjugation, one-point functions of single-trace operators on $\mathbb{RP}^4$ scale as $\propto N_c$ in the large $N_c$ limit \cite{Caetano:2022mus}. However, with charge conjugation, they are $O(1)$ even at large $N_c$. These different scalings result in distinct holographic realizations discussed below.
	\item $\mathbb{RP}^4$ without charge conjugation is dual to a nontrivial back-reacted geometry with an orientifold-like singularity, while $\mathbb{RP}^4$ with charge conjugation corresponds to a probe orientifold plane ($O1$) in AdS$_5\times$S$^5$. 
	\item $\mathbb{RP}^4$ without charge conjugation breaks planar integrability of $\mathcal{N}=4$ SYM while $\mathbb{RP}^4$ with charge conjugation preserves it. In the latter case, one-point functions show a parity selection rule and admit an analytic expression in terms of a ratio of Gaudin determinants.
\end{itemize}

\paragraph{Outline of the paper.} In {\bf Section} \ref{sec:setup}, we explain the setup, including how the supersymmetry (or more precisely the half-BPS condition) can be preserved by combining the conformal inversion with a $\mathbb{Z}_2$ subgroup of $R$-symmetry. We also contrast the setups with and without charge conjugation and explain how the propagators are affected by the charge conjugation. In {\bf Section} \ref{sec:weak}, we study one-point functions of single-trace operators at tree level in various subsectors and show that they can be represented as overlaps in integrable spin chains between energy eigenstates (i.e.~Bethe states) and crosscap states. The results show a parity selection rule, and we present closed-form formulae for these overlaps in terms of Gaudin determinants. Based on the results in the subsectors, we then conjecture a formula for the one-point functions at tree level applicable to general single-trace operators. 

In {\bf Sections}~\ref{sec:classical} and \ref{sec:pp-wave}, we study the strong-coupling regime using
holography. In {\bf Section}~\ref{sec:classical}, we review crosscap states on the classical
string worldsheet in $\AdS_5\times \text{S}^5$, following \cite{Gombor:2022gdj}, and identify the
crosscap relevant to our setup. In {\bf Section}~\ref{sec:pp-wave}, we consider the BMN limit
\cite{Berenstein:2002jq} and study orientifolds of string theory on the pp--wave background.
We construct the corresponding  crosscap states on the pp--wave explicitly and use them
to provide evidence for the proposed holographic duality between
$\mathcal N=4$ SYM on $\mathbb{RP}^4$ with charge conjugation and a probe
orientifold in $\AdS_5\times \text{S}^5$.

In {\bf Section}~\ref{sec:localization}, we turn to finite coupling and compute one-point functions
of half-BPS operators on $\mathbb{RP}^4$ using supersymmetric localization.
This extends the analysis of \cite{Wang:2020jgh} from $\mathbb{RP}^4$ without charge
conjugation to the theory with charge conjugation. By relating the problem to
two-dimensional Yang--Mills theory on $\mathbb{RP}^2$ with charge conjugation,
and restricting to the zero-instanton sector, we derive a matrix model that
computes the non-perturbative one-point functions at finite $N$ and
$g_{\rm YM}$. We then take the planar limit and obtain a closed-form result at
finite 't~Hooft coupling $\lambda$. The answer interpolates between the known
leading weak- and strong-coupling limits and predicts the corresponding
subleading corrections.

Finally, in {\bf Section} \ref{sec:conclusion}, we conclude and discuss future directions. Several appendices are included to explain technical details.
\section{\texorpdfstring{$\mathcal{N}=4$}{N=4} SYM on \texorpdfstring{$\mathbb{RP}^{4}$}{RP4}}\label{sec:setup}
\subsection{CFTs on \texorpdfstring{$\mRP^4$}{RP4}}
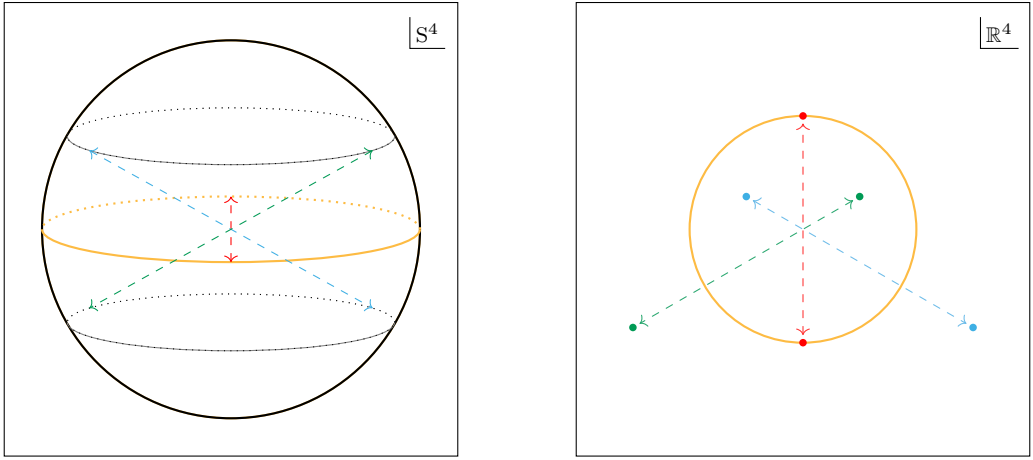
\begin{figure}
    \centering
    \begin{minipage}{0.45\textwidth}
        \centering
\begin{tikzpicture}
    \def\R{2.5}         
    \def\tilt{10}     
    \def\lat{30}      
    \def\offset{30}
    \def\rotb{\offset}
    \def\rotg{\offset+120}
    \def\rotr{\offset+240}
    \def\side{3}

    \pgfmathsetmacro{\rxy}{\R*cos(\lat)}
    \pgfmathsetmacro{\rz}{\R*cos(\lat)*sin(\tilt)}
    \pgfmathsetmacro{\yc}{\R*sin(\lat)*cos(\tilt)}
    \draw[draw=black, fill=white!50] ({-\side},{-\side}) rectangle ({\side},{\side});
    \draw[Dandelion,thick] (0,0) circle(\R);
    \draw[thick] (0,0) circle(\R);

    \draw[Dandelion,dotted,thick] (\R,0) arc(0:180:{\R} and {\R*sin(\tilt)});
    \draw[Dandelion,thick] (\R,0) arc(0:-180:{\R} and {\R*sin(\tilt)});
    \draw[black!50,dotted,thin] (\rxy,\yc) arc(0:180:{\rxy} and {\rz});
    \draw[black!50,thin] (\rxy,\yc) arc(0:-180:{\rxy} and {\rz});
    \draw[black!50,dotted,thin] (\rxy,-\yc) arc(0:180:{\rxy} and {\rz});
    \draw[black!50,thin] (\rxy,-\yc) arc(0:-180:{\rxy} and {\rz});
    \draw[dotted] (0,\yc) ellipse ({\rxy} and {\rz});
    \draw[dotted] (0,-\yc) ellipse ({\rxy} and {\rz});

    \draw[<->,red,dashed] ({\R*cos(\rotr)},{\R*sin(\rotr)*sin(\tilt)})->({-\R*cos(\rotr)},{-\R*sin(\rotr)*sin(\tilt)});

    \draw[<->,CornflowerBlue,dashed] ({-\rxy*cos(\rotb)},{-\rz*sin(\rotb)+\yc})->({\rxy*cos(\rotb)},{\rz*sin(\rotb)-\yc});
    
    \draw[<->,ForestGreen,dashed] ({-\rxy*cos(\rotg)},{-\rz*sin(\rotg)+\yc})->({\rxy*cos(\rotg)},{\rz*sin(\rotg)-\yc});
    \node (box) [
        font=\scriptsize, 
        minimum width=0.4cm, 
        minimum height=0.4cm, 
        inner sep=2pt 
    ] at (2.6,2.6) {\scriptsize $\text{S}^4$};
    
    \draw[] (box.north west) -- (box.south west) -- (box.south east);
\end{tikzpicture}
\end{minipage}
\begin{minipage}{0.45\textwidth}
        \centering
\begin{tikzpicture}
    \def\R{1.5} 
    \def\side{3}
    \def\linR{1.4}
    \def\tilt{10}     
    \def\lat{30}      
    \def\offset{30}
    \def\rotb{\offset}
    \def\rotg{\offset+180}
    \def\rotr{\offset+240}

    \pgfmathsetmacro{\rxy}{\R*cos(\lat)}
    \pgfmathsetmacro{\rz}{\R*cos(\lat)*sin(\tilt)}
    \pgfmathsetmacro{\yc}{\R*sin(\lat)*cos(\tilt)}

    \draw[draw=black, fill=white!50] ({-\side},{-\side}) rectangle ({\side},{\side});
    \draw[Dandelion,thick] (0,0) circle(\R);

    \draw[<->,red!80,dashed] ({\linR*cos(\rotr)},{\linR*sin(\rotr)})->({-\linR*cos(\rotr)},{-\linR*sin(\rotr)});
    \draw[red]  ({\R*cos(\rotr)},{\R*sin(\rotr)}) node[circle,fill,inner sep=1 pt]{};
    \draw[red]  ({-\R*cos(\rotr)},{-\R*sin(\rotr)}) node[circle,fill,inner sep=1 pt]{};

    \draw[<->,CornflowerBlue!80,dashed] (-{(\R*tan((90-\lat)/2)-0.1)*cos(\rotb)},{(\R*tan((90-\lat)/2)-0.1)*sin(\rotb)})->({(\R*tan((90+\lat)/2)-0.1)*cos(\rotb)},{-(\R*tan((90+\lat)/2)-0.1)*sin(\rotb)}) ;
    \draw[CornflowerBlue]  ({-\R*tan((90-\lat)/2)*cos(\rotb)},{\R*tan((90-\lat)/2)*sin(\rotb)}) node[circle,fill,inner sep=1 pt]{};
    \draw[CornflowerBlue]  ({\R*tan((90+\lat)/2)*cos(\rotb)},{-\R*tan((90+\lat)/2)*sin(\rotb)}) node[circle,fill,inner sep=1 pt]{};

    \draw[<->,ForestGreen!80,dashed] ({-(\R*tan((90-\lat)/2)-0.1)*cos(\rotg)},{-(\R*tan((90-\lat)/2)-0.1)*sin(\rotg)})->({(\R*tan((90+\lat)/2)-0.1)*cos(\rotg)},{(\R*tan((90+\lat)/2)-0.1)*sin(\rotg)});
    \draw[ForestGreen]  ({-\R*tan((90-\lat)/2)*cos(\rotg)},{-\R*tan((90-\lat)/2)*sin(\rotg)}) node[circle,fill,inner sep=1 pt]{};
    \draw[ForestGreen]  ({\R*tan((90+\lat)/2)*cos(\rotg)},{\R*tan((90+\lat)/2)*sin(\rotg)}) node[circle,fill,inner sep=1 pt]{};

    \node (box) [
        font=\scriptsize, 
        minimum width=0.4cm, 
        minimum height=0.4cm, 
        inner sep=2pt 
    ] at (2.6,2.6) {\scriptsize $\mathbb{R}^4$};
    
    \draw[] (box.north west) -- (box.south west) -- (box.south east);
\end{tikzpicture}
\end{minipage}
\caption{The identification of points necessary when defining theories on projective space: antipodal identification on $\text{S}^4$
(left) and its stereographic image $x\sim-x/|x|^2$
(right). The yellow equator maps to the unit three-sphere in the flat conformal
frame. Dashed segments connect pairs of identified points.}
    \label{fig:RP4}
\end{figure}

Let us first briefly review the basics of CFTs on $\mathbb{RP}^4$, which has previously been considered in \cite{Nakayama:2016cim,Hasegawa:2016piv,Hogervorst:2017kbj,Hasegawa:2018yqg,Giombi:2020xah}. One can obtain $\mRP^4$ from $\text{S}^4$ by identifying points by the antipodal $\bZ_2$ map $\iota$. Using spherical coordinates with azimuthal angle $\theta\in[0,2\pi)$ and polar angles $\varphi_i\in[0,\pi)$, $\iota$  acts as the map
\begin{equation}
	\iota(\theta)=\theta+\pi,\qquad \iota(\varphi_i)=\pi-\varphi_i\comma \qquad \mRP^4=S^4/\iota\,.\label{eq:roundiota}
\end{equation}
This map is illustrated in Figure \ref{fig:RP4}. 

In (Euclidean) conformal field theories, which are invariant under 
$\text{O}^+(5,1)$,\footnote{Here, $\text{O}^+$ denotes the \textit{orthochronous} component of the disconnected group of special orthogonal transformations, namely the elements that preserve the positivity of the time-direction. This includes in particular parity and thus also conformal inversion.} one can use stereographic projection to map $\text{S}^{4}$ to flat space $\mathbb{R}^4$. The antipodal map
\eqref{eq:roundiota} becomes the conformal inversion on $\mathbb{R}^4$, which we also denote by $\iota$ (see Figure \ref{fig:RP4})
\begin{equation}
	x_\mu\mapsto\iota(x_\mu)= x_\mu'=-\frac{x_\mu}{|x|^2}\,.\label{eq:flatiota}
\end{equation}
Since the analysis in $\mathbb{R}^4$ is more familiar, we work with $\mathbb{R}^4/\iota$ in the rest of this paper.
The theory on $\mathbb{RP}^4$ no longer has the full conformal symmetry $\text{O}^{+}(5,1)$, and the symmetry group is reduced to the isometries of $\mathbb{RP}^4$, namely $\SO(5)$. This breaking of symmetry relaxes some of the constraints that are normally placed on the correlation functions, in a very similar manner to the more familiar case of boundary CFTs \cite{Liendo:2012hy,Gliozzi:2015qsa,Billo:2016cpy}. One-point functions are no longer zero for scalar operators, instead taking the form
\begin{equation}
    \left<\mathcal{O}_i(x)\right>=\frac{A_i}{(1+x^2)^{\Delta_i}}\,,\label{eq:1ptRP}
\end{equation}
where $A_i$'s are new CFT data for a theory on $\mathbb{RP}^4$.
Furthermore, a \textit{cross ratio} $\eta$ appears already at the level of the two-point function, which for scalar operators $\mathcal{O}_{i,j}$ takes the form
\begin{equation}
    \left<\mathcal{O}_i(x)\mathcal{O}_j(y)\right>=\frac{G_{ij}(\eta)}{(1+x^2)^{\Delta_i}(1+y^2)^{\Delta_j}}\label{eq:1pt2pt}
\end{equation}
where
\begin{equation}
\eta(x,y)=\frac{(x-y)^2}{(1+x^2)(1+y^2)}\label{eq:etadef}\,,
\end{equation}
is an invariant under $\SO(5)$ and transforms under the inversion as $\eta\left(\iota(x),y\right)=1-\eta(x,y)$..

Note that, when defining the action of $\iota$, one can combine it with other $\mathbb{Z}_2$ (internal) symmetries of the theory (cf.\cite{Drukker:2026nvv}). This leads to a richer class of theories on $\mathbb{RP}^{4}$ and different choices of $\iota$ lead to different conformal data \eqref{eq:1pt2pt}. In this generalized setting, $\iota$ acts not only on spacetime but also on the operators themselves. For example, for a scalar operator $\mathcal{O}$ with conformal dimension $\Delta$, we have
\begin{equation}
    \iota(\mathcal{O}(x))=\frac{\mathcal{O}'(x')}{|x|^{2\Delta}}\,.\label{eq:iotaonO}
\end{equation}
We require $\iota^2=1$ and thus $\mathcal{O}''=\mathcal{O}$, which gives a $\mathbb{Z}_2$-grading of the operators into $\iota$-even/odd sectors.
\subsection{\texorpdfstring{$\mathcal{N}=4$}{N=4} SYM with \texorpdfstring{$SU(N)$}{SU(N)} gauge group on \texorpdfstring{$\mathbb{RP}^4$}{RP4}}\label{sec:N=4SYM}
With the general overview out of the way, we are now ready to focus on Euclidean $\mathcal{N}=4$ SYM in particular. Since we will need to discuss the symmetries of the theory on $\mathbb{RP}^4$, we will first review the fields and symmetries of Euclidean $\mathcal{N}=4$ SYM on $\mathbb{R}^4$ in some detail.

Just like its more familiar Lorentzian cousin, Euclidean $\mathcal{N}=4$ SYM can be derived from a dimensional reduction of $9+1$ dimensional $\mathcal{N}=1$ SYM. The difference is the choice of dimensions that are reduced: in the Lorentzian theory, we reduce over 6 space-like directions, while to get Euclidean $\mathcal{N}=4$ SYM we reduce $(5+1)$ dimensions and the $R$-symmetry rotating these is thus $\text{Spin}(5,1)_R\cong \SU^{\ast}(4)_R$. The remaining 4 extended directions have a rotation group $\text{Spin}(4)\cong \SU(2)\times\dot{\SU}(2)$. Since the theory is conformal, the latter is promoted to the Euclidean conformal group $\SO(5,1)$. With the theory furthermore being supersymmetric, the total symmetry supergroup of the theory is $\text{PSU}^{\ast}(4|4)$. There is an additional $\SU(N)$ gauge symmetry: since we want to eventually use integrability, we will be interested in taking the limit $N\rightarrow \infty$ .

The field content thus consists of 1 gauge field $A_\mu$, $6$ scalars $\phi^I$, $4$ chiral fermions $\psi^a_\alpha$ and $4$ anti-chiral fermions $\bar{\psi}_{\dot{\alpha}a}$, all in the adjoint representation of the gauge group $\SU(N)$. Here, $\mu=1,\dots 4$ is an index transforming in the fundamental of $\text{SO}(4)$, $\alpha=1,2$ transforms in the fundamental of $\SU(2)$ and $\dot{\alpha}=1,2$ transforms in the fundamental of $\dot{\SU}(2)$. The additional labels transform under the $R$-symmetry:
$I=0,5,\dots,9$ transforms in the fundamental of $\text{SO}(5,1)_R$ and $a=0,\dots 3$ as a raised/lowered index transforms in the fundamental/anti-fundamental of $\text{SU}^*(4)_R$. We will sometimes collect the fermions together into the original 10D chiral fermion $\Psi$, with the location of the indices indicating which four-dimensional field component we are considering. The action, in the absence of a $\theta$-term, is given by
\begin{equation}
    S=\frac{1}{2g_{\rm YM}^2}\int\Dd^4x\,\text{Tr}\left(\frac{1}{2}F_{\mu\nu}F^{\mu\nu}+D_\mu\phi_I D^\mu\phi^I+i\bar{\Psi}\Gamma_\mu D^\mu\Psi-\frac{1}{2}[\phi_I,\phi_J][\phi^I,\phi^J]+i\bar{\Psi}\Gamma_I [\phi^I,\Psi]\right)\label{eq:N4SYMAction}
\end{equation}
where the trace is taken over $\SU(N)$ and $\Gamma_\mu/\Gamma_I$ are $\text{Spin}(4)/\text{Spin}(5,1)_R$ $\Gamma$-matrices.
\paragraph{Discrete symmetries of $\mathcal{N}=4$ SYM.}
As we will be taking $\mathbb{Z}_2$ quotients of the theory, it is important to understand its discrete symmetries. These were beautifully summarized in \cite{Gomis:2025gzb}, and we will refer to that work for additional details. 

The Lorentz-invariant $\mathcal{N}=4$ SYM is, like any QFT, $\mathsf{CPT}$ symmetric. However, the symmetry of the theory under each of the conjugations individually is more subtle \cite{Gomis:2025gzb}. For the purpose of this paper, $\mathsf{T}$-symmetry is not relevant, and we focus on $\mathsf{C}$- and $\mathsf{P}$-symmetries:
\begin{itemize}
    \item $\mathsf{C}$-symmetry: The existence of charge conjugation symmetry relies on having a non-trivial outer automorphism of the gauge group: for $\text{SU}(N)$, this leads to a $\mathsf{C}$-symmetry. As argued in \cite{Henning:2021ctv}, there are, in fact, two distinct charge conjugations in $\text{SU}(N)$ gauge theories. As we shall see later, both will play an important role.
    \item $\mathsf{P}$-symmetry: $\mathcal{N}=4$ SYM is, \textit{not} parity invariant in the usual sense. The simplest way to see this is that $\mathcal{N}=4$ SYM is a dimensional reduction of the ten-dimensional $\mathcal{N}=1$ SYM theory, which is chiral and thus not parity invariant. Since parity and inversion differ only by a conformal transformation, this also implies that $\mathcal{N}=4$ SYM is not inversion invariant.
    
    However, the connection to ten dimensions also suggests a way to restore the inversion invariance \cite{Gomis:2025gzb}. Consider an $R$-symmetry reflection $\mathsf R$ that flips an odd number of the six $R$-symmetry directions. Like the inversion $\mathsf I$, such a reflection is not a symmetry of $\mathcal{N}=4$ SYM. The combined transformation $\mathsf I\times\mathsf R$, however, is a symmetry both in the ten-dimensional theory and in $\mathcal{N}=4$ SYM. At this stage, the argument fixes only the number of reflected $R$-symmetry directions to be odd; as we will see below, supersymmetry selects the specific choice of $\mathsf R$. In the rest of this paper, we call the combined transformation $\mathsf{I}\times \mathsf{R}$ ``parity'', and denote it by $\mathsf{P}$.
\end{itemize}
Using these two discrete symmetries, one can define two different $\mathbb{RP}^4$ for $\mathcal{N}=4$ SYM; either choosing the quotient map $\iota$ to be just $\mathsf{P}=\mathsf{I}\times \mathsf{R}$ or to be $\mathsf{CP}=\mathsf{C}\times \mathsf{I} \times \mathsf{R}$. The former choice (``$\mathsf{P}$-$\mathbb{RP}^4$'') was discussed in \cite{Caetano:2022mus} and has been shown to break integrability and to be holographically dual to a nontrivial backreacted geometry. In this paper, we instead focus on the latter (``$\mathsf{CP}$-$\mathbb{RP}^4$'').

Below, we discuss each of the $\mathbb{Z}_2$ maps in $\mathsf{CP}=\mathsf{C}\times \mathsf{I}\times\mathsf{R}$ and clarify their action on fundamental fields of $\mathcal{N}=4$ SYM.

\paragraph{Space-time inversion.}
First, we consider the transformation of the space-time indices under $\mathsf{I}$. Since the scalars do not carry such indices, they transform according to
\begin{equation}\mathsf{I}\Big(\phi^I(x)\Big)=\frac{\phi^I(x')}{|x|^2}\,,\label{eq:sctr}
\end{equation}
where $x'$ is given by the inversion $\eqref{eq:flatiota}$ and the specific power of $|x|$ follow from the conformal dimension of $\phi$. Next, the transformation of the gauge field is given by \cite{Wang:2020jgh}
\begin{equation}
	\mathsf{I}\Big(A_\mu(x)\Big)=-\frac{1}{|x|^2}I_{\mu\nu}(x)\,A_\nu(x')\,,\quad I_{\mu\nu}(x)\equiv\delta_{\mu\nu}-\frac{2x_\mu x_\nu}{|x|^2}\,.\label{eq:vetr}
\end{equation}
Lastly, the action on the fermions can be found by requiring that, if $\psi$ and $\chi$ are two spinors, the bi-spinor $\bar{\chi}\Gamma_\mu\psi$ should transform as a vector with conformal weight $3$. When acting on the 4D chiral fermion components $\Psi^a_{\alpha}$ of the ten-dimensional fermion, the resulting action is
\begin{equation}
	\mathsf{I}\Big(\Psi^a_{\alpha}(x)\Big)=i\frac{x^\mu}{|x|^4}[\Gamma_\mu]_{\alpha\dot{\beta}}\,\Psi^{a\dot{\beta}}(x')\label{eq:sptr}
\end{equation} 
where $\Gamma_\mu$ is one of the four-dimensional $\Gamma$-matrices. Similar expressions are true for other constellations of the indices. Note that in the ten-dimensional theory the pullback of $\mathsf{I}$ can be seen as a type of parity transformation and thus flips the chirality of $\Psi$, which does not preserve the field content of the theory. In four dimensions, this corresponds to the fact that the fermion on the right-hand side does not have the correct placement of indices to correspond to either a chiral or anti-chiral fermion.
\paragraph{$R$-symmetry reflection.}
Next, we specify the action $\mathsf{R}$ on the internal $R$-symmetry labels. In order for the combined action of inversion and the $R$-symmetry action to be a symmetry, $\mathsf{R}$ must be in the disconnected component of $\text{O}^+(5,1)$ \cite{Gomis:2025gzb}.
This still leaves us with a plethora of choices of $\mathsf{R}$, but the resulting symmetry group will in general have no supersymmetry.
In this paper, we are interested in finding $\mathbb{RP}^4$ that preserves as much supersymmetry as possible. Since we know from the breaking of conformal symmetry that the supersymmetry group should contain an $\SO(5)$ factor, we can look for 1/2-BPS subgroups of $\text{PSU}^{\ast}(4|4)$ whose bosonic subgroup has such a factor. There is precisely one, namely $\text{OSp}(4^*|4)$, whose corresponding $R$-symmetry is $\SO(2,1)\times \SO(3)$ \cite{DHoker:2008wvd}. While it is not obvious that this supergroup can actually be achieved as the normalizer of some $\mathbb{Z}_2$ element $\mathsf{I}\times\mathsf{R}$, this turns out to be the case as already found in \cite{Wang:2020jgh,Caetano:2022mus}: $\mathsf{R}\in \text{O}^+(5,1)$ can be chosen to be represented in the fundamental representation as the matrix $\diag(1,-1,1,-1,1,-1)$, whose normalizer is precisely $\SO(2,1)\times \SO(3)$. This choice of $\mathsf{R}$ implies that the scalars transform under $\mathsf{R}$ as
\begin{equation}
    \mathsf{R}\Big(\phi_I\Big)=(-1)^I\phi_I\,.\label{eq:scP}
\end{equation}
The vector $A_\mu$ is scalar under $R$-symmetry and thus is not affected by $\mathsf{R}$. Lastly, the fermions transform according to
\begin{equation}
    \mathsf{R}\Big(\Psi^a_{\alpha}\Big)=[\Gamma_\mathsf{R}]^{ab}_{\phantom{ab}\alpha\beta}\Psi_{b}^{\beta}\comma \Gamma_{\mathsf{R}}\equiv\Gamma_0\,\Gamma_6\,\Gamma_8,\label{eq:spP}
\end{equation}
where $\Gamma_I$ are the $R$-symmetry $\Gamma$-matrices. 
\paragraph{Charge conjugation.}
Lastly, we specify how $\iota$ acts on $\SU(N)$ gauge labels. We will let $\iota$ act by charge conjugation $\mathsf{C}$, \textit{i.e.} $\iota$ is an outer automorphism of $\SU(N)$. For odd $N$, there is a unique choice of such charge conjugation on elements $M\in\mathfrak{su}(N)$, given by
\begin{equation}
	M^\mathsf{C}\equiv-M^{\mathsf{T}}\label{eq:CO}
\end{equation}
where $\mathsf{T}$ denotes the transpose. If $N$ is even, there is an additional possible choice of charge conjugation given by \cite{Henning:2021ctv}
\begin{equation}
	M^{\mathsf{C}}\equiv-\mathbb{J}_NM^{\mathsf{T}}\mathbb{J}_N^{-1},\label{eq:Cp}
\end{equation}
where $\mathbb{J}_N$ is the symplectic $N\times N$ given by 
\begin{equation}
	\mathbb{J}_N=\begin{pmatrix}
		0&+\mathbb{I}_{N/2}\\-\mathbb{I}_{N/2}&0
	\end{pmatrix}\comma \qquad \mathbb{J}_N^2=-\mathbb{I}_{N},\label{eq:defJ}
\end{equation}
and $\mathbb{I}_N$ is the $N\times N$ identity matrix. For brevity, we write both charge conjugations in the form 
\begin{equation}
    M^{\mathsf{C}}=-\mathbb{K}M^{\mathsf{T}}\mathbb{K}^{-1}\,,\qquad \mathbb{K}^\mathsf{T}=\mathbb{K}^{-1}=\varepsilon_{\mathbb{K}}\mathbb{K}\label{eq:C}
\end{equation}
with $\mathbb{K}$ given by either $\mathbb{I}_N$ or $\mathbb{J}_N$, corresponding to \eqref{eq:CO} and \eqref{eq:Cp} respectively, and $\varepsilon_{\mathbb{I}}=-\varepsilon_{\mathbb{J}}=1$. Note that although $\frac{\SU(N)}{\text{charge conjugation}}$ leads to the groups $\SO(N)$ for \eqref{eq:CO} and $\text{USp}(N)$ for \eqref{eq:Cp} \cite{Henning:2021ctv}, the identification of fields above is happening at different points in space. Thus, the local gauge algebra is still $\mathfrak{su}(N)$, but the global bundle of $\mathcal{N}=4$ SYM on $\mathbb{RP}^4$ is twisted by $\iota$.

To define the complete action of $\iota$ on the fields, we combine the three transformations we have just defined $\iota=\mathsf{C}\times \mathsf{I}\times\mathsf{R}$. For the scalar, we have to perform \eqref{eq:sctr}, \eqref{eq:scP} and \eqref{eq:C} at the same time, such that
\begin{equation}\iota\Big(\phi_I(x)\Big)=-\frac{1}{|x|^2}(-1)^I\mathbb{K}\phi^{\mathsf{T}}_I(x')\mathbb{K}^{-1}\,.\label{eq:iotaphi}
\end{equation}
For the vectors, the same transformation takes the form
\begin{equation}
    \iota\Big(A_\mu(x)\Big)=\frac{1}{|x|^2}I_{\mu}^{\phantom{\mu}\nu}\mathbb{K}A_\nu^{\mathsf{T}}(x')\mathbb{K}^{-1}\,,\label{eq:iotaA}
\end{equation}
and for the fermions it is given by
\begin{equation}
    \iota\Big(\Psi_\alpha^a(x)\Big)=-i\frac{x_\mu}{|x|^4}[\Gamma_\mu\Gamma_\mathsf{R}]_{\phantom{ab}\alpha}^{ab\phantom{\alpha}\dot{\beta}}\mathbb{K}\Psi_{b\dot{\beta}}^{\mathsf{T}}(x')\mathbb{K}^{-1}\,,\label{eq:iotapsi}
\end{equation}
where $\Gamma_{\mathsf{R}}$ was defined in \eqref{eq:spP}. Indeed, this shows that by combining \eqref{eq:sptr} and \eqref{eq:spP}, we get a fermion with the right index structure. In particular, $\iota$ flips the chirality of the four-dimensional fermions while also switching the $R$-symmetry representation to its dual and thus maps $\psi$ to $\bar{\psi}$ and vice versa. 
\subsection{Propagators on \texorpdfstring{$\mRP^4$}{RP4}}
Now let us understand what the mappings \eqref{eq:iotaphi}, \eqref{eq:iotaA} and \eqref{eq:iotapsi} tell us about the structure of the propagators in the quotiented theory. Recall that on $\mathbb{R}^4$, the propagators take the form\
\begin{equation}
\begin{split}
\left<\left[\phi_I(x)\right]_{pq},\left[\phi^J(y)\right]_{rs}\right>_{\mathbb{R}^4}&=\frac{g_{\text{YM}}^2}{4\pi^2} \delta_I^J\,\frac{\delta_{ps}\delta_{qr}-\frac{1}{N}\delta_{pq}\delta_{rs}}{(x-y)^2}\,,
\\
	\left<\left[A_\mu(x)\right]_{pq},\left[A_\nu(y)\right]_{rs}\right>_{\mathbb{R}^4}&=\frac{g_{\text{YM}}^2}{8\pi^2} \left\{\frac{(\partial_\mu^x\eta)(\partial_\nu^y\eta)}{\eta^2}-\frac{2(\partial_\mu^x\partial_\nu^y\eta)}{\eta}\right\}\left[\delta_{ps}\delta_{qr}-\frac{1}{N}\delta_{pq}\delta_{rs}\right]\,,
\\\left<\left[\Psi(x)\right]_{pq},\big[\bar{\Psi}(y)\big]_{rs}\right>_{\mathbb{R}^4}&=-\frac{g_{\text{YM}}^2}{2\pi^2} (x-y)\cdot\Gamma\,\frac{\delta_{ps}\delta_{qr}-\frac{1}{N}\delta_{pq}\delta_{rs}}{(x-y)^4}
\end{split}\label{eq:propbefore}
\end{equation}
where the $p,q,r,s$ indices transform in (anti)fundamental representation of $\SU(N)$, $\eta$ was defined in \eqref{eq:etadef} and $x\cdot \Gamma\equiv x^\mu\Gamma_\mu$. We will be interested in the $N\rightarrow \infty$ limit in the following, and we will thus drop the $1/N$ term from now on. As reviewed in Appendix \ref{sec:Gauge}, we are choosing to work in spherical Fried-Yennie gauge\footnote{See \cite{Ciccone:2024guw,Ankur:2026ylr,Moga:2025gdy} for recent applications of Fried-Yennie gauge in other contexts.} \cite{Fried:1958zz,Yennie:1961ad} for the gauge fields, since that choice leads to a simpler expression for the propagators on $\mathbb{RP}^4$.

To determine the propagators on $\mRP^4$, we use ``the method of images''; namely, we add to the corresponding propagators on $\mathbb{R}^4$  their images under $\iota$. Concretely, for the propagators between two fields $\chi_1,\chi_2$, we write
 \begin{equation}
	\left<\chi_1(x),\chi_2(y)\right>_{\mathbb{RP}^4}=\left<\chi_1(x),\chi_2(y)\right>_{\mathbb{R}^4}+\left<\iota\Big(\chi_1(x)\Big),\chi_2(y)\right>_{\mathbb{R}^4}\,.\label{eq:methodofim}
\end{equation}
The resulting expressions satisfy all the properties required for them to be propagators on $\mathbb{RP}^4$: they reduce to propagators on $\mathbb{R}^4$ when $x\sim y$, and they are invariant under the action of $\iota$. 

Using this and the cross ratio defined in \eqref{eq:etadef}, the propagators on $\mRP^4$ can be written down as follows:
\begin{equation}\label{eq:propagators}
	\begin{split}
\left<\left[\phi_I(x)\right]_{pq},\left[\phi^J(y)\right]_{rs}\right>_{\mathbb{RP}^4}=& \frac{g_{\text{YM}}^2}{4\pi^2} \frac{\delta_I^J}{(1+x^2)(1+y^2)}\Bigg\{\frac{\delta_{ps}\delta_{qr}}{\eta}-(-1)^I\frac{\mathbb{K}_{pr}\mathbb{K}_{sq}^{-1}}{1-\eta}\Bigg\}\,,
\\\left<\left[A_\mu(x)\right]_{pq},\left[A_\nu(y)\right]_{rs}\right>_{\mathbb{RP}^4}=&\frac{g_{\text{YM}}^2}{8\pi^2}\Bigg\{\left[\frac{\delta_{ps}\delta_{qr}}{\eta^2}-\frac{\mathbb{K}_{pr}\mathbb{K}_{sq}^{-1}}{(1-\eta)^2}\right](\partial^x_\mu\eta)(\partial_\nu^y\eta)\\&-2\left[\frac{\delta_{ps}\delta_{qr}}{\eta}+\frac{\mathbb{K}_{pr}\mathbb{K}_{sq}^{-1}}{1-\eta}\right](\partial_\mu^x\partial_\nu^y\eta)\Bigg\}\,,
\\
\left<\left[\Psi(x)\right]_{pq},\big[\bar{\Psi}(y)\big]_{rs}\right>_{\mathbb{RP}^4}=&-\frac{g_{\text{YM}}^2}{2\pi^2}\frac{1}{(1+x^2)^{3/2}(1+y^2)^{3/2}}\Bigg\{\frac{(x-y)\cdot\Gamma}{|x-y|}\frac{\delta_{ps}\delta_{qr}}{\eta^{3/2}}
\\
&-i\frac{1+x\cdot \Gamma\,y\cdot \Gamma}{\sqrt{1+x^2y^2+2x\cdot y}}\Gamma_{\mathsf{R}}\frac{\mathbb{K}_{pr}\mathbb{K}_{sq}^{-1}}{(1-\eta)^{3/2}}\Bigg\}\,,
 \end{split}
\end{equation}
To avoid having to write the gauge indices explicitly, we will use double line notation, in which the transposition in the charge conjugation \eqref{eq:C} acts by twisting the two lines, such that the propagator of two scalar fields can, for example, be written as
\begin{equation}
\langle \phi_I(x), \phi^J(y) \rangle = \frac{g^2_{\text{YM}}}{4\pi^2} \frac{\delta_I^J}{(1+x^2)(1+y^2)} \left\{ 
    \frac{1}{\eta} \,
\propDirect
    -(-1)^I \frac{1}{1-\eta}\,\label{eq:proptwist}
    \propCross
\right\}\,,
\end{equation}
where we should remember that we should act with $\mathbb{K}$ after the twist. 
\subsection{One-point functions on \texorpdfstring{$\mathbb{RP}^4$}{RP4}}
We now compute the vacuum expectation values (VEVs) of single-trace operators at the lowest order in the coupling $g_{\text{YM}}$, which are the simplest observables in the theory on $\mathbb{RP}^4$. As mentioned earlier, the symmetry is enough to fully determine the spacetime dependence of the one-point functions \eqref{eq:1ptRP}. Since we are working at tree level, the computation only involves Wick contractions, whose contributions can be read off from the propagators \eqref{eq:propagators} by taking the limit $x\rightarrow y$. The first term in each propagator diverges like $1/(x-y)^{\alpha}$ for some $\alpha\geq 2$. However, this divergence is identical to the coincident-point divergence of the propagator in flat space (since locally the theory looks the same) and is removed by normal orderings. It is instead the term with crossed lines \eqref{eq:proptwist} in each of the propagators that gives a nontrivial contribution. We choose a matrix realisation in which $i[\Gamma_{\mathsf{R}}]^{ab}_{\phantom{ab}\alpha\beta}=(-1)^a\delta^{ab}\varepsilon_{\alpha\beta}$, and we get
\begin{equation}
\begin{split}
    \left<:\phi_I\phi^J:(x)\right>=&-\frac{g^2_{\text{YM}}}{4\pi^2(1+x^2)^2}\delta_I^J(-1)^I\propCross\,,
    \\
    \left<:A_\mu A_\nu:(x)\right>=&\frac{g^2_{\text{YM}}}{2\pi^2(1+x^2)^2}\delta_{\mu\nu}\propCross\,,
    \\
    \left<:\psi^a_\alpha\psi^b_{\beta}:(x)\right>=&\frac{g^2_{\text{YM}}}{2\pi^2(1+x^2)^3}\delta^{ab}(-1)^a\varepsilon_{\alpha\beta}\propCross\,.\label{eq:Wick}
    \end{split}
\end{equation}

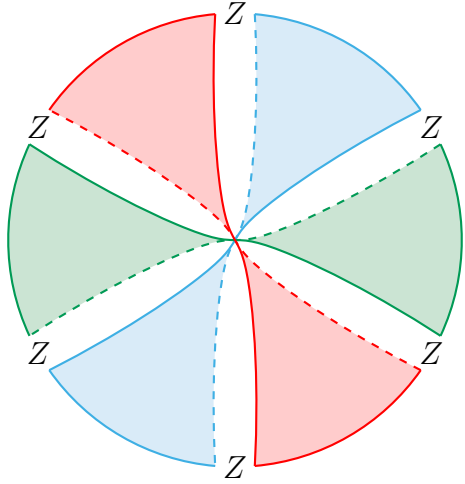
\begin{figure}[t]
	\centering
	\begin{tikzpicture}
		\def\t{60}
        \def\toff{5}
        \def\troff{15}
        \def\roff{0.8}
       \def\R{3}
		\draw ({\R*sin(0)},{\R*cos(0)}) node[]{$Z$};
		\draw ({\R*sin(\t)},{\R*cos(\t)})
		node[]{${Z}$};
		\draw ({\R*sin(2*\t)},{\R*cos(2*\t)})
		node[]{$Z$};
		\draw ({\R*sin(3*\t)},{\R*cos(\R*\t)})
		node[]{${Z}$};
		\draw ({\R*sin(4*\t)},{\R*cos(4*\t)})
		node[]{$Z$};
		\draw ({\R*sin(5*\t)},{\R*cos(5*\t)})
		node[]{${Z}$};

          \fill[red!20] 
        plot [smooth, tension=1] coordinates {
            ({\R*sin(-\toff)},{\R*cos(-\toff)})
            ({\roff*sin(-\troff)},{\roff*cos(-\troff)})
            ({\roff*sin(3*\t-\troff)},{\roff*cos(3*\t-\troff)})
            ({\R*sin(3*\t-\toff)},{\R*cos(3*\t-\toff)})
        }
        arc (-85:-35:3) 
        plot [smooth, tension=1] coordinates{
        ({\R*sin(2*\t+\toff)},{\R*cos(2*\t+\toff)})
        ({\roff*sin(2*\t+\troff)},{\roff*cos(2*\t+\troff)})
        ({\roff*sin(5*\t+\troff)},{\roff*cos(5*\t+\troff)})
        ({\R*sin(5*\t+\toff)},{\R*cos(5*\t+\toff)})}
        arc (145:95:3) 
        ;

                 \fill[CornflowerBlue!20] 
        plot [smooth, tension=1] coordinates {
            ({\R*sin(1*\t-\toff)},{\R*cos(1*\t-\toff)})
            ({\roff*sin(1*\t-\troff)},{\roff*cos(1*\t-\troff)})
            ({\roff*sin(4*\t-\troff)},{\roff*cos(4*\t-\troff)})
            ({\R*sin(4*\t-\toff)},{\R*cos(4*\t-\toff)})
        }
        arc (-145:-95:3) 
        plot [smooth, tension=1] coordinates{
        ({\R*sin(3*\t+\toff)},{\R*cos(3*\t+\toff)})
        ({\roff*sin(3*\t+\troff)},{\roff*cos(3*\t+\troff)})
        ({\roff*sin(0*\t+\troff)},{\roff*cos(0*\t+\troff)})
        ({\R*sin(0*\t+\toff)},{\R*cos(0*\t+\toff)})}
        arc (85:35:3) 
        ;
                  \fill[ForestGreen!20] 
        plot [smooth, tension=1] coordinates {
            ({\R*sin(2*\t-\toff)},{\R*cos(2*\t-\toff)})
            ({\roff*sin(2*\t-\troff)},{\roff*cos(2*\t-\troff)})
            ({\roff*sin(5*\t-\troff)},{\roff*cos(5*\t-\troff)})
            ({\R*sin(5*\t-\toff)},{\R*cos(5*\t-\toff)})
        }
        arc (-205:-155:3) 
        plot [smooth, tension=1] coordinates{
        ({\R*sin(4*\t+\toff)},{\R*cos(4*\t+\toff)})
        ({\roff*sin(4*\t+\troff)},{\roff*cos(4*\t+\troff)})
        ({\roff*sin(1*\t+\troff)},{\roff*cos(1*\t+\troff)})
        ({\R*sin(1*\t+\toff)},{\R*cos(1*\t+\toff)})}
        arc (25:-25:3) 
        ;

        \draw[thick, red] plot [smooth, tension=1] coordinates{
         ({\R*sin(-\toff)},{\R*cos(-\toff)})
        ({\roff*sin(-\troff)},{\roff*cos(-\troff)})
        ({\roff*sin(3*\t-\troff)},{\roff*cos(3*\t-\troff)})
         ({\R*sin(3*\t-\toff)},{\R*cos(3*\t-\toff)})
         };
        \draw[thick, CornflowerBlue] plot [smooth, tension=1] coordinates{
         ({\R*sin(\t-\toff)},{\R*cos(\t-\toff)})
        ({\roff*sin(\t-\troff)},{\roff*cos(\t-\troff)})
        ({\roff*sin(4*\t-\troff)},{\roff*cos(4*\t-\troff)})
         ({\R*sin(4*\t-\toff)},{\R*cos(4*\t-\toff)})
         };
        \draw[thick, ForestGreen] plot [smooth, tension=1] coordinates{
         ({\R*sin(2*\t-\toff)},{\R*cos(2*\t-\toff)})
        ({\roff*sin(2*\t-\troff)},{\roff*cos(2*\t-\troff)})
        ({\roff*sin(5*\t-\troff)},{\roff*cos(5*\t-\troff)})
         ({\R*sin(5*\t-\toff)},{\R*cos(5*\t-\toff)})
         };

        \draw[thick, CornflowerBlue,dashed] plot [smooth, tension=1] coordinates{
         ({\R*sin(+\toff)},{\R*cos(+\toff)})
        ({\roff*sin(+\troff)},{\roff*cos(+\troff)})
        ({\roff*sin(3*\t+\troff)},{\roff*cos(3*\t+\troff)})
         ({\R*sin(3*\t+\toff)},{\R*cos(3*\t+\toff)})
         };
        \draw[thick, ForestGreen,dashed] plot [smooth, tension=1] coordinates{
         ({\R*sin(\t+\toff)},{\R*cos(\t+\toff)})
        ({\roff*sin(\t+\troff)},{\roff*cos(\t+\troff)})
        ({\roff*sin(4*\t+\troff)},{\roff*cos(4*\t+\troff)})
         ({\R*sin(4*\t+\toff)},{\R*cos(4*\t+\toff)})
         };
        \draw[thick, red,dashed] plot [smooth, tension=1] coordinates{
         ({\R*sin(2*\t+\toff)},{\R*cos(2*\t+\toff)})
        ({\roff*sin(2*\t+\troff)},{\roff*cos(2*\t+\troff)})
        ({\roff*sin(5*\t+\troff)},{\roff*cos(5*\t+\troff)})
         ({\R*sin(5*\t+\toff)},{\R*cos(5*\t+\toff)})
         };

        \draw[thick,red] ({\R*sin(-\toff)}, {\R*cos(-\toff)}) arc (95:145:3);
        \draw[thick,CornflowerBlue] 
        ({\R*sin(1*\t-\toff)},{\R*cos(1*\t-\toff)}) arc (35:85:3);
        \draw[thick,ForestGreen] ({\R*sin(2*\t-\toff)},{\R*cos(2*\t-\toff)})  arc (-25:25:3);  
		\draw[thick,red] ({\R*sin(3*\t-\toff)},{\R*cos(3*\t-\toff)}) arc (-85:-35:3);
		\draw[thick,CornflowerBlue] ({\R*sin(4*\t-\toff)},{\R*cos(4*\t-\toff)}) arc (-145:-95:3);
		\draw[thick,ForestGreen] ({\R*sin(5*\t-\toff)},{\R*cos(5*\t-\toff)}) arc (155:205:3);
		
	\end{tikzpicture}
   \caption{An example of the maximally twisted diagram in the case of six scalar fields. The three different coloured regions each contribute with factor of $N$, precisely cancelling the denominator in \eqref{eq:OL}. This diagram is dominant in the $\mathsf{CP}-\mathbb{RP}^4$ theory.} \label{fig:maxtwist}
\end{figure}

With these rules in place, we can then compute one-point functions for single-trace operators in the planar limit and at tree level. Consider an operator built only from a single complex scalar field $Z=(\phi_5+i\phi_6)/\sqrt{2}$, 
\begin{align}
	\mathcal{O}_{(L)}=\left(\frac{4\pi^2}{\lambda}\right)^{\frac{L}{2}} \frac{\Tr Z^L}{\sqrt{L}}\comma \qquad L\in2\mathbb{Z}\period\label{eq:OL}
\end{align}
The normalisation of the operator comes from the requirement that the two-point function of the operator on $\mathbb{R}^4$ should be unit normalised. As shown in Appendix \ref{sec:maxtwist}, when computing the corresponding one-point function $A_{\mathcal{O}_{(L)}}$, there is a unique diagram that contributes in the planar limit: the \textit{maximally twisted diagram}, an example of which is shown in Figure \ref{fig:maxtwist}, which contracts the fields on opposite ends of the trace in pairs using the Wick contraction rules in \eqref{eq:Wick}.

Note that (as seen by the appropriate colours) this diagram has exactly $L/2$ loops, providing the necessary factor of $N^{L/2}$ that cancels the corresponding factor in the normalisation \eqref{eq:OL}.
Since the dominant diagram is unique, we can readily compute the VEV, which is done in \ref{sec:AOLN}. The result is given by
\begin{align}
	A_{\mathcal{O}_{(L)}}=\frac{\varepsilon_{\mathbb{K}}}{\sqrt{L}}+O(N^{-1})=\begin{cases}
		\frac{+1}{\sqrt{L}}+O(N^{-1})\text{ for the charge conjugation in }\eqref{eq:CO}\\
		\frac{-1}{\sqrt{L}}+O(N^{-1})\text{ for the charge conjugation in }\eqref{eq:Cp}
	\end{cases}\label{eq:vacVEV}
\end{align}
where we recall the definition of $\varepsilon_{\mathbb{K}}$ in \eqref{eq:C}, $\varepsilon_{\mathbb{I}}=-\varepsilon_{\mathbb{J}}=1$. The subleading orders in $N$, which include additional diagrams, are computed in \ref{sec:AOLsubN}.
\\
Next, we consider configurations built from two complex scalars $Z$ and $X=(\phi_7+i\phi_8)/\sqrt{2}$. Consider for example replacing two of the $Z$'s in \eqref{eq:OL} by $X$, obtaining an operator in the form
\begin{equation}
    \tilde{\mathcal{O}}=\sum_{n_1<n_2}^Lc_{n_1,n_2}\Tr\left(Z^{n_1-1}XZ^{n_2-n_1-1}XZ^{L-n_2}\right)\,.\label{OL2}
\end{equation}
The precise form and normalization of $\tilde{\mathcal{O}}$ should be chosen such that it is an eigenfunction of the dilation operator and has a unit-normalised two-point function. As shown in \ref{sec:maxtwist}, in the planar limit there is still only one possible contribution to the one-point function $A_{\tilde{\mathcal{O}}}$ that comes from maximally twisted diagrams where the two $X$'s are on opposite sides, \textit{i.e.} terms with $n_2=n_1+\frac{L}{2}$. This is a \textit{selection rule}, with operators with vanishing coefficient $c_{n_1,n_1+\frac{L}{2}}$ having vanishing one-point functions. This pattern holds for any number of insertions and any fields: the only non-vanishing diagrams that contribute are maximally twisted, with the contraction rules given by the structure of the propagators \eqref{eq:Wick}.

\begin{figure}[t]
	\centering
	\begin{tikzpicture}[cross/.style={path picture={ 
				\draw[ultra thick]
				(path picture bounding box.south east) -- (path picture bounding box.north west) (path picture bounding box.south west) -- (path picture bounding box.north east);
		}}]
		\def\t{60}
        \def\toff{5}
        \def\troff{15}
        \def\roffin{1.5+pi/12}
        \def\roffout{1.5-pi/12}
        \def\roff{0.8}
       \def\R{3}
		\draw ({\R*sin(0)},{\R*cos(0)}) node[]{$Z$};
		\draw ({\R*sin(\t)},{\R*cos(\t)})
		node[]{${Z}$};
		\draw ({\R*sin(2*\t)},{\R*cos(2*\t)})
		node[]{$Z$};
		\draw ({\R*sin(3*\t)},{\R*cos(\R*\t)})
		node[]{${Z}$};
		\draw ({\R*sin(4*\t)},{\R*cos(4*\t)})
		node[]{$Z$};
		\draw ({\R*sin(5*\t)},{\R*cos(5*\t)})
		node[]{${Z}$};

       \filldraw[fill=CornflowerBlue!20,draw=CornflowerBlue,thick]
        plot[smooth, tension=1] coordinates{
        (\toff+30:\R)
        (\t/2+30:\roffin)
        (\t-\toff+30:\R)}
        arc (\t-\toff+30:\toff+30:\R)
        --cycle
        ;
        \filldraw[fill=ForestGreen!20,draw=ForestGreen,thick] 
        plot[smooth, tension=1] coordinates{
        (\toff+150:\R)
        (\t/2+150:\roffin)
        (\t-\toff+150:\R)
        }
        arc (\t-\toff+150:\toff+150:\R)
        --cycle
        ;
        \filldraw[fill=red!20,draw=red,thick]
        plot[smooth, tension=1] coordinates{
        (\toff+270:\R)
        (\t/2+270:\roffin)
        (\t-\toff+270:\R)}
        arc (\t-\toff+270:\toff+270:\R)
        --cycle
        ;

        \filldraw[fill=Dandelion!20,draw=Dandelion,thick,even odd rule]
        plot[smooth, tension=1] coordinates{
        (360-\toff+30:\R)
        (\t/2+30:\roffout)
        (\t+\toff+30:\R)
        }
        arc (\t+\toff+30:-\toff+150:\R)
        -- plot[smooth, tension=1] coordinates{
        (-\toff+150:\R)
        (\t/2+150:\roffout)
        (\t+\toff+150:\R)
        }
        arc (\t+\toff+150:-\toff+270:\R)
        -- plot[smooth, tension=1] coordinates{
        (-\toff+270:\R)
        (\t/2+270:\roffout)
        (\t+\toff+270:\R)
        }
        arc (\t+\toff+270:360-\toff+30:\R)
        --cycle
        ;
	\end{tikzpicture}
    \caption{An example of a diagram that contributes to the 1-point function of $Z^6$ in the $\mathsf{P}-\mathbb{RP}^4$ theory. The contribution from the diagram goes like $O(N^1)$ at large $N$ due to having 4 closed cycles, as marked by the different colours.}\label{fig:onepointN}
\end{figure}
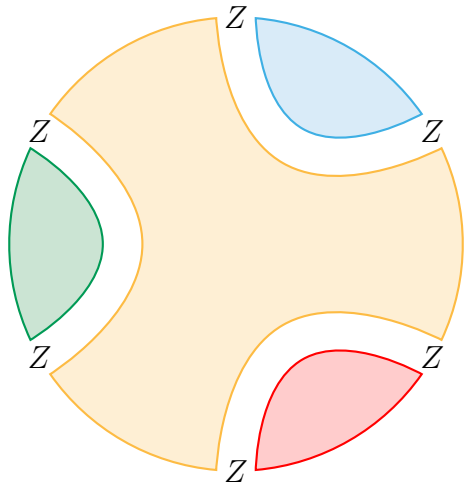

So far, we have focused on $\mathbb{RP}^4$ with charge conjugation included in $\iota$ ($\mathsf{CP}$-$\mathbb{RP}^4$). if we instead consider $\mathbb{RP}^4$ without charge conjugation ($\mathsf{P}$-$\mathbb{RP}^4$) as discussed in  \cite{Caetano:2022mus}, the diagrams that contribute at large $N$ are standard planar diagrams (see Figure 
\ref{fig:onepointN}), and the one-point function blows up in the limit $N\to \infty$. This divergence reflects the fact that $\mathsf{P}$-$\mathbb{RP}^4$ is holographically dual to a backreacted geometry, instead of a probe orientifold in AdS$_5\times$S$^5$.
\section{One-point functions from spin-chain crosscaps}\label{sec:weak}
We now compute these one-point functions for single-trace operators in the planar limit. Recall that planar $\mathcal N=4$ SYM on $\mathbb R^4$ admits an
integrable spin-chain description: the dilatation operator is mapped to the
spin-chain Hamiltonian, and single-trace operators correspond to its energy
eigenstates. Thus, computing scaling dimensions amounts to solving the
spin-chain spectral problem, for example through the Bethe Ansatz Equations
 (BAE). 

This raises a natural question: do one-point functions on $\mathbb{RP}^4$
admit a similarly simple spin-chain interpretation? In related settings --- such
as the $D3$-$D5$ defect CFT \cite{deLeeuw:2015hxa,Komatsu:2020sup,Gombor:2020kgu,Gombor:2020auk}, structure constants of two giant gravitons and a single-trace operator \cite{Jiang:2019xdz,Jiang:2019zig}, and $\mathcal{N}=4$ SYM on the Coloumb branch \cite{Ivanovskiy:2024vel,Coronado:2025xwk} --- one-point functions
are expressed as overlaps between Bethe eigenstates and so-called \textit{integrable boundary states}. 

\begin{figure}
    \centering
    \begin{minipage}{0.45\textwidth}
        \centering
\begin{tikzpicture}
    \def\R{1} 
    \def\side{3}
    \def\linR{0.9}  
    \def\tilt{10}     
    \def\lat{30}      
    \def\offset{90}
    \def\rotb{\offset}
    \def\rotg{\offset +120}
    \def\rotr{\offset+240}

    \pgfmathsetmacro{\rxy}{\R*cos(\lat)}
    \pgfmathsetmacro{\rz}{\R*cos(\lat)*sin(\tilt)}
    \pgfmathsetmacro{\yc}{\R*sin(\lat)*cos(\tilt)}

    \draw[draw=black, fill=white!50] ({-\side},{-\side}) rectangle ({\side},{\side});
    \draw[draw=Dandelion, thick, pattern= north east lines] (0,0) circle(\R);

    \draw[<->,red,dashed] ({\linR*cos(\rotr)},{\linR*sin(\rotr)})->({-\linR*cos(\rotr)},{-\linR*sin(\rotr)}); 
    \draw[red]  ({\R*cos(\rotr)},{\R*sin(\rotr)}) node[circle,fill,inner sep=1 pt]{};
    \draw[red]  ({-\R*cos(\rotr)},{-\R*sin(\rotr)}) node[circle,fill,inner sep=1 pt]{};

    \draw[<->,CornflowerBlue,dashed] (-{\linR*cos(\rotb)},{\linR*sin(\rotb)})->({\linR*cos(\rotb)},{-\linR*sin(\rotb)}) ;
    \draw[CornflowerBlue]  ({-\R*cos(\rotb)},{\R*sin(\rotb)}) node[circle,fill,inner sep=1 pt]{};
    \draw[CornflowerBlue]  ({\R*cos(\rotb)},{-\R*sin(\rotb)}) node[circle,fill,inner sep=1 pt]{};
    \node (box) [
        font=\scriptsize, 
        minimum width=0.4cm, 
        minimum height=0.4cm, 
        inner sep=2pt 
    ] at (2.6,2.6) {\scriptsize $\mathbb{R}^2$};
    \draw[] (box.north west) -- (box.south west) -- (box.south east);
\end{tikzpicture}
\end{minipage}
\begin{minipage}{0.45\textwidth}
        \centering
\begin{tikzpicture}
    \def\R{1} 
    \def\side{3}
    \def\linR{0.9}  
    \def\tilt{10}     
    \def\lat{30}      
    \def\offset{90}
    \def\rotb{\offset}
    \def\rotg{\offset +120}
    \def\rotr{\offset+240}

    \pgfmathsetmacro{\rxy}{\R*cos(\lat)}
    \pgfmathsetmacro{\rz}{\R*cos(\lat)*sin(\tilt)}
    \pgfmathsetmacro{\yc}{\R*sin(\lat)*cos(\tilt)}

    \draw[draw=black, fill=white!50] ({-\side},{-\side}) rectangle ({\side},{\side});
    \draw (0,0) node[circle, fill=Dandelion, text=black, inner sep=2pt] {$\ket{\mathcal{C}}$};

    \node (box) [
        font=\scriptsize, 
        minimum width=0.4cm, 
        minimum height=0.4cm, 
        inner sep=2pt 
    ] at (2.6,2.6) {\scriptsize $\mathbb{R}^2$};
    \draw[] (box.north west) -- (box.south west) -- (box.south east);
\end{tikzpicture}
\end{minipage}
\caption{Crosscap geometry and its state representation in a two-dimensional CFT. Left: remove the interior of the unit circle and identify antipodal points on its
boundary; after compactifying infinity, the resulting surface is $\mathbb{RP}^2$
. Right: radial quantization represents the identified circle by a crosscap state $\ket{\mathcal{C}}$, which implements the corresponding antipodal
gluing of fields.}\label{fig:2dcross}
\end{figure}
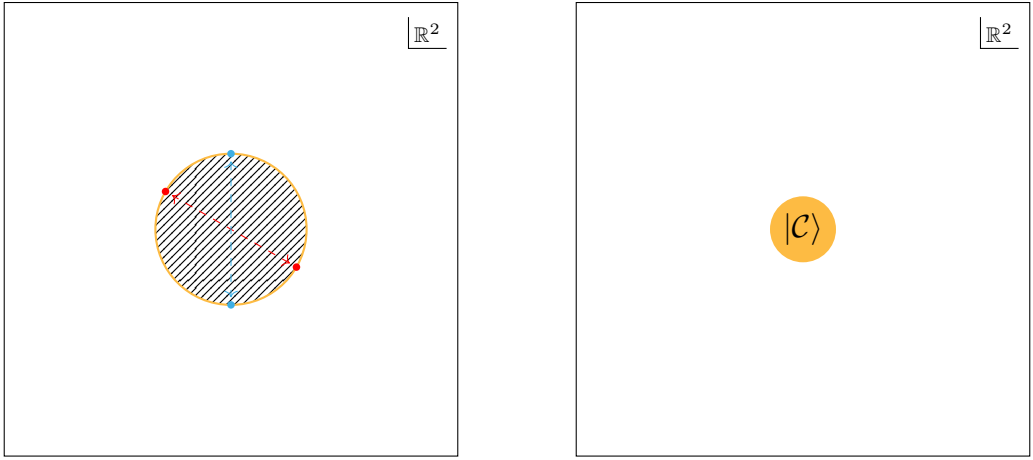

For $\mathbb{RP}^4$, the natural candidate is to consider the overlap with the {\it integrable crosscap state}, introduced
in \cite{Caetano:2021dbh}. It is the spin-chain analogue of a crosscap
in two-dimensional CFT, which is associated with placing the theory on
$\mathbb{RP}^2$; see Figure~\ref{fig:2dcross}. We will test this proposal in
several closed subsectors of $\mathcal N=4$ SYM, working at leading order in
$g_{\rm YM}$, where the spin chains are nearest-neighbour.

We find that the one-point functions are indeed reproduced by overlaps with
the crosscap state. In scalar sectors, these overlaps are given by the ratio of the determinants of certain matrices \cite{Caetano:2021dbh,Gombor:2022deb}, and we
construct the corresponding matrices explicitly. For non-scalar sectors, the
overlaps are more involved; we derive the available results and
conjecture their general form. Note that to stick to the notation in the literature, we will use the names of the Lorentzian subgroups throughout: thus, we use $\SO(6) $ to denote the scalar sector, even if the scalars actually transform in a representation of $\SO(5,1)$ and so on.
\subsection{\texorpdfstring{$\SU(2)$}{SU(2)}-sector}
This subsector consists of the fields $Z$ and $X$ and corresponds to the XXX-spin chain. On each site of the spin chain lives a spin $1/2$ representation of $\text{SU}(2)$, and we will denote the total number of sites by $L$, the Hilbert space by $\mathcal{H}$, and the Hamiltonian by $H$. We will assume that $L$ is even, which will be important for the construction of the croscap state later. We then have
\begin{align}
	\mathcal{H}^{\SU(2)}=\bigotimes_{j=1}^{L}\text{Span}\Big\{\ket{\!\!\uparrow}_j, \ket{\!\!\downarrow}_j\Big\}\comma\quad  H^{\SU(2)}=\sum_{j=1}^L\Big(1-\mathbb{P}_{j,j+1}\Big),\label{eq:HamilSU2}
\end{align}
where $\mathbb{P}_{ij}$ is the permutation operator between site $i$ and site $j$ and $\mathbb{P}_{L,L+1}=\mathbb{P}_{L,1}$. We will denote the lowering operator on the $k$-th site by $S^-_k$, such that $S^-_k\ket{\!\!\uparrow}_k=\ket{\!\!\downarrow}_k$. The spin chain is integrable, so there exists a large set of charges $\mathcal{Q}_n$. They can be systematically extracted from the \textit{transfer-matrix} $\mathcal{T}(u)$, namely
\begin{equation}
    \mathcal{Q}_n=-\frac{i}{(n-1)!}\partial_u^{n-1}\log \mathcal{T}(u)\Big|_{u=0}
\end{equation}
The first two charges can be identified with the total momentum and energy
\begin{align}
	\mathcal{Q}_1=P_{\text{tot}}\comma \quad \mathcal{Q}_2=H\,.
\end{align}
The crucial identity is $[\mathcal{T}(u),\mathcal{T}(v)]=0$, which implies $[\mathcal{Q}_n,\mathcal{Q}_m]=0$, \textit{i.e.} the charges are conserved and mutually commuting. Since the spin chain has a global $\SU(2)$ symmetry, it is customary to include an additional charge $\mathcal{Q}_0=2S^z_{\text{tot}}$, which similarly commutes with the entire tower of charges. The vacuum of the spin-chain is degenerate: we choose it to be given by a sequence of $\ket{\!\!\uparrow}$
\begin{equation}
	\ket{\emptyset}\equiv\bigotimes_{j=1}^{L}\ket{\!\!\uparrow}_j\,,
\end{equation}
for which $S^z_{\text{tot}}=L/2$ while $P_{\text{tot}}=E_{\text{tot}}=0$. 
\paragraph{Bethe states.}
Excited states of the spin-chain can be constructed using the Bethe Ansatz. They are labelled by Bethe roots $\bm{u}=\left\{u_j\right\}_{j=1}^M$, with each root corresponding to a magnon, which have momenta $p_j$ and energies $E_j$ given by
\begin{equation}
        e^{ip(u)}=\frac{u+\frac{i}{2}}{u-\frac{i}{2}}\comma \quad E(u)=\frac{1}{u^2+\frac{1}{4}}\,.\label{eq:Bethemom}
\end{equation}
The Bethe state can then be written in the form
\begin{equation}\label{eq:BetheState}
\begin{split}
    \ket{\bm{u}}=\sum_{\bm{n}}\Psi(\bm{u}|\bm{n})\,S^-_{n_1}\dots S^-_{n_M}\,\ket{\emptyset}_{(L)}\,,\quad 
     &\Psi(\bm{u}|\bm{n})=\sum_{\sigma\in S_M}k(\bm{u},\sigma)\left(\prod_{j=1}^Me^{in_jp_{\sigma_j}}\right)\,,\\&k(\bm{u},\sigma)=\text{sign}(\sigma)\prod_{j<k}\left(e^{ip_{\sigma_k}+ip_{\sigma_j}}-2e^{ip_{\sigma_j}}+1\right)\,,
\end{split}
\end{equation}
where $S_M$ is the permutation group on $M$ elements and the coefficients $k$ are independent of the positions $n_j$. In order for the Bethe state \eqref{eq:BetheState} to be an eigenstate of the Hamiltonian \eqref{eq:HamilSU2}, the Bethe roots must satisfy the corresponding Bethe equations. Let us first introduce the \textit{Baxter Q-functions}
\begin{equation}
    Q_0(x)= x^L\,,\quad Q_1(x)=\prod_{j=1}^M(x-u_j)\,,\label{eq:QSU2}
\end{equation}
where the roots of the latter are by construction precisely the Bethe roots. Then the Bethe equations can be expressed simply in the form
\begin{equation}
    R(u_j|\bm{u})=1\,,\quad j=1,\dots ,M\,,\qquad R(x|\bm{u})=-\frac{Q_0\left(x-\frac{i}{2}\right)}{Q_0\left(x+\frac{i}{2}\right)}\frac{Q_1(x+i)}{Q_1(x-i)}\label{eq:BetheEQ}
\end{equation}
Once the Bethe equations are satisfied, the conserved charges of the state $\ket{u_1,\dots,u_M}$ are given by simply adding the charges of the individual magnons \eqref{eq:Bethemom}
\begin{equation}
    S^z_{\text{tot}}=L/2-M\comma\quad  P_{\text{tot}}=\sum_{j=1}^Mp_j\comma \quad E_{\text{tot}}=\sum_{j=1}^ME_j\,.\label{SPE}
\end{equation}
The higher charges $\mathcal{Q}_n$ can similarly be expressed as sums over contributions from the individual magnons $\mathcal{Q}_n(u)$. Let us note that the even/odd charges $q_n$ are even/odd functions of $u$.
\paragraph{Crosscap state.}
Let us return to the construction of the crosscap state $\ket{\mathcal{C}}$. The radial-quantization picture in Figure \ref{fig:2dcross} motivates the spin chain construction: the crosscap
entangles spins at antipodal sites of the chain
\begin{equation}
	\ket{\mathcal{C}}\equiv\bigotimes_{j=1}^{L/2} \ket{c}_j\comma \qquad \ket{c}_j=\ket{\!\!\uparrow}_j\otimes\ket{\!\!\uparrow}_{j+L/2}+\ket{\!\!\downarrow}_j\otimes\ket{\!\!\downarrow}_{j+L/2}\,.\label{eq:SU2C}
\end{equation}
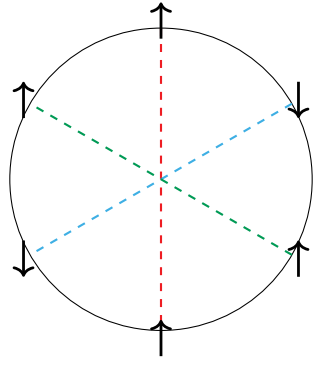
\begin{figure}
    \centering
  \begin{tikzpicture}
      \draw[] (0,0) circle[radius=2];
  \foreach \a/\c in {90/Red,30/CornflowerBlue,-30/ForestGreen}
    \draw[\c,thick,dashed] (\a:2)--(\a+180:2);
  \foreach \a/\c/\s in {90/CornflowerBlue/1,30/Red/-1,-30/ForestGreen/1,-90/CornflowerBlue/1,-150/Red/-1,150/ForestGreen/1}{%
    \coordinate (S) at (\a:2.1);
    \draw[line width=1.05pt,->] ($(S)+(0,{-.24*\s})$)--($(S)+(0,{.24*\s})$);
    }
  \end{tikzpicture}
    \caption{One state in the superposition defining the crosscap state \eqref{eq:SU2C} for $L=6$. Coloured dashed
lines connect the antipodal pairs $(j, j +L/2)$. The antipodal pairing corresponds to the maximally twisted
Wick contraction in Figure \ref{fig:maxtwist}.}
    \label{fig:crosscapspinchain}
\end{figure}
Figure \ref{fig:crosscapspinchain} illustrates the antipodal pairing for a
chain of length six.
It is clear that $\ket{\mathcal{C}}$ has vanishing momentum and is a superposition of states with integer spin. However,
unlike the Bethe state \eqref{eq:BetheState}, the crosscap state is not an eigenstate of all the charges, including the Hamiltonian.
It is instead a special state satisfying \cite{Caetano:2021dbh}
\begin{equation}
    \mathcal{T}(u)\ket{\mathcal{C}}=\mathcal{T}(-u)\ket{\mathcal{C}}\Rightarrow 
    \mathcal{Q}_{2n+1}\ket{\mathcal{C}}=0\,\quad n=1,2,\dots\,.\label{eq:QonC}
\end{equation}
Since $\ket{\mathcal{C}}$ is not itself an energy eigenstate, it has non-vanishing overlap with many different Bethe states. Given \eqref{eq:QonC}, such overlaps must vanish unless $\mathcal{Q}_{2n+1}\ket{\bm{u}}=0$ for all $n\in\mathbb{N}$, which acts as a selection rule. Since these odd charges are odd under $\bm{u}\mapsto-\bm{u}$, this condition is satisfied if $\bm{u}=-\bm{u}$. Furthermore, since every element of the superposition of states that make up the crosscap state has manifestly even spin, there must be an even number of magnons. These conditions enforce that $\bm{u}$ takes the form
\begin{equation}
    \left\{u_j\right\}_{j=1}^{2m}=\{u_1,\dots,u_m,-u_1,\dots -u_m\}\,.\label{eq:PairedStates}
\end{equation}
We call states with such configurations of Bethe roots \textit{parity-symmetric states}. The inner product of the corresponding Bethe state with the crosscap state then becomes a new observable in the theory, the \textit{crosscap overlap}
\begin{equation}
    \frac{\left<\mathcal{C}|\bm{u}\right>}{\left<\bm{u}|\bm{u}\right>^{\frac{1}{2}}}\,,\label{eq:OverlapA}
\end{equation}
which is defined to be invariant under rescaling of $\ket{\bm{u}}$.

Remarkably, one can find closed-form expressions for these overlaps \cite{Caetano:2021dbh,Gombor:2022deb}. Firstly, the norm of the corresponding Bethe state can be written in the form
\begin{equation}
    \left<\bm{u}|\bm{u}\right>=\left(\prod_{j<k}^M(u_j-u_k)^2+1\right)\left(\prod_{j=1}^M(u_j^2+1/4)^{2-M}\right)\det G\,,\quad S(u,v)=\frac{u-v+i}{u-v-i}\,.
\end{equation}
Here, $G$ is the $M\times M$ \textit{Gaudin matrix}, which is given by
\begin{equation}
    G_{jk}=-i\frac{\partial}{\partial u_j}\log R(u_k|\bm{u})\,,\label{eq:Gaudin}
\end{equation}
where $R$ was given in \eqref{eq:BetheEQ}. For parity-symmetric states \eqref{eq:PairedStates}, $\det G$ factorises \cite{deLeeuw:2015hxa}. We can consider the sum and difference of the row corresponding to the root $u_j$ and $-u_j$
\begin{equation}
    G^\pm_{i,j}=G_{i,j}\pm G_{i+\frac{M}{2},j}\,,\label{eq:Gpm}
\end{equation}
where $1\leq j\leq m$, such that $G^\pm$ is $\frac{M}{2}\times\frac{M}{2}$. Then one can show that the determinant of the Gaudin matrix factorises $\det G=\det G^+\det G^-$. Using this factorisation, the crosscap overlap can be written in the form 
\begin{equation}
    \left<\mathcal{C}|\bm{u}\right>=\left(\prod_{j<k}^MS(u_j,u_k)\right)^{1/2}\left(\prod_{j=1}^M\frac{1}{\partial_{u_j} p_j}\right)^{1/2}\det G^+\,,
\end{equation}
which leads to an exceedingly simple expression for the crosscap overlap \cite{Caetano:2021dbh}
\begin{equation}
    \frac{\left<\mathcal{C}|\bm{u}\right>}{\left<\bm{u}|\bm{u}\right>^{\frac{1}{2}}}=\sqrt{\frac{\det G^+}{\det G^-}}\,.\label{eq:SU2overlap}
\end{equation}
Let us compare this to an integrable boundary state $\ket{\mathcal{B}}$ \cite{Piroli:2017sei,deLeeuw:2015hxa}, a prototypical example of which is 
\begin{equation}
|\mathcal{B}\rangle \equiv \bigotimes_{j=1}^{L/2} \ket{b}_j\comma \qquad \ket{b}_j=\ket{\!\!\uparrow}_{2j-1}\otimes\ket{\!\!\uparrow}_{2j}+\ket{\!\!\downarrow}_{2j-1}\otimes\ket{\!\!\downarrow}_{2j}\,.
\end{equation}
Integrable boundary states are also annihilated by the odd charges $\mathcal{Q}_{2n+1}\ket{\mathcal{B}}=0$, and hence their overlap with Bethe states vanishes unless the Bethe roots come in parity-symmetric pairs. Much like the crosscap state, the normalized overlap of such a parity-symmetric state with the boundary state can be written in terms of the ratio $\det G_+/\det G_-$. However, a crucial difference is that the overlap with boundary states typically includes extra prefactors multiplying this ratio. These prefactors are $\bm{u}$-dependent and are given by ratios of products of Baxter functions evaluated at various special values of $x$ (see e.g.\cite{deLeeuw:2015hxa}).
\paragraph{Implications for $\mathcal{N}=4$ SYM.}
We now relate the above discussion of crosscap states on the $\SU(2)$ spin chain to computations in $\mathcal{N}=4$ SYM. The $\SU(2)$ sector of single-trace operators in $\mathcal{N}=4$ SYM consists of operators built from  $Z=(\phi_5+i\phi_6)/\sqrt{2}$ and $X=(\phi_7+i\phi_8)/\sqrt{2}$. The seminal work \cite{Minahan:2002ve} then showed that one can interpret the dilatation operator $D$ acting on operators in this sector to first order in $\lambda$ as the Hamiltonian of the spin chain\eqref{eq:HamilSU2}
\begin{equation}
    D=L+\frac{\lambda }{8\pi^2}H+O(\lambda^2)\,,\label{eq:DandH}
\end{equation}
where the $O(\lambda^0)$ term is the engineering dimension. Thus, there is a map between eigenstates of $D$ and energy eigenstates of the Hamiltonian. The simplest example of this map is the state we encountered in \eqref{eq:OL}, which is in correspondence with the spin-chain vacuum
\begin{equation}
\ket{\emptyset}=\bigotimes_{j=1}^{L}\ket{\!\!\uparrow}_k\quad \Leftrightarrow\quad \mathcal{O}_{(L)}=\left(\frac{4\pi^2}{\lambda}\right)^{\frac{L}{2}}\frac{\Tr Z^L}{\sqrt{L}}\\,. \label{eq:SU2Vac}
\end{equation}
Indeed, this operator is $\frac{1}{2}$-BPS and its conformal dimension thus does not receive perturbative corrections. More generally, we identify $\ket{\!\!\uparrow}$ with $Z$ fields and $\ket{\!\!\downarrow}$ with $X$ fields. The operators $\tilde{\mathcal{O}}$ considered in \eqref{OL2} should then be identified with a two magnon state $\ket{\bm{u}}$, with the coefficients $c_{n_1,n_2}$ identified with $\Psi(u_1,u_2|n_1,n_2)$. There is an important caveat: cyclicity of the trace implies that only Bethe states with vanishing total momentum \eqref{SPE} can correspond to single-trace operators.

Can we use this mapping to compute the one-point functions $A_i$ on $\mathbb{RP}^4$? Recall that the only contributions in the computation of the one-point function came from the maximally twisted diagrams \ref{fig:maxtwist}, and this required us to identify fields on opposite sides of the trace. Translating this to the spin chain, this corresponds to projecting the spin-chain wave function to the one in which states at antipodal sites are identical. This can be implemented precisely by taking the overlap with the crosscap state \eqref{eq:SU2C}. Fixing also the normalization, this provides the map
\begin{equation}\label{eq:1ptvscross}
    \left|A_{\mathcal{O}_{\bm{u}}}\right|=\frac{\varepsilon_{\mathbb{K}}}{\sqrt{L}}\left|\frac{\left<\mathcal{C}|\bm{u}\right>}{\left<\bm{u}|\bm{u}\right>^{\frac{1}{2}}}\right|\,,
\end{equation}
where $\mathcal{O}_{\bm{u}}$ is the single-trace operator corresponding to the Bethe state $|\bm{u}\rangle$. 

Before concluding this subsection, let us make three comments.

First, although \eqref{eq:1ptvscross} was derived here in the $SU(2)$ subsector, its structure is more general.
The prefactor $\varepsilon_{\mathbb K}/\sqrt{L}$ follows only from the Wick-contraction rule and the normalization of the operators, and therefore persists in other sectors as well. 

Second, the one-point function of a local operator is defined only up to an
overall phase. Indeed, the redefinition
\begin{equation}
    \mathcal O \to e^{i\theta}\mathcal O
\end{equation}
leaves the normalization of the two-point function
$\langle \mathcal O\bar{\mathcal O}\rangle$ unchanged while it modifies the one-point function by a phase. This is why
\eqref{eq:1ptvscross} is naturally written in terms of an absolute value.
Consequently, the overall sign associated with $\varepsilon_{\mathbb K}$ has no
invariant meaning by itself. What is meaningful, however, is the relative sign
between the $\mathbb K=\mathbb I$ and $\mathbb K=\mathbb J$ crosscaps, and we
therefore retained $\varepsilon_{\mathbb K}$ in the formula.

Third, the resemblance between the spin-chain crosscap state, which pairs
opposite sites, and the antipodal identification defining $\mathbb{RP}^4$ is
not itself the origin of the correspondence. Rather, the crucial ingredient is
the twisted Wick contraction \eqref{eq:Wick}, which forces the corresponding
spin-chain state to entangle sites at opposite ends of the chain.

\subsection{\texorpdfstring{$\SO(6)$}{SO(6)}-sector}
We now go to a larger subsector of $\mathcal{N}=4$ SYM, consisting of single-trace operators built from 6 scalar letters $X,Y,Z,\bar{X},\bar{Y},\bar{Z}$, transforming in the fundamental representation of $\SO(5,1)$. This subsector is closed at one-loop and is mapped to an $\text{SO}(6)$ spin chain. The corresponding Hilbert space and Hamiltonian are given by 
\begin{equation}
    \mathcal{H}^{\SO(6)}=\bigotimes_{j=1}^L\text{Span}\Big\{\ket{X}_j, \ket{Y}_j,\ket{Z}_j,\ket{\bar{X}}_j, \ket{\bar{Y}}_j,\ket{\bar{Z}}_j\Big\}\comma H^{\SO(6)}=\sum_{j=1}^{L}\Big(1-\mathbb{P}_{j,j+1}+\frac{1}{2}\mathbb{K}_{j,j+1}\Big),\label{eq:HSO6}
\end{equation}
where $\mathbb{K}$ is the contraction operator. We then define a vacuum and a cross-cap state
\begin{equation}
\begin{split}
    \ket{\emptyset}=&\bigotimes_{j=1}^L \ket{Z}_j\,,\quad\ket{\mathcal{C}}=\bigotimes_{j=1}^{L/2} \ket{c}_j\comma 
    \\
    \ket{c}_j=&\ket{X}_j\otimes\ket{X}_{j+L/2}+\ket{Y}_j\otimes\ket{Y}_{j+L/2}+\ket{Z}_j\otimes\ket{Z}_{j+L/2}
    \\
    +&\ket{\bar{X}}_j\otimes\ket{\bar{X}}_{j+L/2}+\ket{\bar{Y}}_j\otimes\ket{\bar{Y}}_{j+L/2}+\ket{\bar{Z}}_j\otimes\ket{\bar{Z}}_{j+L/2}\,.\label{eq:SO6C}
\end{split}
\end{equation}
Note that the crosscap state defined here is not invariant under the full SO$(6)$.

The excited energy eigenstates are described by the nested Bethe ansatz. Since the explicit construction of the wave function based on the Bethe ansatz appeared many times in the literature (see e.g.\cite{Basso:2017khq}), we simply summarize the minimum details needed for presenting the final result.
In the SO(6) sector, the Bethe roots come in three sets $\bm{u}=(\bm{u}^1,\bm{u}^2,\bm{u}^3)$, corresponding to the three nodes in the Dynkin diagram of SO(6)
\begin{align}
    \begin{tikzpicture}[cross/.style={path picture={ 
			\draw[black,thick]
			(path picture bounding box.south east) -- (path picture bounding box.north west) (path picture bounding box.south west) -- (path picture bounding box.north east);
	}}]
	\draw [thick](0,0)--(1,0)--(2,0)--(3,0)--(4,0);
	\draw (1,0)node[circle,draw=black,fill=white]{};
    \draw (2,0)node[circle,draw=black,fill=white]{};
    \draw (3,0)node[circle,draw=black,fill=white]{};
    \end{tikzpicture}
\end{align}
Only the middle node roots $\bm{u}^2$ are needed to compute the charges of the integrable hierarchy and are called \textit{momentum-carrying}, such that the Bethe state $\ket{\bm{u}}$ has energy and momentum given by
\begin{equation}
    P_{\text{tot}}=\sum_{j=1}^{M_2}p_{2,j}\,,\quad E_{\text{tot}}=\sum_{j=1}^{M_2}E_{2,j}\,,
\end{equation}
where $p_{2,j}$ and $E_{2,j}$ are given by \eqref{eq:Bethemom}. The number of roots should satisfy $M_2\geq 2M_1,2M_3$ and $L\geq 2M_2-M_1-M_3$: the SO(6) global charges $(J^1_{\text{tot}},J^2_{\text{tot}},J^3_{\text{tot}})$ are then given by $(L-M_2,M_2-M_1-M_3,M_1-M_3)$, such that the vacuum (which has no excitations and thus no Bethe roots) has $J^1_{\text{tot}}=L,J^2_{\text{tot}}=J^3_{\text{tot}}=0$. The magnons $\bm{u}$ have to satisfy the Bethe equations. These are given by \eqref{eq:GenBethe}, 
where $C_{ab}$ and $q_a$ are the Cartan matrix of $\SO(6)$ and Dynkin label of the vector representation respectively
\begin{equation}
    C=\begin{pmatrix}
        2&-1&0\\-1&2&-1\\0&-1&2
    \end{pmatrix}\,,\qquad q=\begin{pmatrix}
        0\\1\\0
    \end{pmatrix}\,.
\end{equation}
The latter also explains why the middle node carries momentum: only nodes with non-vanishing Dynkin labels carry momentum, which depends on our choice of representation.
The relationship between the dilatation operator and the spin chain Hamiltonian \eqref{eq:DandH} remains intact, and there is then again a correspondence between zero-momentum Bethe states and operators in the scalar sector of $\mathcal{N}=4$ SYM that are eigenstates of the dilatation operator. The map is straightforward, replacing the spins by letters: denoting 
\begin{equation}
\begin{split}
    (Z,\bar{Z},X,\bar{X},Y,\bar{Y})&=(X_1,X_2,X_3,X_4,X_5,X_6)
    \\
    &=\frac{1}{\sqrt{2}}(\phi_5+i\phi_6,\phi_5-i\phi_6,\phi_7+i\phi_8,\phi_7-i\phi_8,\phi_9+i\phi_{0},\phi_9-i\phi_{0})
    \end{split}
\end{equation}
we have
\begin{equation}
    \bigotimes_{j=1}^L\ket{X_{n_j}}_j\quad \Leftrightarrow \quad \left(\frac{4\pi^2}{\lambda}\right)^{L/2}\frac{\Tr \left(X_{n_1}X_{n_2}\dots X_{n_L}\right)}{\sqrt{L}}\,.
\end{equation}
Indeed, just like the Bethe states, the scalar operators in $\mathcal{N}=4$ SYM also carry $\SO(6)_R$ charges $(J_1,J_2,J_3)$. Using this identification, it is then straightforward to see that the crosscap state \eqref{eq:SO6C} again perfectly implements the maximally twisted diagram, and we have the correspondence 
    \begin{equation}
    \left|A_{\mathcal{O}_{\bm{u}}}\right|=\frac{\varepsilon_{\mathbb{K}}}{\sqrt{L}}\left|\frac{\left<\mathcal{C}|\bm{u}\right>}{\left<\bm{u}|\bm{u}\right>^{\frac{1}{2}}}\right|\,.
\end{equation}
Note that the crosscap state is again an integrable boundary state satisfying $\mathcal{T}(u)\ket{\mathcal{C}}=\mathcal{T}^\mathfrak{C}(-u)\ket{\mathcal{C}}$, which implies that the overlap on the right-hand side vanishes unless the state is parity-symmetric $\bm{u}=-\bm{u}$. The ``charge-conjugation'' $\mathfrak{C}$ is related to the charge-conjugation on the world-sheet that we consider in Section \ref{sec:classical} and originates from the fact that the scalar crosscap state is \textit{twisted} in the language of \cite{Gombor:2022deb},\footnote{The crosscap state \eqref{eq:SO6C} is given as \cite[Eq.~(4.1)]{Gombor:2022deb}.} which was not manifest when looking only at the SU(2) sector.
This leads to a factorisation of the $(M_1+M_2+M_3)\times(M_1+M_2+M_3)$ Gaudin matrix $G$. Remarkably, it was found in \cite{Gombor:2022deb} that the right-hand side of the above equation is again given simply in terms of the ratio of determinants of the Gaudin matrix
\begin{equation}
        \frac{\left<\mathcal{C}|\bm{u}\right>}{\left<\bm{u}|\bm{u}\right>^{\frac{1}{2}}}=\sqrt{\frac{\det G_+}{\det G_-}}\,,\label{eq:SO6overlap}
\end{equation}
which provides us with a straightforward way to compute the one-point functions of scalar operators. 
\subsection{\texorpdfstring{$\SU(1|2)$}{SU(1|2)}-sector}\label{sec:SU12}
We now turn our attention to a superspin-chain sector, corresponding to single-trace operators built from one scalar $Z$ and one two-component fermion $\psi^1_{\alpha}$. On the spin chain, we denote the states $\ket{0},\ket{\!\!\uparrow},\ket{\!\!\downarrow}$, the latter two of which are fermionic. The correspondence with a state in the gauge theory is straightforward, replacing the 0-state by the boson $Z$ and the spins $\ket{\!\!\uparrow}$ and $\ket{\!\!\downarrow}$ by the components $\psi_1^1$ and $\psi^1_2$. We will keep the discussion rather brief in the main text and refer the reader to Appendix \ref{sec:SU12} for further details.

The Hilbert space and the Hamiltonian are given by
\begin{align}
    \mathcal{H}^{\SU(1|2)}=\bigotimes_{k=1}^{L}\Big\{\ket{0}_j,\ket{\!\!\uparrow}_j,\ket{\!\!\downarrow}_j\Big\}\comma \quad H^{\SU(1|2)}=\sum_{j=1}^L\left(1-\hat{\mathbb{P}}_{j,j+1}\right),\label{eq:0updown}
\end{align}
where $\hat{\mathbb{P}}$ is a graded permutation operator that accounts for the non-commuting nature of the fermions $\ket{\!\!\uparrow}$ and $\ket{\!\!\downarrow}$. The crosscap state is given by
\begin{equation}
    \ket{\mathcal{C}}=\bigotimes_{j=1}^{L/2} \ket{c}_j\comma 
    \ket{c}_j=\ket{0}_j\otimes\ket{0}_{j+L/2}+\ket{\!\!\uparrow}_j\otimes\ket{\!\!\downarrow}_{j+L/2}-\ket{\!\!\downarrow}_j\otimes\ket{\!\!\uparrow }_{j+L/2}\,.\label{eq:SU126C}
\end{equation}
Notice that the identification of fermions across the chain forms the singlet state $\ket{\!\!\uparrow\downarrow}-\ket{\!\!\downarrow\uparrow}$, which reflects the two-point functions \eqref{eq:Wick} of the fermions.

To study this spin-chain using the Bethe ansatz, we first need to choose a vacuum. Unlike bosonic spin-chains, there are distinct pseudovacua, corresponding to different gradings of the Lie superalgebra. One choice is to take the vacuum to be the all-$0$ state
\begin{equation}
     \ket{\emptyset}_{(L)}=\bigotimes_{j=1}^L \ket{0}_j\,,
\end{equation}
which corresponds to the $\SU(1|2)$ Dynkin diagram
\begin{align}
\begin{tikzpicture}[cross/.style={path picture={ 
			\draw[black,thick]
			(path picture bounding box.south east) -- (path picture bounding box.north west) (path picture bounding box.south west) -- (path picture bounding box.north east);
	}}]
	\draw [thick](0,0)--(1,0);
	\draw[thick,dashed] (1,0)--(2,0)--(3,0);
	\draw (1,0)node[circle,draw=black,fill=white,cross]{};
	\draw(2,0) node[circle,draw=black,fill=white]{};
\end{tikzpicture}\label{eq:SU12Dynkin}
\end{align}
and Dynkin labels $q=(1,0)$.
We refer the reader to Appendix \ref{sec:Dynkin} for an introduction to Dynkin diagrams and labels for supergroups and their implications for the superspin chain. The upshot is that we should consider two sets of roots $\bm{u}=(\bm{u}^1,\bm{u}^2)$ or equivalently their associated $Q$-functions $Q_1,Q_2$ \eqref{eq:GenQ}, with the $\bm{u}^1$ roots being momentum-carrying. We denote the corresponding state $\ket{\bm{u}}$, whose wave function is given by \eqref{eq:SU12WaveFunc}. The momentum-carrying magnons excite the 0-state $\ket{0}$ into $\ket{\!\!\uparrow}$, while the auxiliary magnons, whose rapidities are $\bm{u}^2$, flip this spin $\ket{\!\!\uparrow}\mapsto\ket{\!\!\downarrow}$. In this case, the number of roots must satisfy $M_1\geq M_2$. 

Another choice of the vacuum is to take the all-$\uparrow$ state, with the momentum-carrying magnons corresponding to $\ket{\!\!\uparrow}\rightarrow \ket{0}$ and the auxiliary magnons corresponding to $\ket{0}\rightarrow \ket{\!\!\downarrow}$. This corresponds to a different grading of the superalgebra, given by 
\begin{align}
\begin{tikzpicture}[cross/.style={path picture={ 
			\draw[black,thick]
			(path picture bounding box.south east) -- (path picture bounding box.north west) (path picture bounding box.south west) -- (path picture bounding box.north east);
	}}]
	\draw [thick](1,0)--(2,0);
	\draw[thick,dashed] (0,0)--(1,0) (2,0)--(3,0);
	\draw (1,0)node[circle,draw=black,fill=white,cross]{};
	\draw(2,0) node[circle,draw=black,fill=white,cross]{};
\end{tikzpicture}\label{eq:SU12DynkinAlt}
\end{align}
We can then again construct eigenstates by using two sets of roots $\td{\bm{u}}=(\td{\bm{u}}_1,\td{\bm{u}}_2)$ and obtain the state $\ket{\td{\bm{u}}}$, whose wave function is again given by \eqref{eq:SU12WaveFunc}, but now with a different Cartan matrix $C$ and Dynkin label $q$.
Of course, the choice of vacuum does not change the space of eigenstates. Thus, for each state $\ket{\bm{u}}$ in the grading \eqref{eq:SU12Dynkin}, there is a corresponding state $\ket{\td{\bm{u}}}$ in the grading \eqref{eq:SU12DynkinAlt}.\footnote{Note that both states will generically not be highest weight, but they will be members of the same $\SU(1|2)$ multiplet.} The relationship between the roots, or rather their corresponding $Q$-functions, in different gradings is known as the {\it fermionic duality}. For the case at hand, it reads
\begin{equation}
\begin{split}
    Q_1(x)\td{Q}_1(x)&=\frac{1}{i(L-M_2)}\left[\left(x+\frac{i}{2}\right)^LQ_2\left(x-\frac{i}{2}\right)-\left(x-\frac{i}{2}\right)^LQ_2\left(x+\frac{i}{2}\right)\right]\,,
    \\
    Q_2(x)&=\td{Q}_2(x)\,.\label{eq:duality}
\end{split}
\end{equation}
Importantly, parity-symmetric states \eqref{eq:PairedStates} are mapped to parity-symmetric states under a change of vacuum. If there are $M_1$ momentum-carrying roots in the grading \eqref{eq:SU12Dynkin}, there will be $\td{M}_1=L+M_2-M_1-1$ roots in the grading \eqref{eq:SU12DynkinAlt}: this implies that if there are no zero roots in the original grading, there will be a zero root in the new grading.

Since the Bethe equations depend on the choice of gradings, so do the Gaudin matrix \eqref{eq:genG} and, for parity-symmetric states, its superdeterminant \eqref{eq:genGpGm}.
The transformation of the superdeterminant was conjectured in  \cite{Kristjansen:2020vbe}. If $Q_1(0)\neq 0$, the superdeterminant in the new grading is conjectured to be given by
\begin{equation}
    \frac{\det \td{G}_+}{\det \td{G}_-}=(2i)^L(L-M_2)\frac{Q_1(0)\td{Q}_1^{\circ}(0)}{Q_2(i/2)}\frac{\det G_+}{\det G_-}\,,\label{eq:overlaptransform}
\end{equation}
where we used the notation $\td{Q}^{\circ}(x)$ to denote that the root at $x=0$ of the dual Q-function has been excluded from the product in \eqref{eq:GenQ}.
In this sector, a closed-form expression for the overlap between the Bethe state and the crosscap state is not known in the literature. One can anticipate that the simple relationship between the overlap and the ratio of determinants \eqref{eq:SU2overlap} and \eqref{eq:SO6overlap} cannot carry over directly to the current case, since the ratio $\det G_+/\det G_-$ is grading dependent while the overlap should not depend on the choice. 

Instead of trying to derive an analytic formula, we numerically computed the overlaps for various states up to length $L=16$ and compared the result against the superdeterminant in the grading \eqref{eq:SU12Dynkin}. The outcomes of this analysis can be summarized as follows:
\begin{itemize}
\item In order for the overlap to be nonzero, the multiplicities of the magnons need to satisfy $M_1=2M_2$. This is just a simple consequence of the structure of the crosscap state \eqref{eq:SU126C}. 
\item As in other sectors, the overlap is nonzero only if the eigenstate is parity-symmetric $\bm{u}=-\bm{u}$, suggesting integrability of the crosscap state. Unlike for bosonic spin chains, this does not imply that there must be an even number of roots on each node: instead, there can be roots at $x=0$, but only on the second node, in line with the statement $M_1=2M_2$.
\item For states without such zero-roots, we find that the numerical results agree with the following analytic formula:
\begin{equation}
\frac{\langle\bm{u}|\mathcal{C}\rangle}{{\langle\bm{u}|\bm{u}\rangle}^{\frac{1}{2}}}= \sqrt{\frac{Q_{1}(0)}{Q_2(i/2)^2}\frac{\det G_+}{\det G_-}}\,.\label{eq:su12overlap}
\end{equation}
\item \eqref{eq:su12overlap} does not hold for states with a zero root (which, as explained earlier, can appear only on the second node). Instead, we find numerically that such configurations satisfy
\begin{equation}
    \frac{\langle\bm{u}|\mathcal{C}\rangle}{\langle\bm{u}|\bm{u}\rangle^{\frac{1}{2}}}= \sqrt{\frac{Q_{1}(i/2)}{Q^{\circ}_2(0)Q^{\circ}_2(i)}\frac{\det G_+}{\det G_-}}\,,\label{eq:su12overlap0}
\end{equation}
where $Q^\circ$ was defined above and $G_\pm$ are defined in Appendix \ref{sec:parstate}.
That the overlap formula depends on whether or not zero roots are present is in stark contrast to the corresponding formulae for integrable boundary states \cite{Piroli:2017sei,deLeeuw:2015hxa}, where the same formula works regardless of the presence of zero roots. We will leave the explanation of the origin of this difference as an open question for the future.
\item The equations \eqref{eq:su12overlap} and \eqref{eq:su12overlap0} have the attractive feature that it transforms nicely under the fermionic duality transformation \eqref{eq:overlaptransform}. Under the duality transformation \eqref{eq:SU12DynkinAlt}, the formula \eqref{eq:su12overlap}, which originally holds for a state without zero roots, instead describes a state with a zero root at a momentum-carrying node. When written in terms of the new roots, the overlap formula takes the form\footnote{This differs from what we would have obtained using \eqref{eq:overlaptransform} by some kinematical factors that are independent of $u$. This is because the state $\ket{\bm{u}}$ is strictly speaking a descendant in the dual grading \cite{leeuw_two-point_2017,Kristjansen:2020vbe}.}
\begin{equation}
    \frac{\langle\bm{\td{u}}|\mathcal{C}\rangle}{\langle\td{\bm{u}}|\td{\bm{u}}\rangle^{\frac{1}{2}}}=\sqrt{\frac{1}{\td{Q}^{\circ}_1(0)\td{Q}_2(i/2)}\frac{\det\td{G}_+}{\det\td{G}_-}}\,,
\end{equation}
where the tildes denote that we are now evaluating the expressions in the dual grading and with the dual Bethe roots. We emphasise that the fact that this expression is written only in terms of new Bethe roots is a non-trivial property. It required $Q_1$ to only show up as $Q_1(0)^1$ in \eqref{eq:su12overlap}, with no other powers possible. Similarly, \eqref{eq:su12overlap0}, which describes a setup with a zero root at the auxiliary level in either grading, can be written in terms of the new roots in the form
\begin{equation}
    \frac{\langle\bm{\td{u}}|\mathcal{C}\rangle}{\langle\td{\bm{u}}|\td{\bm{u}}\rangle^{\frac{1}{2}}}=\sqrt{\frac{1}{\td{Q}_1^{\bullet}(i/2)\td{Q}^{\circ}_2(i)}\frac{\det\td{G}_+}{\det\td{G}_-}}\,,
\end{equation}
where the roots of $\td{Q}^\bullet_1(x)$ at $x=+i/2$ and $x=-i/2$ (whose existence follows from \eqref{eq:duality}) have been removed.
\end{itemize}
\subsection{Full \texorpdfstring{$\PSU(2,2|4)$}{PSU(2,2|4)}-sector}\label{sec:PSU224}
Let us end by discussing the complete sector of the theory, which we denote here by $\text{PSU}(2,2|4)$, as is appropriate in the Lorentzian theory. States can be described using seven sets of nested Bethe roots, or alternatively seven $Q$ functions $Q_1,\dots ,Q_7$. Furthermore, since we are dealing with a superalgebra, there is the subtlety mentioned above on the choice of vacuum. We choose to focus on the ``Beauty'' grading, whose Dynkin diagram takes the form
\begin{align}
\begin{tikzpicture}[cross/.style={path picture={ 
			\draw[black,thick]
			(path picture bounding box.south east) -- (path picture bounding box.north west) (path picture bounding box.south west) -- (path picture bounding box.north east);
	}}]
	\draw [thick](-2,0)--(2,0);
	\draw[thick,dashed] (-4,0)--(-2,0) (2,0)--(4,0);
	\draw (-2,0) node[circle,draw=black,fill=white,cross]{};
	\draw(2,0) node[circle,draw=black,fill=white,cross]{};
    \draw(-3,0) node[circle,draw=black,fill=white]{};
    \draw(-1,0) node[circle,draw=black,fill=white]{};
    \draw(0,0) node[circle,draw=black,fill=white]{};
    \draw(1,0) node[circle,draw=black,fill=white]{};
    \draw(3,0) node[circle,draw=black,fill=white]{};
\end{tikzpicture}\label{eq:Beauty}
\end{align}
where only the middle node is momentum-carrying.
This choice is distinguished by having a central SO(6) sector and two $\text{SU}(1|2)$ ``wings'' on either side.

As we did in the SU$(1|2)$ sector, ideally one should compute the overlap for various states in this sector and infer a closed-form expression. However, performing such a computation that is sensitive to the full details of $\PSU(2,2|4)$ is rather laborious. Therefore, we take a shortcut and present an educated guess for the overlap based on the results obtained in other sectors. We conjecture that the overlaps vanish unless the Bethe state is parity symmetric $\bm{u}=-\bm{u}$, whose properties are briefly reviewed in Appendix \ref{sec:parstate}. Given the different nature of states with and without zero-roots seen in Section \ref{sec:SU12}, we will focus for simplicity on states without zero-roots, for which we conjecture that the overlap is given by
\begin{equation}
    \frac{\langle\bm{u}|\mathcal{C}\rangle}{\langle\bm{u}|\bm{u}\rangle^{\frac{1}{2}}}= \sqrt{\frac{Q_2(0)Q_6(0)}{Q_1(i/2)^2Q_7(i/2)^2}\frac{\det G_+}{\det G_-}}\,.\label{eq:psu224overlap}
\end{equation}
This formula was motivated by the following consideration:
\begin{enumerate}
    \item In the absence of excitations on the wings, the result should readily reduce to that in the SO(6)-sector. Given the SO(6) overlap formula \eqref{eq:SO6overlap}, this requires that the corresponding $\text{\PSU}^(2,2|4)$ formula does not depend on $Q_3,Q_4$ or $Q_5$.
    \item Since the wings have the structure of $\text{SU}(1|2)$ sectors, it is natural to expect that the dependence on $Q_1,Q_2,Q_6$ and $Q_7$ agrees with the ratios of $Q$-functions in \eqref{eq:su12overlap}. This property is indeed satisfied in a related setup of the D3-D5 overlap formula in the Beauty grading \cite{Kristjansen:2020vbe}. We should, however, note that this is an educated guess rather than a requirement: there is no subsector in this choice of grading that reduces to SU$(1|2)$ and hence we cannot impose the reduction to the SU$(1|2)$ formula as a logical constraint.
\end{enumerate}
This conjecture also passes more refined consistency checks/sanity checks:
\begin{enumerate}
    \item[3.] We can relate the conjecture more directly to the SU$(1|2)$ formula \eqref{eq:su12overlap} by performing fermionic dualities and converting the Beauty grading to a grading in which the middle sector is $\text{SU}(1|2)$. Such a grading is given by
    \begin{align}
\begin{tikzpicture}[cross/.style={path picture={ 
			\draw[black,thick]
			(path picture bounding box.south east) -- (path picture bounding box.north west) (path picture bounding box.south west) -- (path picture bounding box.north east);
	}}]
	\draw [thick](-2,0)--(0,0) (2,0)--(4,0);
	\draw[thick,dashed] (-4,0)--(-2,0) (0,0)--(2,0);
	\draw (-2,0) node[circle,draw=black,fill=white,cross]{};
	\draw(2,0) node[circle,draw=black,fill=white,cross]{};
    \draw(-3,0) node[circle,draw=black,fill=white]{};
    \draw(-1,0) node[circle,draw=black,fill=white]{};
    \draw(0,0) node[circle,draw=black,fill=white,cross]{};
    \draw(1,0) node[circle,draw=black,fill=white]{};
    \draw(3,0) node[circle,draw=black,fill=white]{};
\end{tikzpicture}\label{eq:fermionic}
\end{align}
Indeed, nodes 4 and 5 now take the form of the $\text{SU}(1|2)$ Dynkin diagram \eqref{eq:SU12Dynkin}. Using the fermionic duality rules \cite{Kristjansen:2020vbe} \eqref{eq:overlaptransform} and the QQ-relations, which relate the $Q$-functions in different choices of grading, we obtain the overlap formula
\begin{equation}
    \frac{\langle\td{\bm{u}}|\mathcal{C}\rangle}{\langle\td{\bm{u}}|\td{\bm{u}}\rangle^{\frac{1}{2}}}=\sqrt{\frac{\td{Q}_2(0)\td{Q}_4(0)\td{Q}_6(0)}{\td{Q}_1(i/2)^2\td{Q}_5(i/2)^2}\frac{\det \td{G}_+}{\det \td{G}_-}}\,,\label{eq:psu224overlapferm}
\end{equation}
where $\td{u}$, $\td{Q}$ and $\td{G}$ denote the roots, the $Q$-functions and the Gaudin matrix in the new grading. In the case where only $\td{Q}_4$ and $\td{Q}_5$ are non-trivial, this precisely reproduces \eqref{eq:su12overlap}.
    \item[4.] Lastly, we can perform a series of dualities taking us from the Beauty grading to the alternating grading, whose Dynkin diagram is
        \begin{align}
\begin{tikzpicture}[cross/.style={path picture={ 
			\draw[black,thick]
			(path picture bounding box.south east) -- (path picture bounding box.north west) (path picture bounding box.south west) -- (path picture bounding box.north east);
	}}]
	\draw [thick](-4,0)--(-3,0) (-1,0)--(1,0) (3,0)--(4,0);
	\draw[thick,dashed] (-3,0)--(-1,0) (1,0)--(3,0);
	\draw (-2,0) node[circle,draw=black,fill=white]{};
	\draw(2,0) node[circle,draw=black,fill=white]{};
    \draw(-3,0) node[circle,draw=black,fill=white,cross]{};
    \draw(-1,0) node[circle,draw=black,fill=white,cross]{};
    \draw(0,0) node[circle,draw=black,fill=white]{};
    \draw(1,0) node[circle,draw=black,fill=white,cross]{};
    \draw(3,0) node[circle,draw=black,fill=white,cross]{};
\end{tikzpicture}\label{eq:alternating}
\end{align}
Performing the necessary fermionic dualities, one finds that  the ansatz \eqref{eq:psu224overlap} becomes
\begin{equation}
    \frac{\langle\td{\bm{u}}|\mathcal{C}\rangle}{\langle\td{\bm{u}}|\td{\bm{u}}\rangle^{\frac{1}{2}}}=\sqrt{\frac{\td{Q}_1(i/2)\td{Q}_3(i/2)\td{Q}_5(i/2)\td{Q}_7(i/2)}{\td{Q}_2^\circ(0)\td{Q}_2(i)\td{Q}_6^\circ(0)\td{Q}_6(i)}\frac{\det \td{G}_+}{\det \td{G}_-}}\,.\label{eq:psu224overlapalt}
\end{equation}
Like in the D3-D5 overlap formula \cite{Kristjansen:2020vbe}, this formula is perfectly symmetric in $(1,3,5,7)$ and in $(2,6)$, reflecting the structure of the Dynkin diagram.
\end{enumerate}
The fact that it passes all these sanity checks makes \eqref{eq:psu224overlap} a plausible formula for the full $\PSU(2,2|4)$ sector. However, we emphasize that these checks are by no means a proof. It remains a very interesting open problem to check the formula more extensively by computing the overlaps directly and ultimately to derive it analytically.  
\section{Comparison with crosscaps on the worldsheet}\label{sec:classical}
So far we have worked at weak coupling in $\mathcal N=4$ SYM. We now turn to the strong-coupling regime, where the theory is described holographically by type IIB string theory on $\AdS_5\times \text{S}^5$. In this section we focus on operators whose dimensions scale as
 $\Delta =O( \sqrt{\lambda})$  as $\lambda\to\infty$. Such operators are dual to macroscopic semiclassical strings in $\AdS_5\times \text{S}^5$, whose dynamics is governed by the classical integrable sigma model on this background \cite{Metsaev:1998it,Bena:2003wd}. Given the well-known relation between integrable spin chains at weak coupling and semiclassical strings at strong coupling \cite{Frolov:2002av,Kazakov:2004qf}, it is natural to ask for the worldsheet counterpart of the spin-chain crosscap states encountered above. The relevant objects are integrable crosscap states of the $\AdS_5\times \text{S}^5$ sigma model. These have been classified in \cite{Gombor:2022deb}. Below we briefly review this classification and identify the particular crosscap corresponding to $\mathcal N=4$ SYM on ${\sf CP}$-$\mathbb{RP}^4$ considered in this paper.
 
\subsection{Crosscaps in integrable sigma models: review}
A two-dimensional  sigma model on the world-sheet $\Sigma=\mathbb{R}_\tau\times S^1_\sigma$ is classically integrable if there exists a connection $L(z)$, which depends on an auxiliary ``spectral parameter'' $z\in \mathbb{CP}^1$, for which the corresponding Wilson line, called the monodromy matrix,
\begin{equation}
    U(\gamma|z)=\text{Pexp}\left[\int_\gamma -L(z)\right]\label{eq:Monodromy}
\end{equation}
is independent of the chosen path $\gamma$, for any value of the spectral parameter. Indeed, one can then construct the transfer matrix
\begin{equation}
    T(\tau_0|z)=\Tr U(\gamma_\sigma(\tau_0)|z)\,,\label{eq:Transfer}
\end{equation}
where $\gamma_\sigma(\tau)$ is the path wrapping $S^1_\sigma$ at a fixed time $\tau=\tau_0$. Thanks to the path independence, $T(\tau_0|z)$ is independent of $\tau_0$, and hence we drop the argument $\tau_0$ in what follows. If we expand it as a function of $z$, each coefficient of $T(z)$ is a conserved quantity. The path-independence of $U$ normally originates in the local form of the equations of motion of the sigma model, which can be recast in the form
\begin{equation}
    \forall z\in \mathbb{CP}^1:[D_\mu(z),D_\nu(z)]=0\,,\qquad D_\mu(z)=\partial_\mu+L_\mu(z)\,.\label{eq:LaxFlat}
\end{equation}

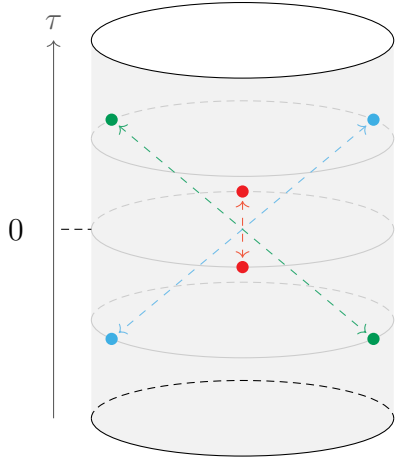
\begin{figure}
    \centering
        \begin{tikzpicture}
              \def\rx{2}\def\ry{.5}\def\T{2.5}
    arc[start angle=180,end angle=0,x radius=\rx,y radius=\ry]
    --(\rx,0) arc[start angle=0,end angle=-180,x radius=\rx,y radius=\ry];
  \fill[black!05] (-\rx,-\T)--(-\rx,\T)
    arc[start angle=180,end angle=0,x radius=\rx,y radius=\ry]
    --(\rx,-\T) arc[start angle=0,end angle=-180,x radius=\rx,y radius=\ry];
    \draw[fill=white] (0,\T)
  ellipse[x radius=\rx,y radius=\ry];
  \draw[densely dashed] (-\rx,-\T)
    arc[start angle=180,end angle=0,x radius=\rx,y radius=\ry];
  \draw[] (-\rx,-\T)
    arc[start angle=180,end angle=360,x radius=\rx,y radius=\ry];
  \draw[black!20,densely dashed] (-\rx,0)
    arc[start angle=180,end angle=0,x radius=\rx,y radius=\ry];
  \draw[black!20] (-\rx,0)
    arc[start angle=180,end angle=360,x radius=\rx,y radius=\ry];
    \foreach \h in {1.2,-1.2} {
  \draw[black!20,densely dashed] (-\rx,\h)
    arc[start angle=180,end angle=0,
        x radius=\rx,y radius=\ry];
  \draw[black!20] (-\rx,\h)
    arc[start angle=180,end angle=360,
        x radius=\rx,y radius=\ry];
}
  \coordinate (B1) at ({\rx*cos(30)},{1.2+\ry*sin(30)});
  \coordinate (B2) at ({-\rx*cos(30)},{-1.2-\ry*sin(30)});
    \coordinate (G1) at ({\rx*cos(150)},{1.2+\ry*sin(150)});
  \coordinate (G2) at ({-\rx*cos(150)},{-1.2-\ry*sin(150)});
  \coordinate (R1) at ({\rx*cos(90)},{\ry*sin(90)});
  \coordinate (R2) at ({\rx*cos(-90)},{\ry*sin(-90)});
    \draw[<->,CornflowerBlue!80,dashed,shorten <=3.5pt,shorten >=3.5pt] (B1)--(B2);
  \fill[CornflowerBlue] (B1) circle[radius=2.3pt];
  \fill[CornflowerBlue] (B2) circle[radius=2.3pt];
    \draw[<->,ForestGreen!80,dashed,shorten <=3.5pt,shorten >=3.5pt] (G1)--(G2);
  \fill[ForestGreen] (G1) circle[radius=2.3pt];
  \fill[ForestGreen] (G2) circle[radius=2.3pt];
\draw[<->,Red!80,dashed,shorten <=3.5pt,shorten >=3.5pt] (R1)--(R2);
  \fill[Red] (R1) circle[radius=2.3pt];
  \fill[Red] (R2) circle[radius=2.3pt];
  \draw[->,black!65] (-2.5,-2.5)--(-2.5,2.5) node[above] {$\tau$};
  \draw[densely dashed] (-2.4,0)--(-2,0);
  \node[anchor=east] at (-2.75,0) {$0$};
\end{tikzpicture}
    \caption{Crosscap identification on the cylinder $(\tau, \sigma)\sim
(-\tau,\sigma +\pi)$. On the $\tau=0$ circle, the identification pairs antipodal points.}
    \label{fig:WorldSheetCrosscap}
\end{figure}

Let us now discuss crosscaps. To illustrate the basic idea in a simple setup, we consider a principal chiral model whose target space is $G$. (As discussed in \cite{Gombor:2022gdj}, one can readily generalize it to a coset manifold relevant for AdS$_5\times$S$^{5}$.) In this case, the Lax connection is given by
\begin{equation}
    L_\mu(z)=\frac{J_\mu}{1+z^2}+\frac{z\,\varepsilon_{\mu \nu}J^\nu}{1+z^2}\,,\label{eq:PCMLax}
\end{equation}
where $J\in\mathfrak{g}$ is a $G$-current built from the fields of the theory. The theory has a global $G$-symmetry, with $g\in G$ acting as
\begin{equation}
    L\mapsto gLg^{-1}\label{eq:LgLg}
\end{equation}
To define the crosscap, we identify the worldsheet points $(\tau,\sigma)$ under the involution
$\iota(\tau,\sigma)=(-\tau,\sigma+\pi)$, as illustrated in Figure~\ref{fig:WorldSheetCrosscap}. After Wick rotation to the Euclidean worldsheet and mapping the cylinder to the radial coordinate of $\mathbb{R}^2$, this becomes precisely the conformal inversion that defines a crosscap in two dimensions. As discussed in Section~\ref{sec:setup}, in even-dimensional CFTs conformal inversion is equivalent, up to a conformal transformation, to parity. In the present case, this is the worldsheet parity $\mathfrak{P}$.
As in $\mathcal{N}=4$ SYM, the involution $\iota$ may also be combined with an internal $\mathbb{Z}_2$ symmetry $\eta$, satisfying $\eta^2=1$. The current must then obey
\begin{equation}
J_\mu(\tau,\sigma)=\iota\left[J_\mu(\tau,\sigma)\right]=(-1)^\mu J^{\eta}_\mu(-\tau,\sigma+\pi)\,,
\end{equation}
where $(-1)^x=-(-1)^t=1$, and $J_\mu^\eta$ denotes the current after the action of $\eta$. Using the Lax connection \eqref{eq:PCMLax}, this implies
\begin{equation}
    \iota\left[L_\mu(\tau,\sigma|z)\right]
    =
    (-1)^\mu L_\mu^\eta(-\tau,\sigma+\pi|-z)\,.
    \label{eq:LC}
\end{equation}
The sign flip $z\to -z$ arises because $\varepsilon_{\mu\nu}$ is odd under the orientation-reversing involution. Hence, on the crosscap slice $t=0$, the Lax connection satisfies
\begin{align}
L_\sigma(0,\sigma|z)=L_\sigma^{\eta}(0,\sigma+\pi|-z)\,.
\label{eq:kLLk}
\end{align}

Generally, the $\mathbb{Z}_2$ action $\eta$ may contain both inner and outer automorphisms of the worldsheet symmetry group. We therefore write
\begin{align}
L_\sigma^{\eta}=\kappa^{-1}L_\sigma^{\mathfrak{C}}\kappa\,,
\end{align}
where $\kappa$, often called the reflection matrix, implements the inner automorphism, while $\mathfrak{C}$ denotes the outer automorphism, which we refer to as worldsheet charge conjugation. 
Then, the condition \eqref{eq:kLLk} leads to the following condition on the transfer matrix on the crosscap slice $\tau=0$:
\begin{equation}
T(z)=T^{\mathfrak{C}}(-z)\,,
\end{equation}
which is the classical analogue of the spin-chain crosscap condition \eqref{eq:QonC}.

Applying \eqref{eq:kLLk} twice gives
\begin{equation}
    L_\sigma(0,\sigma_0|z)
    =
    L_\sigma^{\eta^2}(0,\sigma_0|z)
    =
    \kappa^{-1}
    \left[
        \kappa^{-1}
        L_\sigma^{\mathfrak{C}}(0,\sigma_0|z)
        \kappa
    \right]^{\mathfrak{C}}
    \kappa\,.
    \label{eq:LkkLkk}
\end{equation}
Since $\eta^2=1$, the first and last expressions must agree identically. This therefore imposes a consistency condition on the allowed reflection matrices $\kappa$.
Finally, recall that the theory has a global $G$ symmetry \eqref{eq:LgLg}. The crosscap generally reduces this symmetry to the subgroup that preserves the involution. Thus only those $g\in G$ satisfying
\begin{equation}
    g^{-1}\kappa^{-1}(gLg^{-1})^{\mathfrak{C}}\kappa g
    =
    \kappa^{-1}L^{\mathfrak{C}}\kappa
    \label{eq:gkgLgkg}
\end{equation}
remain unbroken. A nontrivial charge conjugation $\mathfrak{C}$ and/or reflection matrix $\kappa$ therefore generically breaks part of the original global symmetry. When both are trivial, $\kappa=\mathfrak{C}=1$, the full symmetry is preserved as can be readily verified from \eqref{eq:gkgLgkg}.

Before concluding this subsection, let us emphasize one point. On the worldsheet,
the crosscap is defined by combining worldsheet parity $\mathfrak{P}$ with a
$\mathbb{Z}_2$ involution $\eta$, where $\eta$ itself may contain both an
inner automorphism $\kappa$ and charge conjugation $\mathfrak{C}$. The correspondence with $\mathcal{N}=4$ SYM exchanges the roles of
parity and charge conjugation. The charge conjugation $\mathsf{C}$ of
$\mathcal{N}=4$ SYM, which produces the antipodal pairing on the spin chain,
is mapped to worldsheet parity $\mathfrak{P}$. Conversely, the parity
$\mathsf{P}$ of $\mathcal{N}=4$ SYM, which controls the breaking of the global
symmetry, is mapped to the worldsheet involution
$\eta=\kappa\circ\mathfrak{C}$:
\begin{align}
    \mathsf{C}
    &\leftrightarrow
    \mathfrak{P},
    \qquad
    \mathsf{P}
    \leftrightarrow \eta=\kappa\circ\mathfrak{C}.
\end{align}

\subsection{The coset crosscap}
We now focus on the specific sigma model relevant for $\mathcal{N}=4$ SYM, namely the coset model with target space $\frac{\text{PSU}^*(4|4)}{\SO(1,4)\times \SO(5)}$. The bosonic target space is the Wick-rotated ${\rm AdS}_5\times {\rm S}^{5}$, i.e.~${\rm H}_5\times {\rm dS}_{5}$, where ${\rm H}_{5}$ is the Euclidean hyperbolic space and ${\rm dS}_{5}$ is the de Sitter space. 

The classification of integrable crosscaps (that is, enumerating all possible choices of $\mathfrak{C}$ and $\kappa$) in this background was performed in \cite{Gombor:2022deb}, and we will not repeat it here. Instead, we focus on the specific choice that corresponds to $\mathcal{N}=4$ SYM on $\mathbb{RP}^4$ with charge conjugation. The relevant choice of the $\mathbb{Z}_2$ involution $\sigma$ is given by a combination of
\begin{equation}
    \kappa=\left(
\begin{array}{c|c}
\mathbb{J}_4 & 0 \\
\hline
0 & -\mathbb{J}_4
\end{array}
\right)\,,
\end{equation}
and the charge conjugation $\mathfrak{C}$, given by\footnote{The supertransposition itself is an order $4$ involution rather than order 2. Only after one combines it with the conjugation by $M$ does it become a $\mathbb{Z}_2$ involution.}
\begin{equation}
    L^\mathfrak{C}=-M^{-1}L^{\mathsf{ST}}M\,.
\end{equation}
Here $\mathsf{ST}$ denotes supertransposition, which acts as
\begin{equation}
\left(
\begin{array}{c|c}
A&B\\
\hline
C&D
\end{array}
\right)^\mathsf{ST}=\left(
\begin{array}{c|c}
A^\mathsf{T}&-C^\mathsf{T}\\
\hline
B^\mathsf{T}&D^\mathsf{T}
\end{array}
\right)\,,
\end{equation}
where $A,B,C,D$ are $4\times 4$ matrices and $M$ is given by
\begin{equation}
M=\left(
\begin{array}{c|c}
\mathbb{I}_4 & 0 
 \\
\hline
0 & \mathbb{J}_4
\end{array}
\right)\,,
\end{equation}
(Recall the definition of the symplectic matrix in \eqref{eq:defJ}.)
One can verify that the consistency condition \eqref{eq:LkkLkk} follows from 
\begin{equation}
    M^\mathsf{ST}=M^{-1}\,,\quad \kappa^{\mathsf{ST}}=\kappa^{-1}=-\kappa\,,\quad M\kappa=\kappa M=\left(
\begin{array}{c|c}
\mathbb{J}_4&0\\
\hline
0&\mathbb{I}_4
\end{array}
\right)\,.\label{eq:MkkM}
\end{equation}

What is the global symmetry in the presence of this crosscap? \eqref{eq:gkgLgkg} implies that $g\in \text{PSU}^*(4|4)$ is a symmetry iff
\begin{equation}
    g^\mathsf{ST} M\kappa g=M\kappa \,.\label{eq:gMkgMk}
\end{equation}
Let us focus on bosonic symmetries and write $g$ in the form
\begin{equation}
    g=\left(
\begin{array}{c|c}
g_1&0\\
\hline
0&g_2
\end{array}
\right)\,.
\end{equation}
Then \eqref{eq:MkkM} implies that
\begin{equation}
    g_1^\mathsf{T} \mathbb{J}_4 g_1=\mathbb{J}_4\,,\qquad g_2^\mathsf{T} \mathbb{I}_4 g_2=\mathbb{I}_4\,,
\end{equation}
which implies that the bosonic part of the symmetry group is a real slice of $\text{USp}(4,\mathbb{C})\times \text{Spin}(4,\mathbb{C})\cong \text{Spin}(5,\mathbb{C})\times \text{Spin}(3,\mathbb{C})\times \text{Spin}(3,\mathbb{C})$, just like we saw in Section \ref{sec:N=4SYM}. And indeed, upon taking care of the reality conditions and including the fermionic symmetry, \eqref{eq:gMkgMk} implies that the symmetry is $\text{OSP}(4^*|4)$ \cite{Gombor:2022gdj}.

As is well known, a worldsheet crosscap has a spacetime interpretation as an
orientifold \cite{Sagnotti:1987tw}. In the present case, the crosscap condition
$L=L^\eta$, together with the preserved supergroup
$\mathrm{OSP}(4^\ast|4)$, is sufficiently restrictive to identify the
corresponding orientifold:
the breaking of the Euclidean AdS isometry
$\mathrm{SO}(5,1)\to \mathrm{SO}(5)$ implies that the $\mathbb Z_2$
involution has a fixed point in AdS. At the same time, the breaking of the
R-symmetry to $\mathrm{SO}(2,1)\times\mathrm{SO}(3)$ implies a
two-dimensional fixed locus inside the internal space. Together, these fixed
loci identify the relevant object as an O1-plane. In the next section, we will recover the same conclusion independently from
the pp-wave limit.

\section{BMN limit and crosscap on pp-wave}\label{sec:pp-wave}

Another tractable strong-coupling regime is the pp-wave, or BMN, limit \cite{Berenstein:2002jq}. The
starting point is a half-BPS state with large R-charge $J$, dual to a null
geodesic in AdS$_5\times$S$^5$ that carries large angular momentum along a
great circle of S$^5$. One then studies small string fluctuations around this
geodesic. In the large 't~Hooft-coupling limit, equivalently at large string
tension, this reduces to the spectrum of the string sigma model
in the local geometry near the geodesic, which is precisely the maximally supersymmetric pp-wave
background.

A key advantage of this limit is that, after fixing light-cone gauge, the
worldsheet theory becomes quadratic and its spectrum can be solved exactly by
elementary methods. In this section, we extend this construction to the
presence of an orientifold. We identify the specific orientifold that corresponds to $\mathcal N=4$ SYM on $\mathbb{RP}^4$ with charge
conjugation, and construct the corresponding crosscap state on the 
worldsheet.

Let us first review how the pp-wave geometry arises by zooming in on a null geodesic. As compared to what has been discussed in the literature, there is one point worth mentioning in the present setting. Since we are studying Euclidean $\mathcal{N}=4$ SYM, we use the complexified bulk theory on the $\mathrm{H}_5\times\mathrm{dS}_5$ slice associated with the non-compact $R$-symmetry real form described in Section~\ref{sec:N=4SYM}, where $\mathrm{H}_5$ denotes five-dimensional hyperbolic space, or Euclidean AdS, which is holographically dual to Euclidean $\mathcal{N}=4$ SYM. However, as we see later, since the orientifold sits at the intersection of the $\mathrm{H}_5\times\mathrm{dS}_5$ and Lorentzian $\mathrm{AdS}_5\times \mathrm{S}^5$ real slices, we can study it in the standard $\mathrm{AdS}_5\times \mathrm{S}^5$ and its pp-wave limit.

We describe $\mathrm{AdS}_5$ by embedding coordinates $X_M$, $M=-1,0,1,\ldots,4$, and $\text{S}^5$ by $Y_A$, $A=5,\ldots,10$. We reserve $\mu,\nu=1,\ldots,4$ for the AdS transverse directions and $a,b=7,\ldots,10$ for the sphere transverse directions. Their constraints are
\begin{equation}\label{eq:embedH}
 -X_{-1}^2-X_0^2+X_\mu X^\mu=-R^2,\qquad
 Y_{5}^2+Y_{6}^2+
 Y_a Y^a=R^2\,,
\end{equation}
We use the parametrization
\begin{equation}
\begin{split}\label{eq:embedpara}
 X_{-1}&=R\cosh\rho\cos t,\qquad
 X_0=R\cosh\rho\sin t,\qquad
 X_\mu=R\sinh\rho\,\Omega_\mu,\\
 Y_{5}&=-R\cos\theta\sin\psi,\qquad
 Y_{6}=R\cos\theta\cos\psi,\qquad
 Y_a=R\sin\theta\,\hat{\Omega}_a\,,
\end{split}
\end{equation}
where $\Omega_\mu\Omega_\mu=\hat{\Omega}_a\hat{\Omega}_a=1$ describe unit three-spheres. The metric is
\begin{equation}\label{eq:globalAdSS}
\begin{split}
 \Dd s^2=R^2\big[&-\cosh^2\rho\,\Dd t^2+\Dd\rho^2+
 \sinh^2\rho\,\Dd\Omega^2\\
 &+\cos^2\theta\,\Dd\psi^2+\Dd\theta^2+
 \sin^2\theta\,\Dd\hat{\Omega}^2\big]\,.
\end{split}
\end{equation}
We then take the Penrose-G\"uven limit of this space-time. We consider the limit $R\rightarrow\infty$ keeping $R\rho$ and $R\theta$ finite: this ensures that the $X_\mu/{Y}_{a}$ coordinates scale as $O(R^0)$, while the $X_{-1},X_0/{Y}_5,Y_6$ scale as $O(R)$. One can then observe that the $X_\mu/T_{a}$ will span four-dimensional flat space with $R\rho/R\theta$ playing the role of the radius
\begin{equation}
 X_\mu\longrightarrow x_\mu=R\rho\,\Omega_\mu,\qquad
 Y_a\longrightarrow y_a
 =R\theta\,\hat{\Omega}_a\,.
\end{equation}
Introducing
\begin{equation}\label{eq:lightcone}
 x^+=\frac{t+\psi}{2m},\qquad
 x^-=\frac{mR^2}{2}(t-\psi),\qquad m>0\,,
\end{equation}
and keeping $x^\pm$ and the transverse coordinates fixed gives the pp-wave metric
\begin{equation}\label{eq:PPmet}
 \Dd s^2=-4\Dd x^+\Dd x^- -m^2x_Ix_I(\Dd x^+)^2+
 \Dd x_I\Dd x_I\,.
\end{equation}
Here, $x^I=(x^\mu,y^a)$ denotes the eight transverse coordinates.

The spacetime is supported by a five-form flux. In the pp-wave limit, the five-form flux takes
the form
\begin{equation}\label{eq:PPRR}
 F_5=2m\,\Dd x^+\wedge\big[
 \Dd x^1\wedge\Dd x^2\wedge\Dd x^3\wedge\Dd x^4
 +\Dd y^{7}\wedge\Dd y^{8}
 \wedge\Dd y^{9}\wedge\Dd y^{10}\big]\,.
\end{equation}

\paragraph{Symmetries.}
The bosonic symmetries of the pp-wave follow by contracting the $\SO(4,2)\times\SO(6)$ isometry algebra of $\mathrm{AdS}_5\times \text{S}^5$. In the embedding basis, the metric takes the form $\eta_{MN}=\text{diag}(-1,-1,+1,+1,+1,+1)$, and the generators $J_{MN}$ satisfy
\begin{equation}
    \left[J_{MN},J_{KL}\right]=-i\left(\eta_{NK}J_{ML}+\eta_{MK}J_{LN}+\eta_{NL}J_{KM}+\eta_{ML}J_{NK}\right)\,,
\end{equation}
The sphere generators $J_{AB}$ satisfy the analogous relation instead using the Euclidean metric $\delta_{AB}$, and commute with all $J_{MN}$. In the contraction $R\to\infty$, each factor is organized into transverse $\SO(4)$ rotations, two $\SO(4)$ vectors and one scalar
\begin{equation}
    J_{MN}\mapsto J_{\mu\nu},P_{\mu},J_{+\mu},U_X\comma \qquad J_{AB}\mapsto J_{ab},P_{a},J_{+a},U_Y\,.
\end{equation}
The two scalar generators combine into the light-cone time translation and a central null translation. The Hermitian generators are given by (where again $I\in\{\mu,a\}$)
\begin{equation}
\begin{split}
 P_-&=-i\partial_{x^+}\,,\qquad P_+=\frac i2\partial_{x^-}\,,\\
 P_I&=-i\left[\cos(mx^+)\partial_I-\frac m2\sin(mx^+)x_I\partial_{x^-}\right]\,,\\
 J_{+I}&=-i\left[\frac1m\sin(mx^+)\partial_I+\frac12\cos(mx^+)x_I\partial_{x^-}\right]\,,\\
 J_{IJ}&=-i\left(x_I\partial_J-x_J\partial_I\right)\,.
\end{split}
\end{equation}
The last definition applies when $I,J$ lie in the same transverse four-plane. The light-cone Hamiltonian is $H_{\rm LC}=-P_-=i\partial_{x^+}$. Introducing $i\in\{+,\mu,a\}$, these 30 bosonic symmetry generators satisfy the algebra
\begin{equation}\label{eq:PPbos}
\begin{aligned}
\relax
 [J_{ij},J_{kl}]&=-i\left(\td\delta_{jk}J_{il}+\td\delta_{ik}J_{lj}
 +\td\delta_{jl}J_{ki}+\td\delta_{il}J_{jk}\right)\,,\\
 [P_i,P_j]&=0,\qquad
 [P_i,J_{jk}]=-i\left(\td\delta_{ij}P_k-\td\delta_{ik}P_j\right),\\
 [P_-,P_I]&=im^2J_{+I},\qquad [P_-,J_{+I}]=-iP_I\,,\\
 [P_-,J_{\mu\nu}]&=0\,,\qquad [P_-,J_{ab}]=0,
 \qquad [P_-,P_+]=0\,.
\end{aligned}
\end{equation}
Here $\td\delta_{ij}=\delta_{ij}-\delta_{+i}\delta_{+j}$, and $J_{\mu b}$ are excluded. Furthermore, the 32 supercharges $Q$ of $\mathrm{AdS}_5\times \text{S}^5$ are also carried over in the limit. Denoting the ten-dimensional tangent-space gamma matrices by $\Gamma_\pm,\Gamma_I$, the supercharges naturally split in two
\begin{equation}
    Q=Q_{+}+Q_{-}\comma \qquad Q_{\pm}\Gamma_{\mp}\equiv 0\,.
\end{equation}
Just like $x_+$ and $x_-$, $Q_+$ and $Q_-$ scale differently under the Penrose-G\"uven limit. For the following brackets, we use a complex basis of supercharges, with brackets with the Hermitian-conjugate supercharges following from $[B,Q^\dagger]=-[B,Q]^\dagger$ for a Hermitian bosonic generator $B$. One then finds their commutators with the bosonic symmetry generators to be of the form
\begin{equation}\label{eq:PPfer}
\begin{aligned}
\relax
 [J_{\mu\nu},Q_\pm]&=-\frac i2 Q_\pm\gamma_{\mu\nu}\,,\qquad
 [J_{ab},Q_\pm]=-\frac i2 Q_\pm\gamma_{ab}\,,\\
 [J_{+I},Q_-]&=-\frac i2 Q_+\gamma_{+I}\,,\qquad [J_{+I},Q_+]=0\,,\\
 [P_I,Q_-]&=\frac m2 Q_+\Pi\gamma_{+I}\,,\qquad [P_I,Q_+]=0\,,\\
 [P_-,Q_+]&=mQ_+\Pi\,,\qquad [P_-,Q_-]=0,\qquad [P_+,Q_\pm]=0\,.
\end{aligned}
\end{equation}
where $\Pi=\Gamma_1\Gamma_2\Gamma_3\Gamma_4$ and $\gamma_{ij}$ are the generators induced by the ten-dimensional Clifford algebra on chiral spinors. The presence of $\Pi$ explains why we have $\SO(4)\times\SO(4)$ rotational symmetry instead of $\SO(8)$: the metric would seemingly have an additional set of rotational symmetry generated by $J_{\mu b}$, but this is broken by the RR-field \eqref{eq:PPRR} and $\Pi$.

Notice that since $\gamma_{++}$ and $J_{++}$ vanish, $P_+$ is a central element of both \eqref{eq:PPbos} and \eqref{eq:PPfer}: it can thus be fixed to a constant
\begin{equation}
    P_+=\frac{\ell}{2\pi}\,,\label{eq:PL}
\end{equation}
where $\ell$ plays the role of the length of a string in lightcone gauge. Here $\alpha'=1$; restoring the string length gives $\ell=2\pi\alpha'P_+$.

\paragraph{$\mathbb{RP}^4$ in the pp-wave limit.}

As explained in Section \ref{sec:setup}, $\mathcal{N}=4$ SYM on $\mathbb{RP}^4$ arises by identifying points on $\mathbb{R}^4$ by $\mathsf{P}$ or $\mathsf{CP}$. As already discussed in \cite{Caetano:2022mus}, translating the action of $\mathsf{P}$ to the bulk is straightforward if one uses embedding space \eqref{eq:embedH}. 
First, we identify the sphere labels with the compactly continued scalar directions as
\begin{equation}\label{eq:spherelabels}
 (5,6,7,8,9,10)_{\text{S}^5}\longleftrightarrow (5,6,7,8,9,0)_{\mathrm{scalar}}\,.
\end{equation}
Concretely, $\mathsf{P}=\mathsf{I}\times \mathsf{R}$ considered in \eqref{eq:roundiota} and \eqref{eq:scP} naturally becomes
\begin{equation}\label{eq:AdSori}
\begin{split}
 &X_{-1}\mapsto X_{-1},\qquad X_0\mapsto-X_0,\qquad
 X_\mu\mapsto-X_\mu,\\
 &Y_A\mapsto(-1)^{A}Y_A,
 \qquad A=5,\ldots,10.
\end{split}
\end{equation}
In \eqref{eq:embedpara} this sends $T\mapsto-T$ and $\Psi\mapsto-\Psi$, preserving the reference geodesic. At $T=0$ its fixed locus is the AdS centre times the two-sphere
\begin{equation}\label{eq:fixedS2}
 Y_{5}=Y_{7}=Y_{9}=0\,,\qquad
 Y_{6}^2+Y_{8}^2+Y_{10}^2=R^2\,.
\end{equation}
In the pp-wave limit, the involution becomes
\begin{equation}\label{eq:ppori}
 x^\pm\mapsto-x^\pm\,,\qquad x_\mu\mapsto-x_\mu\,,\qquad
 y_a\mapsto
 \begin{cases}
 -y_a,&a=7,9,\\
 +y_a,&a=8,10.
 \end{cases}
\end{equation}
Its fixed two-plane is $x^+=x^-=x_\mu=y_{7}=y_{9}=0$, parametrized by $y_{8},y_{10}$. It preserves the rotations $\SO(4)\times\SO(2)\times\SO(2)$, together with $J_{+\mu},J_{+7},J_{+9}$ and $P_{8},P_{10}$.

Let us make a few comments. First, unlike the action of $\mathsf{P}$, the action of $\mathsf{C}$ is not geometric; it becomes the orientation reversal of the worldsheet as we see later. Second, the bulk theory obtained after the identification is expected to preserve 16 supercharges, just like the boundary theory described in Section~\ref{sec:N=4SYM}: we will return to these supercharges later.

\subsection{String theory on the pp-wave}

We now follow \cite{Berenstein:2002jq,metsaev_exactly_2002} in constructing the string theory on the pp-wave. We denote its Lorentzian worldsheet time by $\tau$, choose $x^+=\tau$, and rescale the transverse fields to absorb the string tension $T=(2\pi\alpha')^{-1}$. We retain the same symbols for these canonically normalized transverse fields. We obtain the action of 8 massive bosons $\{x_I\}$ and 8 massive two-dimensional Majorana fermions, whose two real eight-component fields are denoted by $\theta$ and $\bar\theta$
\begin{equation}
\begin{split}
 S=\int d\tau\int_0^\ell d\sigma\bigg\{&
 \frac12\left(\dot x_I^2-\acute{x}_I^2-m^2x_I^2\right)
 \\ +&\frac i2\left[\theta^{\mathsf T}\partial_+\theta+
 \bar\theta^{\mathsf T}\partial_-\bar\theta-2m\theta^{\mathsf T}\Pi\bar\theta\right]\bigg\}\,,
\end{split}
\end{equation}
where $\acute{x}$ denotes the derivative with respect to $\sigma$ and $\partial_\pm=\partial_\tau\pm\partial_\sigma$. The length of the string is related to $P_+$ by \eqref{eq:PL}, as is natural in light-cone gauge, and $\Pi$ was defined below \eqref{eq:PPfer}. The EOMs arising from the action are given by
\begin{equation}
 (\partial_\tau^2-\partial_\sigma^2+m^2)x^I=0\,,
 \qquad \partial_+\theta-m\Pi\bar\theta=0\,,
 \qquad \partial_-\bar\theta+m\Pi\theta=0\,,
\end{equation}
while the longitudinal coordinate $x^-$ is determined by the light-cone constraint
\begin{equation}
 -\frac{1}{\pi}\acute{x}^-+\dot x_I \acute{x}_I+
 \frac i2\left(\theta^{\mathsf T}\theta'+\bar\theta^{\mathsf T}\bar\theta'\right)=0\,.
 \label{eq:xm}
\end{equation}
The corresponding integrated constraint imposes closed-string level matching.
The closed string spectrum on this background is easily computed to be given by
\begin{equation}\label{eq:xtonshell}
\begin{split}
 x^I(\sigma,\tau)&={\frac{1}{\sqrt\ell}}\left[\cos(m\tau)x_0^I+\frac{\sin(m\tau)}m p_0^I\right]
 +i\sum_{n\ne0}a_n\left(\varphi_n\alpha_n^I+\bar\varphi_n\bar\alpha_n^I\right)\,,\\
 \theta(\sigma,\tau)&={\frac1{\sqrt\ell}}\left\{\cos(m\tau)\theta_0+
 \sin(m\tau)\Pi\bar\theta_0+
 \sum_{n\ne0}b_n\left(\varphi_n\beta_n+{i d_n}\bar\varphi_n\Pi\bar\beta_n\right)\right\}\,,\\{
 \bar\theta(\sigma,\tau)}&{=\frac1{\sqrt\ell}\left\{\cos(m\tau)\bar\theta_0-
 \sin(m\tau)\Pi\theta_0+
 \sum_{n\ne0}b_n\left(\bar\varphi_n\bar\beta_n-i d_n\varphi_n\Pi\beta_n\right)\right\}\,.
}\end{split}
\end{equation}
Here we have defined
\begin{equation}
\begin{split}
 \quad k_n=\frac{2\pi n}{\ell}\,,\quad&\omega_n=\sgn(n)\sqrt{k_n^2+m^2}\,,\quad  \varphi_n=e^{-i(\omega_n\tau-k_n\sigma)}\,,\quad
 {\bar\varphi_n=e^{-i(\omega_n\tau+k_n\sigma)}}\,,\\
 &a_n=\frac{\sgn(n)}{\sqrt{2\ell|\omega_n|}}\,,\qquad
 d_n=\frac{\omega_n-k_n}{m}\,,\qquad b_n=\frac1{\sqrt{1+d_n^2}}\,,
\end{split}
\end{equation}
and we have chosen to normalise the creation and annihilation operators such that
\begin{equation}
	\begin{split}
  &[x_0^I,p_0^J]=i\delta^{IJ}\,,\quad [\alpha_m^I,\alpha_n^J]=[\bar\alpha_m^I,\bar\alpha_n^J]
 =\sgn(m)\delta_{m+n,0}\delta^{IJ}\,,\\
 &\{\theta_0^{\alpha},\theta_0^{\beta}\}=\{\bar\theta_0^{\alpha},\bar\theta_0^{\beta}\}=\delta^{\alpha\beta}\,,\quad
 \{\beta_m^{\alpha},\beta_n^{\beta}\}=\{\bar\beta_m^{\alpha},\bar\beta_n^{\beta}\}
 =\delta_{m+n,0}\delta^{\alpha\beta}\,.
\end{split}
\end{equation}
For the zero modes, it will later prove useful to define
\begin{equation}
 a_0^I=\sqrt{\frac m2}\,x_0^I+\frac{i}{\sqrt{2m}}p_0^I\,,
 \qquad \chi_0=\frac{\theta_0+i\Pi\bar\theta_0}{\sqrt2}\,.\label{eq:zeromodes}
\end{equation}
Indeed, the supersymmetric vacuum $|0,\ell\rangle$ is annihilated by $a_0,\chi_0$ and all positive-frequency oscillators. We distinguish the bosonic and fermionic Fock vacua by the labels $\mathrm B$ and $\mathrm F$, respectively, so that $\ket{0,\ell}=\ket{0,\ell}_{\rm B}\otimes\ket{0,\ell}_{\rm F}$.

\paragraph{BMN dictionary.}

An important aspect of string theory on the pp-wave background is that the map to the boundary $\mathcal N=4$ SYM theory is straightforward. The physical light-cone energy and momentum obey
\begin{equation}
 H_{\rm LC}=-P_-=m(\Delta-J)\,,\qquad P_+=\frac{\Delta+J}{2mR^2}\,.
\end{equation}
Notably, in order for these charges to be non-vanishing in the $R\rightarrow\infty$ limit, we need $\Delta+J\sim O(R^2)$ but $\Delta-J\sim O(R^{0})$. $\Delta$ is naturally identified with the conformal dimension of an operator in $\mathcal N=4$ SYM, while $J$ is the R-charge selected by the complex scalar $Z$ in \eqref{eq:OL}. Concretely, the states of the string theory with zero energy are states in the CFT with $\Delta=J$. At a fixed value of $P_+=\ell/(2\pi)$, these states are straightforward to match:
\begin{align}\label{eq:stringvacua}
 \ket{0,\ell}_{\text{String}}\leftrightarrow\mathcal O_{(mR^2\ell/(2\pi))}\,.
\end{align}
Here $\mathcal O_{(L)}$ is the half-BPS operator with $J=\Delta=L$ defined in \eqref{eq:OL}. Excited states $\ket{E,\ell}$ can then be obtained by acting on the vacuum with raising operators $\alpha_{-n}/\beta_{-n}$ and their barred partners, subject to level matching, and similarly on the gauge-theory side by replacing some factors of $Z$ with other fields.

\subsection{Crosscap on the pp-wave}
\begin{figure}
	\centering
\begin{tikzpicture}[
	geometry/.style={draw=CornflowerBlue,line width=0.9pt},
	wire/.style={draw=CornflowerBlue!75,line width=0.55pt},
	hidden/.style={dash pattern=on 2.4pt off 2pt},
	locus/.style={draw=red,line width=1.0pt},
	font=\rmfamily,line join=round,line cap=round
	]
	\path[use as bounding box] (-1.12,-1.98) rectangle (6.13,1.98);
	
	\fill[CornflowerBlue!7]
	(-1.05,-1.65) -- (-1.05,1.65)
	arc[start angle=180,end angle=0,x radius=1.05,y radius=0.30]
	-- (1.05,-1.65)
	arc[start angle=0,end angle=-180,x radius=1.05,y radius=0.30]
	-- cycle;
	\draw[wire,hidden]
	(1.05,-1.65)
	arc[start angle=0,end angle=180,x radius=1.05,y radius=0.30];
	\draw[geometry]
	(-1.05,-1.65)
	arc[start angle=180,end angle=360,x radius=1.05,y radius=0.30];
	\draw[geometry] (-1.05,-1.65) -- (-1.05,1.65);
	\draw[geometry] (1.05,-1.65) -- (1.05,1.65);
	\draw[geometry] (0,1.65) ellipse[x radius=1.05,y radius=0.30];
	
	\draw[wire,hidden]
	(1.05,0)
	arc[start angle=0,end angle=180,x radius=1.05,y radius=0.30];
	\draw[wire]
	(-1.05,0)
	arc[start angle=180,end angle=360,x radius=1.05,y radius=0.30];
	
	\node at (2.25,0) {\Large $\times$};
	
	\begin{scope}[shift={(4.45,0)}]
		\fill[CornflowerBlue!7] (0,0) circle[radius=1.28];
		\draw[wire,hidden]
		(1.28,0)
		arc[start angle=0,end angle=180,x radius=1.28,y radius=0.36];
		\draw[wire]
		(-1.28,0)
		arc[start angle=180,end angle=360,x radius=1.28,y radius=0.36];
		\draw[geometry] (0,0) circle[radius=1.28];
	\end{scope}
	
	\fill[red] (0,0) circle[radius=1.25pt];
	
	\begin{scope}[shift={(4.45,0)}]
		\draw[locus,hidden]
		(0,1.28)
		arc[start angle=90,end angle=270,x radius=0.46,y radius=1.28];
		\draw[locus]
		(0,-1.28)
		arc[start angle=-90,end angle=90,x radius=0.46,y radius=1.28];
		\draw[red,line width=0.55pt] (0.36,0.80) -- (1.24,0.86);
		\node[red,above right,inner sep=2pt] at (1.24,0.86) {$S^2$};
	\end{scope}
\end{tikzpicture}
\caption{Fixed locus of the orientifold in $\mathrm{AdS}_5\times \text{S}^5$. The red point is the fixed point at the AdS centre, and the red internal locus represents the $\text{S}^2$ in \eqref{eq:fixedS2}. Its tangent plane in the Penrose limit is spanned by $y_{8},y_{10}$.}
\label{fig:ppOrientifold}
\end{figure}
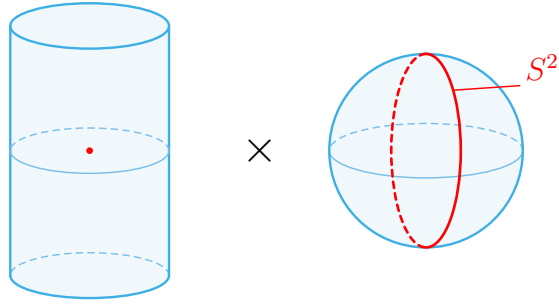

\paragraph{Orientifold geometry on the pp-wave.}
We are now ready to discuss the result of the orientifolding of the pp-wave in string theory. One crucial thing to observe is that while the action of parity $\mathsf P$ \eqref{eq:ppori} leaves the metric \eqref{eq:PPmet} invariant, it reverses the RR-flux \eqref{eq:PPRR}. To resolve this issue, we need to supplement the action of $\mathsf P$ with an orientation reversal on the string worldsheet $\mathfrak{P}(\tau,\sigma)=(-\tau,\sigma+\ell/2)$. Worldsheet parity $\mathfrak{P}$ sends the type-IIB RR four-form potential, and hence $F_5$, to its negative: the two signs therefore cancel in $\iota=\mathfrak{P}\times\mathsf P$. This is precisely how orientifolding is defined in string theory. More precisely, an O$p$-plane is located at the fixed locus of the orientifold action and leaves invariant $p+1$ directions of the bulk spacetime. Given \eqref{eq:ppori}, we have an instantonic O1-plane extending along $y_{8},y_{10}$ at $x^+=x^-=x_\mu=y_{7}=y_{9}=0$. Figure~\ref{fig:ppOrientifold} illustrates the fixed locus before the Penrose limit.
\paragraph{Open string interpretation.}
In an open string picture, if a stack of $N$ D-branes intersects an O$p$-plane, the Chan--Paton labeling of open strings ending on the D-branes is modified. If we denote by $\ket{i,j}$ the massless vector string state starting at the $i$-th D-brane and ending on the $j$-th, it must satisfy
\begin{equation}
 \ket{i,j}=\iota\left(\ket{i,j}\right)\equiv-\mathbb K_{j,j'}\ket{j',i'}\mathbb K^{-1}_{i',i}\,.
\end{equation}
Here $\mathbb K$ is an $N\times N$ matrix and the reversal of the $i$ and $j$ labels comes from orientation reversal on the worldsheet. Requiring $\iota^2=1$ on Chan--Paton matrices gives $\mathbb K^{\mathsf T}=\pm\mathbb K$. Up to an allowed change of Chan--Paton basis, the nondegenerate unitary choices are represented by $\mathbb I$ and, for even $N$, $\mathbb J$. This is precisely the string theory analogue of the charge conjugations \eqref{eq:CO} and \eqref{eq:Cp}.
In a closed string picture, the orientifold corresponds to a \textit{crosscap} state $\ket C$: this is analogous to the spin-chain crosscap state first encountered in \eqref{eq:SU2C}. We now identify an exact expression for this crosscap state in the pp-wave limit, following the similar analysis performed for boundary states in \cite{billo_boundary_2002}.

\paragraph{Closed string: bosonic sector.}

In the bosonic subsector, the orientation-reversing identification is
\begin{equation}
 \left[x^I(\tau,\sigma)-M_{IJ}x^J(-\tau,\sigma+\ell/2)\right]\ket C_{\text{B}}=0\,,
 \qquad
 M_{IJ}=\delta_{IJ}\begin{cases}+1,&I=8,10\,,\\-1\,,&\text{otherwise}\,.
 \end{cases}
\end{equation}
These signs are the same as those appearing in \eqref{eq:ppori}. The crosscap state is inserted at $\tau=0$. Differentiating the identification, including the reversal of time, gives
\begin{equation}\label{eq:cccondition}
 \left[\partial_\pm x^I({0},\sigma){+}M_{IJ}\partial_\mp x^J({0},\sigma+\ell/2)\right]\ket C_{\text{B}}=0\,.
\end{equation}
Plugging in the on-shell mode expansion \eqref{eq:xtonshell}, the condition on the excited modes on the crosscap becomes
\begin{equation}\label{eq:crossbos}
 \left[\alpha_n^I+(-1)^nM_{IJ}\bar\alpha_{-n}^J\right]\ket C_{\text{B}}=0\,.
\end{equation}
For the zero modes it gives instead $x_0^I\ket C_{\text{B}}=0$ when $M_{II}=-1$ and $p_0^I\ket C_{\text{B}}=0$ when $M_{II}=+1$, which are delta-function position and momentum states, respectively. The bosonic solution is then
\begin{equation}\label{eq:crosscapbosoniic}
 \ket C_{\rm B}\propto
 \exp\left[\frac{1}{2}M_{IJ}a_0^{I\dagger}a_0^{J\dagger}-\sum_{n=1}^\infty(-1)^nM_{IJ}
 {\alpha_n^{I\dagger}\bar\alpha_n^{J\dagger}}\right]
 \ket{0,\ell}_{\rm B}\,,
\end{equation}
where we recall that $\ket{0,\ell}_{\rm B}$ is the bosonic Fock vacuum and $a^{I\dagger}_0$ was defined in \eqref{eq:zeromodes}. We will leave the normalization of the state unfixed until we have also considered the fermionic sector.

\paragraph{Closed string: fermionic sector.}

The fermionic sector is slightly harder to pin down. The basic idea is that we want the orientifold to be half-BPS, namely to preserve 16 supercharges. This implies that there should exist a pair of matrices $M^\pm$ such that
\begin{equation}\label{eq:crosssup}
 \left[Q^\pm+M^\pm\bar Q^\pm\right]\ket C=0\,,
\end{equation}
where $Q^\pm$ and $\bar Q^\pm$ denote the two real IIB sets of kinematical ($+$) and dynamical ($-$) supercharges. The kinematical index lies in $8_s$, whereas the dynamical index lies in $8_c$, and $\gamma^I$ maps between these two representations. The broken supercharges are then the combinations $Q^\pm-M^\pm\bar Q^\pm$.

In terms of the mode expansion \eqref{eq:xtonshell}, the supercharges can be written in the form
\begin{equation}\label{eq:superchargecrosscap}
\begin{split}
 Q^+&\propto\theta_0\,,\qquad \bar Q^+\propto\bar\theta_0\,,\\
 Q^-&\propto p_0^I\gamma^I\theta_0-mx_0^I\gamma^I\Pi\bar\theta_0
 +\sum_{n=1}^\infty\sqrt{2\omega_n}\,b_n\gamma^I
 (\alpha_{-n}^I\beta_n+\alpha_n^I\beta_{-n})\\
 &\hspace{4.05 cm}+\sum_{n=1}^\infty\frac{im}{\sqrt{2\omega_n}\,b_n}\gamma^I\Pi
 (\bar\alpha_{-n}^I\bar\beta_n-\bar\alpha_n^I\bar\beta_{-n})\,,\\{
 \bar Q^-}&\propto p_0^I\gamma^I\bar\theta_0+mx_0^I\gamma^I\Pi\theta_0
 +\sum_{n=1}^\infty\sqrt{2\omega_n}\,b_n\gamma^I
 (\bar\alpha_{-n}^I\bar\beta_n+\bar\alpha_n^I\bar\beta_{-n})\\&\hspace{4.05 cm}-\sum_{n=1}^\infty\frac{im}{\sqrt{2\omega_n}\,b_n}\gamma^I\Pi
 (\alpha_{-n}^I\beta_n-\alpha_n^I\beta_{-n})\,.
\end{split}
\end{equation}
The kinematical component of \eqref{eq:crosssup} fixes the zero modes. Combining the dynamical component with the bosonic gluing \eqref{eq:crossbos} fixes the nonzero modes. Together these conditions are
\begin{equation}
 (\theta_0+M^+\bar\theta_0)\ket C=0\,,\quad (\beta_n+(-1)^nM^+\bar\beta_{-n})\ket C=0\,.\label{eq:crossfer}
\end{equation}
The fermionic sector of the crosscap state is then
\begin{equation}\label{eq:crosscapfermionic}
\begin{split}
 \ket C_{{\rm F}}\propto 
 \exp\left[-\frac i2\chi_0^{\alpha\dagger}(M^+\Pi)_{\alpha\beta}\chi_0^{\beta\dagger}-\sum_{n=1}^\infty(-1)^nM^+_{\alpha\beta}
 {\beta_n^{\alpha\dagger}\bar\beta_n^{\beta\dagger}}\right]
 \ket{0,\ell}_{{\rm F}}\,.
\end{split}
\end{equation}
The spinor gluing matrices are determined by the invariant transverse plane $(8,10)$ to be
\begin{equation}\label{eq:spinorlift}
 M^+=i\left.\gamma^{8}\gamma^{10}\right|_{8_s},
 \qquad
 M^-=i\left.\gamma^{8}\gamma^{10}\right|_{8_c}.
\end{equation}
The overall common sign chooses the spinor lift; we use the plus sign. With $\Pi=\gamma^1\gamma^2\gamma^3\gamma^4$ on $8_s$, these matrices obey
\begin{equation}
	M^-\gamma^I=-M_{IJ}\gamma^JM^+\,,\qquad
	[M^+,\Pi]=0\,,\qquad (M^+)^2=1\,,\qquad
	M^+(M^+)^{\mathsf T}=-1\,.
\end{equation}
The first two identities cancel the momentum and mass terms of the dynamical charges, including the zero modes. The last identity ensures that the fermionic gluing constraints mutually anticommute. Thus eight kinematical and eight dynamical supercharges annihilate the crosscap.

Combining the bosonic and fermionic factors, the full crosscap state is
\begin{equation}\label{eq:crosscapfull}
\begin{split}
 \ket C&=c_\ell\exp\left[\frac12M_{IJ}a_0^{I\dagger}a_0^{J\dagger}
 -\frac i2\chi_0^{\alpha\dagger}(M^+\Pi)_{\alpha\beta}\chi_0^{\beta\dagger}\right]\\
 &\quad\times\exp\left[-\sum_{n=1}^\infty(-1)^n
 \left(M_{IJ}\alpha_n^{I\dagger}\bar\alpha_n^{J\dagger}
 +M^+_{\alpha\beta}\beta_n^{\alpha\dagger}\bar\beta_n^{\beta\dagger}\right)\right]
 \ket{0,\ell}\,,
\end{split}
\end{equation}
where $c_\ell$ is the single overall normalization of the state. The creation-operator exponentials have unit vacuum component, so $c_\ell=\langle0,\ell|C\rangle$. Since $P_+$ lies in the center of the pp-wave isometry algebra, symmetry alone cannot relate the normalizations at different $\ell$. We will show instead in Section \ref{subsec:det} that the direct path-integral evaluation leads to the following result
\begin{equation}
c_{\ell}=c_{\infty}\tanh \left(\frac{m \ell}{4}\right)\,,
\end{equation}
where $c_{\infty}$ is an overall $\ell$-independent normalization.

\subsection{Fixing overall coefficients from the path integral}\label{subsec:det}
Let us now fix the overall coefficients $c_{\ell}$ \eqref{eq:crosscapbosoniic}, which were left undetermined by the symmetry analysis. The analysis is conceptually simple: we directly compute the wave functions of the ground state and the crosscap state from the path integral and compute the overlap. We detail the analysis of the bosonic sector with a sketch of the analysis for the fermionic sector and combine the results to determine $c_{\ell}$.

\paragraph{Ground-state wave function.} Let us first recap the ground-state wave function of a massive free boson $\phi$ quantized on $S^1$, whose Hamiltonian is given by
\begin{equation}
H=\frac{1}{2}\int_{0}^{\ell} d\sigma \left(\pi^2 +(\partial_{\sigma}\phi )^2+m^2\phi^2\right)\comma
\end{equation}
where $\pi (\sigma)$ is a momentum density,
\begin{equation}
    \pi (\sigma)=-i\frac{\delta}{\delta \phi (\sigma)}\,.
\end{equation}
After the Fourier decomposition
\begin{equation}
\phi (x)=\frac{1}{\sqrt{\ell}}\sum_{n\in \mathbb{Z}}\phi_ne^{ik_n\sigma}\comma
\end{equation}
with $k_n=\frac{2\pi n}{\ell}$, the Hamiltonian reads
\begin{equation}
    H=\frac{1}{2}\sum_{n}\left[\pi_n \pi_{-n}+\omega_n^2\phi_n\phi_{-n}\right]\,,
\end{equation}
where $\omega_n$ is
\begin{equation}
\omega_n=\sqrt{\left(\frac{2\pi n}{\ell}\right)^2+m^2}\period
\end{equation}

The ground-state wave function of this system is a standard Gaussian wave function,
\begin{align}
    \Psi_{\Omega} [\phi]=\mathcal{N}\exp \left[-\frac{1}{2}\sum_{n}\omega_n \phi_n \phi_{-n}\right]\,.
\end{align}
The overall normalization constant $\mathcal{N}$ is required to make the state unit-normalized $\langle \Psi|\Psi\rangle=1$,
\begin{equation}
    \mathcal{N}=\prod_{n\in \mathbb{Z}}\omega_n^{1/4}\,.
\end{equation}
\paragraph{Crosscap wave function and overlap.} We next consider the wave function for the crosscap state. The crosscap state satisfies
\begin{align}
    \left(\phi(\sigma)-r\phi (\sigma+\ell/2)\right)|C\rangle=0\,,
\end{align}
where $r=\pm 1$ accounts for a possible additional $\mathbb{Z}_2$ action. For $r=+1$, only even Fourier modes $\phi_{2m}$ are allowed while, for $r=-1$, only odd Fourier modes are allowed. Hence the wave function reads
\begin{equation}
\Psi_{C}[\phi]=\begin{cases}\prod_{m\in \mathbb{Z}}\delta (\phi_{2m+1})\qquad &r=+1\,,\\
\prod_{m\in \mathbb{Z}}\delta (\phi_{2m})\qquad &r=-1\,.
\end{cases}
\end{equation}
We can then compute the overlap by performing the integral over $\phi_n$'s,
\begin{align}
\langle C|\Omega\rangle_{\ell}=\begin{cases}\mathcal{N}\prod_{m}(\int d\phi_{2m}) \exp\left[-\frac{1}{2}\sum_{m}\omega_{2m}\phi_{2m}\phi_{-2m}\right]\qquad & r=+1\\
\mathcal{N}\prod_{m}(\int d\phi_{2m+1}) \exp\left[-\frac{1}{2}\sum_{m}\omega_{2m+1}\phi_{2m+1}\phi_{-2m-1}\right]\qquad & r=-1
\end{cases}\,.
\end{align}
Performing the integral, we obtain
\begin{align}
\langle C|\Omega\rangle_{\ell}=c_0\left(\frac{\mathcal{N}_{o}}{\mathcal{N}_{e}}\right)^{r}\,,
\end{align}
where $c_0$ is the length-independent overall factor and $\mathcal{N}_{e,o}$ are given by
\begin{align}
\mathcal{N}_e=\prod_{m}\omega_{2m}^{1/4}\,,\qquad \mathcal{N}_o=\prod_{m}\omega_{2m+1}^{1/4}\period 
\end{align}
The ratio $\mathcal{N}_o/\mathcal{N}_e$ can be computed by taking a logarithm, differentiating with respect to $\ell$, summing over mode numbers, and integrating\footnote{This produces an overall $\ell$-independent ambiguity, but one can absorb it into $c_0$.} with respect to $\ell$. As a result, we get
\begin{align}
\langle C|\Omega\rangle_{\ell}=c_0 \left(\tanh \left(\frac{m\ell}{4}\right)\right)^{-r/4}\,.
\end{align}

Now, applying this formula to our current case, we note that there are two bosonic directions with $r=+1$ and six bosonic directions with $r=-1$ (cf.~\eqref{eq:cccondition}). Therefore, the contribution from the bosonic sector reads
\begin{align}
\left.c_{\ell}\right|_{\rm bosonic}=c_0\tanh \left(\frac{m\ell}{4}\right)\,.
\end{align}
\paragraph{Fermionic sector and full result.} Essentially the same argument applies also to the fermionic sector. However, for the fermions, there are equal numbers of ``$r=+1$ modes'' and ``$r=-1$ modes'', and their contributions cancel out completely. This can be seen straightforwardly by diagonalizing the matrix $M_{ab}^{+}$ that appears in the crosscap condition \eqref{eq:crossfer} and observing that it has equal numbers of $+1$ and $-1$ eigenvalues.

Therefore, the full result is given simply by
\begin{align}\label{eq:finalBMN}
c_{\ell}=c_0\tanh \left(\frac{m\ell}{4}\right)\,.
\end{align}
As we see in the next section, this agrees precisely with the BMN limit of the localization result.
\section{One-point functions from supersymmetric localization}
\label{sec:localization}
We now turn to half-BPS one-point functions and compute them exactly at finite
$N$ and $g_{\rm YM}$ using supersymmetric localization. Localization on
$\mathbb{RP}^4$ without charge conjugation was studied in \cite{Wang:2020jgh}; our goal is
to extend that analysis to $\mathbb{RP}^4$ with charge conjugation.

As in \cite{Wang:2020jgh}, the four-dimensional computation reduces to the zero-instanton
sector of two-dimensional Yang--Mills theory (2d YM) on $\mathbb{RP}^2$. The
only new ingredient is that $\mathbb{RP}^2$ is itself twisted by charge
conjugation. After projecting to the zero-instanton sector, we obtain a matrix
model that computes the one-point functions exactly at finite $N$ and
$g_{\rm YM}$. We then analyze this matrix model at weak and strong coupling
and compare the results with perturbative gauge theory, supergravity, and the
pp-wave string theory.

\subsection{From \texorpdfstring{$\mathcal{N}=4$}{N=4} SYM on \texorpdfstring{$\mathbb{RP}^4$}{RP4} to constrained 2d YM on \texorpdfstring{$\mathbb{RP}^2$}{RP2}}
\paragraph{From 4d to 2d.}
Supersymmetric localization reduces a class of protected observables in
$\mathcal{N}=4$ SYM to two-dimensional Yang--Mills theory \cite{Pestun:2009nn,Drukker:2007yx,Drukker:2007qr,Bassetto:2008yf,Giombi:2009ms,Giombi:2009ds,Giombi:2009ek,Giombi:2012ep,Giombi:2018qox,Giombi:2018hsx,Giombi:2020pdd,Giombi:2020amn,Wang:2020seq,Wang:2020jgh}. This construction
has been used to study Wilson--'t~Hooft loops, correlation functions, and
various supersymmetric defects. The basic mechanism is standard. One chooses a
supercharge $\mathcal Q$ preserved by the observables of interest and deforms
the action by a $\mathcal Q$-exact term,
\begin{align}
    S_{\mathcal N=4}
    \;\longrightarrow\;
    S_{\mathcal N=4}
    +
    t\int d^4x\,\sqrt{g}\,
    |\mathcal Q\Psi|^2 .
\end{align}
Since expectation values of $\mathcal Q$-closed observables are independent of
$t$, one may take $t\to\infty$, where the path integral localizes onto the
solutions of $\mathcal Q\Psi=0$.

For $\mathcal N=4$ SYM on $\mathbb{RP}^4$, this localization was worked out in
detail in \cite{Wang:2020jgh}. The resulting theory is two-dimensional Yang--Mills on
$\mathbb{RP}^2$, with action
\begin{align}
    S_{2d}
    =-\frac{1}{g_{2d}^2}
    \int_{\mathbb{RP}^2}
    d^2x\,\sqrt{g}\,
    {\rm Tr}\!\left((\star\mathcal F)^2\right),
\end{align}
where $\mathcal F$ is the two-dimensional field strength, $g_{2d}$ is the two-dimensional coupling constant,
\begin{align}
g_{2d}^2=-\frac{g_{\rm YM}^2}{2\pi R^2}\,,
\end{align}
and $R$ is the radius
of $S^2$ before the $\mathbb{Z}_2$ quotient. Under the same localization map, the half-BPS operator
${\rm Tr}(Z^L)$ descends to
${\rm Tr}\!\left((\star\mathcal F)^L\right)$.

The only new ingredient in our setup is charge conjugation. The
four-dimensional charge-conjugation twist descends directly to the localized
two-dimensional theory. Thus $\mathcal N=4$ SYM on $\mathbb{RP}^4$ with charge
conjugation is mapped to 2d YM on $\mathbb{RP}^2$, likewise twisted by charge
conjugation.

Strictly speaking, the localized theory is not ordinary 2d YM, but its
zero-instanton sector, often referred to as constrained 2d YM; see, for
example, \cite{Pestun:2009nn}. This projection will be important below: it allows the
discrete sum over 2d YM representations to be rewritten as a multiple integral,
leading naturally to a matrix-model description.
\paragraph{2d YM on $\mathbb{RP}^2$ with charge conjugation.}
Having reduced the four-dimensional problem to 2d YM, we now study the
two-dimensional theory on $\mathbb{RP}^2$ with charge conjugation directly.
The theory without charge conjugation was analyzed already in the original paper by Witten
\cite{Witten:1991we}. We first review the standard gluing construction and
then incorporate the charge-conjugation twist.

\begin{figure}[t]
\centering
\begin{tikzpicture}[font=\large,line cap=round,line join=round]

\def\r{1.0}
\tikzset{
  circ/.style={line width=0.9pt},
  arr/.style={-{Stealth[length=2.4mm,width=2.1mm]},line width=0.9pt},
  dot/.style={circle,fill,inner sep=1.5pt}
}

\begin{scope}[xshift=0cm]
  \draw[circ] (0,0) circle (\r);

  \draw[arr]
    ({\r*cos(180)},{\r*sin(180)})
    arc[start angle=190,end angle=180,radius=\r];

  \draw[arr]
    ({\r*cos(0)},{\r*sin(0)})
    arc[start angle=-10,end angle=0,radius=\r];

  \node[dot] at (0,\r) {};
  \node[dot] at (0,-\r) {};

  \node[left=1.5mm]  at (-\r,0) {$X$};
  \node[right=1.5mm] at (\r,0) {$X^{-1}$};

  \node at (0,-1.55) {$(a)$ $S^{2}$};
\end{scope}

\begin{scope}[xshift=4.5cm]
  \draw[circ] (0,0) circle (\r);

  \draw[arr]
    ({\r*cos(180)},{\r*sin(180)})
    arc[start angle=190,end angle=180,radius=\r];

  \draw[arr]
    ({\r*cos(0)},{\r*sin(0)})
    arc[start angle=0,end angle=-5,radius=\r];

  \node[dot] at (0,\r) {};
  \node[dot] at (0,-\r) {};

  \node[left=1.5mm]  at (-\r,0) {$X$};
  \node[right=1.5mm] at (\r,0) {$X$};

  \node at (0,-1.55) {$(b)$ $\mathbb{RP}^2$};
\end{scope}

\begin{scope}[xshift=9.0cm]
  \draw[circ] (0,0) circle (\r);

  \draw[arr]
    ({\r*cos(180)},{\r*sin(180)})
    arc[start angle=190,end angle=180,radius=\r];

  \draw[arr]
    ({\r*cos(0)},{\r*sin(0)})
    arc[start angle=0,end angle=-5,radius=\r];

  \node[dot] at (0,\r) {};
  \node[dot] at (0,-\r) {};

  \node[left=1.5mm]  at (-\r,0) {$X$};
  \node[right=1.5mm] at (\r,0) {$X^{\mathsf{C}}$};

  \node at (0,-1.55) {$(c)$ $\mathbb{RP}^2_{\mathsf{C}}$};
\end{scope}

\end{tikzpicture}

\caption{Gluing 2d YM on a disk into $(a)$ $S^2$, $(b)$ $\mathbb{RP}^2$ and $(c)$ $\mathbb{RP}^2$ with charge conjugation. Depending on how we identify holonomies of each half circles, we obtain different geometries.}
\label{fig:three-circles}
\end{figure}
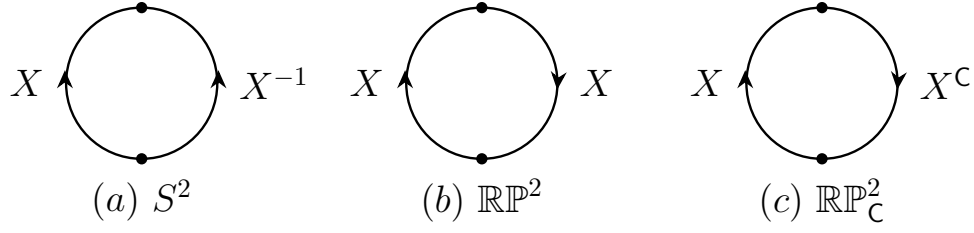

We begin with 2d YM on a disk whose boundary holonomy is $U$. Standard cutting
and gluing arguments, reviewed for example in \cite{Tachikawa:2016kfc}, give
the disk wavefunction
\begin{align}
    \Psi(U)
    =
    \sum_R
    ({\rm dim}R)\,
    e^{-\frac{1}{2}\pi R^2 g_{2d}^2 c_2(R)}
    \chi_R(U)\,,
\end{align}
where the sum runs over irreducible representations $R$ of $U(N)$,
$c_2(R)$ is the quadratic Casimir, and $\chi_R(U)$ is the corresponding
character. Here we chose the area of the disk so that one obtains  
$\mathbb{RP}^2$ after gluing, whose area is half the area of the unit sphere.
Both the $S^2$ and $\mathbb{RP}^2$ partition functions follow from different
gluings of the same disk amplitude; see Figure~\ref{fig:three-circles}. To make this explicit,
split the boundary circle into two intervals with holonomies $X$ and $Y$, so
that
\begin{equation}
    U=XY\,.
\end{equation}
For $S^2$, the two intervals are glued with opposite orientation,
$Y=X^{-1}$. Integrating over the common holonomy gives
\begin{align}
\begin{aligned}
Z_{S^2}
&=
\int_{U(N)} dX\,\Psi(XX^{-1})
\\
&=
\int_{U(N)} dX
\sum_R
({\rm dim}R)\,
e^{-\frac{1}{2}\pi R^2 g_{2d}^2 c_2(R)}
\chi_R(1)
\\
&=
\sum_R
({\rm dim}R)^2
e^{-\frac{1}{2}\pi R^2 g_{2d}^2 c_2(R)}\,,
\end{aligned}
\end{align}
where we used $\chi_R(1)={\rm dim}R$. This is the familiar expression for
2d YM on $S^2$ with area $2\pi$.

For $\mathbb{RP}^2$ without charge conjugation, the appropriate identification
is instead $Y=X$.
The partition function therefore becomes \cite{Witten:1991we,Wang:2020jgh}
\begin{align}
    Z_{\mathbb{RP}^2}
    =
    \sum_R
    ({\rm dim}R)\,
    e^{-\frac{1}{2}\pi R^2 g_{2d}^2 c_2(R)}
    f_R\,,
\end{align}
where
\begin{equation}
    f_R
    =
    \int_{U(N)}dX\,\chi_R(X^2)
\end{equation}
is the Frobenius--Schur indicator.

We now include charge conjugation. The gluing is modified to
\begin{align}
    Y
    =
    X^{\mathsf C}
    =
    \mathbb K\,\overline{X}\,\mathbb K^{-1}\,,
\end{align}
and hence
\begin{equation}
    Z_{\mathbb{RP}_{\mathsf C}^2}
    =
    \sum_R
    ({\rm dim}R)\,
    e^{-\frac{1}{2}\pi R^2 g_{2d}^2 c_2(R)}
    \nu_R\,,
\end{equation}
where
\begin{equation}
    \nu_R
    =
    \int_{U(N)}dX\,
    \chi_R\!\left(XX^{\mathsf C}\right)
\end{equation}
is the twisted Frobenius--Schur indicator.

In the present case, $\nu_R$ can be evaluated explicitly. Using the invariance
of the Haar measure, we redefine $X\to X\mathbb K$. Since
$\mathbb K^2=\varepsilon_{\mathbb{K}}$, this gives
\begin{equation}
    XX^{\mathsf C}
    \mapsto
    X\mathbb K^2\overline X
    =
    \varepsilon_{\mathbb{K}}X\overline X\,.
\end{equation}
Therefore, 
\begin{equation}
\begin{split}
\nu_R
&=
(\varepsilon_{\mathbb{K}})^{Q_R}
\int_{U(N)}dX\,\chi_R(X\overline X)
\nonumber\\
&=
(\varepsilon_{\mathbb{K}})^{Q_R}
\int_{U(N)}dX\,
D^R_{ij}(X)\,
\overline{D^R_{ji}(X)}\,,
\end{split}
\end{equation}
where $D^R_{ij}(X)$ denotes a matrix element of $X$ in the representation $R$. Here we used $\chi_{R}(-g)=(-1)^{Q_R}\chi_R (g)$ with $Q_R$ being the $U(1)$ charge of the representation $R$.
Using the orthogonality relation
\begin{equation}
    \int_{U(N)}dX\,
    D_{ij}^R(X)\,
    \overline{D_{kl}^S(X)}
    =
    \frac{\delta_{RS}}{{\rm dim}R}
    \delta_{ik}\delta_{jl}\,,
\end{equation}
we obtain
\begin{equation}
    \nu_R
    =
    (\varepsilon_{\mathbb{K}})^{Q_R}\,.
\end{equation}

Putting everything together, the partition function of 2d YM on
$\mathbb{RP}^2$ with charge conjugation is
\begin{equation}\label{eq:ZRP2C}
    Z_{\mathbb{RP}_{\mathsf C}^2}
    =
    \sum_R
    (\varepsilon_{\mathbb{K}})^{Q_R}
    ({\rm dim}R)\,
    e^{-\frac{1}{2}\pi R^2 g_{2d}^2 c_2(R)}\,.
\end{equation}
In order to write an explicit representation, we introduce the ``row lengths''\footnote{For the $U(N)$ gauge group, $\lambda_i$'s can be negative.} $(\lambda_1,\ldots, \lambda_N)$ of the representation $R$ satisfying $\lambda_1\geq \lambda_2\geq \cdots \geq \lambda_N$, and define the shifted lengths
\begin{align}
m_i = \lambda_i+\frac{N+1}{2}-i\,.
\end{align}
The shifted lengths $m_i$'s are an $N$-tuple of strictly decreasing (half-)integers $m_1>m_2>\cdots >m_{N}$.  
In terms of $m_i$'s, various quantities appearing in \eqref{eq:ZRP2C} can be written as
\begin{align}
\begin{aligned}
{\rm dim}\, R&=\frac{\Delta(m)}{\Delta(N-i)}=\frac{\Delta(m)}{G(N+1)},\qquad
 c_2(R)=\sum_im_i^2-\frac{N(N^2-1)}{12},
 \\Q_R&=\sum_i\lambda_i=\sum_im_i\,,
\end{aligned}
\end{align}
where $\Delta (x)=\prod_{i<j}(x_i-x_j)$ is the Vandermonde determinant and $G(x)$ is the Barnes-G function. Hence, we obtain the following expression
\begin{equation}\label{eq:ZCRP2q}
Z_{\mathbb{RP}_{\mathsf C}^2}
    =\frac{e^{\frac{N(N^2-1)\pi R^2g_{2d}^2}{24}}}{G(N+1)}\sum_{m_1>\cdots>m_N}\Delta (m)\,e^{-\frac{1}{2}\pi R^2 g_{2d}^2 \sum_i m_i^2}\,e^{i\pi \sigma_{\mathbb{K}}\sum_im_i}\,,
\end{equation}
with ${\sigma_{\mathbb{K}}=(1-\varepsilon_{\mathbb{K}})/2}$.

\subsection{Matrix model at finite \texorpdfstring{$N$}{N} and \texorpdfstring{$g_{\rm YM}$}{gYM}}
\paragraph{Partition function on $\mathbb{RP}^4$.}
The next step is to project onto the zero-instanton sector. As discussed in the
literature (see, for example, \cite{Giombi:2009ek}), this can be implemented by Poisson
resumming the representation sum in \eqref{eq:ZCRP2q}. The resummed expression
is naturally organized as a sum over instanton sectors, and retaining the
zero-instanton contribution gives the constrained 2d YM theory relevant for
localization.
Since our main interest is the large-$N$ limit, the distinction between even
and odd $N$ will not be essential. For simplicity, we therefore take
$N=2n$ below.

There is, however, one technical subtlety. Ordinarily, one replaces the
ordered sum
$m_1>\cdots>m_N$ by an unrestricted sum over the $m_i$, dividing by $N!$.
Here this cannot be done directly: the Vandermonde factor $\Delta(m)$ is
antisymmetric under permutations of the $m_i$, unlike the $\Delta(m)^2$
appearing in the $S^2$ partition function. We must therefore include an
additional antisymmetric factor that compensates this sign. This leads to
\begin{align}
Z_{\mathbb{RP}_{\mathsf C}^2}
    =
    \frac{
        e^{\frac{N(N^2-1)\pi R^2g_{2d}^2}{24}}
    }{
        N!G(N+1)
    }
    \sum_{m_1,\ldots,m_N}
    \Delta(m)\,
    {\rm Pf}\,E\,
    e^{-\frac{1}{2}\pi R^2 g_{2d}^2\sum_i m_i^2}\,
    e^{i\pi\sigma_{\mathbb K}\sum_i m_i}\,,
\end{align}
where ${\rm Pf}\,E$ is the Pfaffian of the antisymmetric sign matrix
\begin{align}
    E_{ij}
    =
    {\rm sgn}(m_i-m_j)\,.
\end{align}
The product $\Delta(m)\,{\rm Pf}\,E$ is symmetric under permutations of the
$m_i$, and hence allows the ordered sum to be replaced consistently by the
unrestricted sum above.

The next step is to Poisson resum the representation sum. Following
\cite{Wang:2020jgh}, we rewrite the Vandermonde factor multiplied by the Gaussian
as
\begin{align}
\Delta(m)
e^{-\frac{1}{2}\pi R^2 g_{2d}^2\sum_i m_i^2}
&=
i^{-\frac{N(N-1)}{2}}
\left(
    \frac{2}{R^2g_{2d}^2}
\right)^{\frac{N^2}{2}}
\int_{\mathbb R^N}d^Nz\,
\Delta(z)\,
e^{-\frac{2\pi}{R^2g_{2d}^2}\sum_i z_i^2}
e^{2\pi i\sum_i z_i m_i}.
\end{align}
Taking into account the two-dimensional counterterm \cite{Wang:2020seq,Wang:2020jgh}, the partition
function becomes
\begin{align}\label{eq:intermediate}
\begin{aligned}
Z_{\mathbb{RP}^2_{\mathsf C}}
=
\frac{i^{\frac{N(1-2N)}{2}}}{N!}
\int d^Nz\,
\Delta(z)\,
e^{-\frac{2\pi}{R^2g_{2d}^2}\sum_i z_i^2}
\sum_{m_1,\ldots,m_N}
{\rm Pf}\,E\,
e^{2\pi i\sum_i z_i m_i}
e^{i\pi\sigma_{\mathbb K}\sum_i m_i}.
\end{aligned}
\end{align}

To perform the sums over the $m_i$, we expand the Pfaffian into pairings,
\begin{align}
    {\rm Pf}\,E
    =
    {\frac{1}{2^nn!}}\sum_{\sigma\in S_{2n}}
    {\rm sgn}(\sigma)\,
    E_{\sigma_1\sigma_2}
    \cdots
    E_{\sigma_{2n-1}\sigma_{2n}}.
\end{align}
Substituting this into \eqref{eq:intermediate} and relabelling the integration
variables $z$'s, all pairings give identical contributions. We therefore obtain
\begin{align}
\begin{aligned}
Z_{\mathbb{RP}^2_{\mathsf C}}
=&\,
i^{\frac{N(1-2N)}{2}}
\int d^Nz\,
\Delta(z)\,
e^{-\frac{2\pi}{R^2g_{2d}^2}\sum_i z_i^2}
\\
&\times
\sum_{m_1,\ldots,m_N}
\left[
    \prod_{k=1}^{n}
    {\rm sgn}(m_{2k-1}-m_{2k})
\right]
e^{2\pi i\sum_i z_i m_i}
e^{i\pi\sigma_{\mathbb K}\sum_i m_i}.
\end{aligned}
\end{align}

The integer sums now factorize into $n (=N/2)$ independent pairs. For each pair,
setting $m_{2k-1}=m_{2k}+\delta$ gives
\begin{align}
\begin{aligned}
&
\sum_{m_{2k-1},m_{2k}}
{\rm sgn}(m_{2k-1}-m_{2k})\,
e^{2\pi i(z_{2k-1}m_{2k-1}+z_{2k}m_{2k})}
e^{i\pi\sigma_{\mathbb K}(m_{2k-1}+m_{2k})}
\\
&=
e^{i\pi\sigma_{\mathbb K}}
\sum_{\delta\in\mathbb Z}
{\rm sgn}(\delta)\,
e^{2\pi i
\left(z_{2k-1}+\frac{\sigma_{\mathbb K}}{2}\right)\delta}
\sum_{m_{2k}\in\mathbb Z+\frac{1}{2}}
e^{2\pi i(z_{2k-1}+z_{2k})m_{2k}}
\\
&=
i\left(\cot(\pi z_{{2k-1}})\right)^{\varepsilon_{\mathbb{K}}}
\sum_{s_k\in\mathbb Z}
(-1)^{s_k}
\delta(z_{2k-1}+z_{2k}-s_k).
\end{aligned}
\end{align}
The delta functions localize one variable in each pair, yielding
\begin{align}\label{eq:identieigen}
\begin{aligned}
Z_{\mathbb{RP}^2_{\mathsf C}}
=&\,
i^{N(1-N)}
\sum_{s_1,\ldots,s_n}
\int d^nz\,
\Delta(z_1,s_1-z_1,\ldots,z_n,s_n-z_n)
\\
&\times
\prod_k
\left(\cot(\pi z_k)\right)^{\varepsilon_{\mathbb{K}}}
e^{-\frac{2\pi}{R^2g_{2d}^2}
\sum_k\left[z_k^2+(s_k-z_k)^2\right]}.
\end{aligned}
\end{align}
The integers $s_k$ label the instanton sectors. The constrained theory relevant
for localization is obtained by retaining only the zero-instanton sector,
$s_k=0$ for all $k$. We therefore find
\begin{align}
\begin{aligned}
\left.
Z_{\mathbb{RP}^2_{\mathsf C}}
\right|_{\text{0-inst}}
=
i^{N(1-N)}
\int d^nz\,
\Delta(z^2)^2\,
e^{-\frac{4\pi}{R^2g_{2d}^2}\sum_k z_k^2}
\prod_k
2z_k
\left(\cot(\pi z_k)\right)^{\varepsilon_{\mathbb{K}}}.
\end{aligned}
\end{align}

Finally, since $g_{2d}^2<0$, we Wick rotate the contour by setting
$a_k=iz_k$. Expressing the result in terms of the four-dimensional coupling
$g_{\rm YM}$ gives
\begin{align}
\begin{aligned}
\mathcal{Z}_{\mathbb{RP}^4_{\mathsf{C}}}
&={\frac{i^{\varepsilon_{\mathbb{K}}n}}{2^n n!}}
\int_{\mathbb R^n}d^na\,
\Delta(a^2)^2\,
e^{-\frac{8\pi^2}{g_{\rm YM}^2}\sum_k a_k^2}
\prod_k
2a_k
\left(\coth(\pi a_k)\right)^{\varepsilon_{\mathbb{K}}}
\\
&=
{\frac{i^{\varepsilon_{\mathbb{K}}n}}{ n!}
}\int_{\mathbb R_+^n}d^nx\,
\Delta(x)^2\,
e^{-\frac{8\pi^2}{g_{\rm YM}^2}\sum_k x_k}
\prod_k
\left(
    \coth(\pi\sqrt{x_k})
\right)^{\varepsilon_{\mathbb{K}}}.
\end{aligned}
\end{align}
In the second line, we introduced $x_k=a_k^2$. The resulting expression is a
Wishart--Laguerre-type matrix integral \cite{altland1997nonstandard}.

\paragraph{One-point functions.}
We now insert the half-BPS operator $\mathcal O_{(L)}$ at the north pole of
$S^4$. Under localization, this operator is represented in the matrix model by
\begin{align}
    \mathcal O_{(L)}
    \;\longrightarrow\;
    \sum_{i=1}^{N} f_L(a_i)\,,
\end{align}
where $f_L$ is a degree-$L$ polynomial that incorporates the mixing of
$\mathcal O_{(L)}$ with lower-dimensional operators $\mathcal O_{(J)}$,
$J<L$, on $S^4$; see, for example, \cite{Gerchkovitz:2016gxx}.

In the present theory, the eigenvalues are paired according to
\eqref{eq:identieigen}. The localized one-point function is therefore
\begin{align}
\begin{aligned}
    \langle \mathcal O_{(L)}\rangle_{\mathbb{RP}^4_{\mathsf C}}
    =
    \frac{1}{\mathcal{Z}_{\mathbb{RP}^4_{\mathsf{C}}}}
    \int_{\mathbb R^n} d^na\,
    \Delta(a^2)^2\,
    e^{-\frac{8\pi^2}{g_{\rm YM}^2}\sum_k a_k^2}
    \prod_k
    2a_k
    \left(\coth(\pi a_k)\right)^{\varepsilon_{\mathbb{K}}}
    \sum_k
    \left[
        f_L(a_k)+f_L(-a_k)
    \right].
\end{aligned}
\end{align}

The polynomial $f_L$ is fixed by removing operator mixing through a
Gram--Schmidt procedure applied to the $S^4$ two-point functions \cite{Rodriguez-Gomez:2016cem}.
The result is
\begin{align}\label{eq:GramSchmidtlargeN}
    f_L(x)
    =
    \frac{i^{L}}{\sqrt{L}}
    \left[
        2T_L\left(\frac{x}{2g}\right)
        +
        \delta_{L2}
    \right],
\end{align}
with
\begin{equation}
    g
    =
    \frac{\sqrt{\lambda}}{4\pi},
\end{equation}
where $T_L$ is the Chebyshev polynomial of the first kind.
Since
\begin{align}
    f_L(-x)=(-1)^L f_L(x),
\end{align}
the one-point function vanishes for odd $L$. Equivalently,
\begin{align}\label{eq:finiteNmatrix}
\begin{aligned}
    \langle \mathcal O_{(L)}\rangle_{\mathbb{RP}^4_{\mathsf C}}
    =
    \frac{1+(-1)^L}{\mathcal{Z}_{\mathbb{RP}^{4}_{\mathsf{C}}}}
    \int_{\mathbb R^n} d^na\,
    \Delta(a^2)^2\,
    e^{-\frac{8\pi^2}{g_{\rm YM}^2}\sum_k a_k^2}
    \prod_k
    2a_k
    \left(\coth(\pi a_k)\right)^{\varepsilon_{\mathbb{K}}}
    \sum_k f_L(a_k)\,.
\end{aligned}
\end{align}
Thus only even-length half-BPS operators acquire nonvanishing one-point
functions on $\mathbb{RP}^4_{\mathsf C}$.

\subsection{Solving the matrix model in the planar limit}We now solve the matrix model at large $N$, following the steps in a related problem \cite{Komatsu:2020sup}.

For this purpose, it is useful to first exponentiate the operator insertion and consider the following integral
\begin{align}
\mathcal{Z}_L=\int_{\mathbb R^n} d^na\,
    \Delta(a^2)^2\,
    \prod_k
    2a_k
    {\left|\coth(\pi a_k)\right|}^{\varepsilon_{\mathbb{K}}}\,e^{-\frac{N}{2g^2}\sum_k a_k^2 -Ng_L\sum_{k}f_L(a_k)}
\end{align}
The one-point function can be extracted from its logarithmic derivative,
\begin{align}\label{eq:1ptderformula}
    \langle \mathcal O_{(L)}\rangle_{\mathbb{RP}^4_{\mathsf C}}=-\frac{1+(-1)^{L}}{N}\left.\partial_{g_L}\log \mathcal{Z}_L\right|_{g_L=0}\,.
\end{align}
Here and below, we always assume $L$ to be even.
\paragraph{Equation for resolvent.} As in the standard large $N$ analysis of matrix models, the first step is to introduce a density of eigenvalues $\rho (u)$ defined by
\begin{align}
\rho (u)=\frac{1}{N}\sum_{k}\left(\delta (u-a_k)+\delta (u+a_k)\right)\,.
\end{align}
Using $\rho$, $\mathcal{Z}_L$ can be expressed as
\begin{align}
\mathcal{Z}_L=\int_{\mathbb{R}^n}\frac{d^na_k}{|\tanh (\pi a_k)|^{\varepsilon_{\mathbb{K}}}}e^{-N^2S[\rho]}\,,
\end{align}
where $S[\rho]$ is given by
\begin{align}
\begin{aligned}
S[\rho]&=\frac{1}{4g^2}\int du \,u^2\rho(u)+\frac{g_L}{2}\int du f_{L}(u)\rho(u)-\frac{1}{2}\int\!\!\!\!\!\!\!\text{---}\,\,\, \log |u-v| \rho(u)\rho(v)\,,
\end{aligned}
\end{align}
where the last term is the principal value integral. The large $N$ saddle point of $\rho$ is determined by taking a variation of $S$ and acting with $\partial_u$, which leads to 
\begin{align}
\frac{u}{2g^2}+\frac{g_L}{2}f_L'(u)=\int\!\!\!\!\!\!\!\text{---}\,\,\, dv \frac{\rho(v)}{u-v}\equiv \frac{1}{2}\int dv \frac{\rho (v)}{u-v+i\epsilon}+\frac{\rho (v)}{u-v-i\epsilon}\,.
\end{align}
Next we introduce the {\it resolvent}
\begin{align}
    R(u)=\frac{1}{N}\sum_k\left(\frac{1}{u-a_k}+\frac{1}{u+a_k}\right)\,,
\end{align}
The two equations that the resolvent satisfies are
\begin{align}
R(u+i\epsilon)-R(u-i\epsilon)&=-2\pi i\rho(u)\,,\\
R(u+i\epsilon)+R(u-i\epsilon)&=\frac{u}{g^2}+g_{L}f'_{L}(u)\,.\label{eq:resoleq}
\end{align}
In addition, $R(u)$ must be anti-symmetric, and because $\int du \,\rho (u)=1$, it needs to decay as $1/u$ at $u\to \infty$:
\begin{equation}
R(u)=-R(-u)\,,\qquad 
R(u)\sim \frac{1}{u}\,,\quad u\to \infty\,.
\end{equation}
\paragraph{Solving for $R(u)$.}
To determine the resolvent, we begin with the Cauchy-type representation
\begin{equation}
    R(u)
    =
    \int_{v\sim u}\frac{dv}{2\pi i}
    \sqrt{\frac{(u-\mu)(u+\mu)}{(v-\mu)(v+\mu)}}
    \frac{R(v)}{u-v}\,,
\end{equation}
where the contour encircles $u$ clockwise, and the interval
$[-\mu,\mu]$ is chosen to coincide with the branch cut generated by the
condensed eigenvalues. Deforming the contour away from $u$, we obtain
\begin{equation}
    R(u)
    =
    \oint_{\mathcal C}\frac{dv}{2\pi i}
    \sqrt{\frac{(u-\mu)(u+\mu)}{(v-\mu)(v+\mu)}}
    \frac{R(v)}{u-v}\,,
\end{equation}
where $\mathcal C$ now encircles the cut $[-\mu,\mu]$
counterclockwise. Using the saddle-point equation \eqref{eq:resoleq}, this
becomes
\begin{equation}
\begin{split}
R(u)
&=
\int_{-\mu-i\epsilon}^{\mu-i\epsilon}
\frac{dv}{2\pi i}
\sqrt{\frac{(u-\mu)(u+\mu)}{(v-\mu)(v+\mu)}}
\frac{R(v+i\epsilon)+R(v-i\epsilon)}{u-v}
\\
&=
\int_{-\mu-i\epsilon}^{\mu-i\epsilon}
\frac{dv}{2\pi i}
\sqrt{\frac{(u-\mu)(u+\mu)}{(v-\mu)(v+\mu)}}
\frac{\frac{v}{g^2}+g_L f_L'(v)}{u-v}
\\
&=
\oint\frac{dv}{4\pi i}
\sqrt{\frac{(u-\mu)(u+\mu)}{(v-\mu)(v+\mu)}}
\frac{\frac{v}{g^2}+g_L f_L'(v)}{u-v}\,.
\end{split}
\end{equation}

Let us first set $g_L=0$. The contour integral can then be evaluated
explicitly, giving
\begin{equation}
    R(u)
    =
    \frac{u-\sqrt{u^2-\mu^2}}{2g^2}\,.
\end{equation}
The asymptotic condition $R(u)\sim 1/u$ as $u\to\infty$ fixes the endpoint of
the cut to be $\mu=2g$. Hence
\begin{equation}
    \left.R(u)\right|_{g_L=0}
    =
    \frac{u-\sqrt{u^2-4g^2}}{2g^2}
    =
    \frac{1}{g\,x(u)}\,,
\end{equation}
where $x(u)$ is the Zhukovsky variable,
\begin{equation}
    u
    =
    g\left(x(u)+\frac{1}{x(u)}\right)
    \qquad\Longleftrightarrow\qquad
    x(u)
    =
    \frac{u+\sqrt{u^2-4g^2}}{2g}\,.
\end{equation}

We next turn on an infinitesimal source $g_L$. Differentiating the resolvent
with respect to $g_L$, and setting $\mu=2g$ only after the differentiation,
gives
\begin{equation}
\begin{split}
\partial_{g_L}R(u)
=&
\oint \frac{dv}{4\pi i}
\sqrt{\frac{u^2-4g^2}{v^2-4g^2}}
\frac{f_L'(v)}{u-v}
\\
&+
\frac{\partial_{g_L}\mu}{g^2}
\oint \frac{v\,dv}{8\pi i}
\sqrt{\frac{u^2-4g^2}{v^2-4g^2}}
\left[
\frac{1}{(v-2g)(u-2g)}
-
\frac{1}{(v+2g)(u+2g)}
\right].
\end{split}
\end{equation}
These integrals are conveniently evaluated in Zhukovsky variables,
$x=x(u)$ and $y=x(v)$, using \cite{Komatsu:2020sup}
\begin{equation}
    f_L(v)
    =
    \frac{i^L}{\sqrt{L}}
    \left[
        y^L+\frac{1}{y^L}+\delta_{L2}
    \right].
\end{equation}
The remaining quantity $\partial_{g_L}\mu$ is fixed by requiring
$\partial_{g_L}R(u)=O(1/u^2)$ as $u\to\infty$. Carrying out the contour
integrals then gives \cite{Komatsu:2020sup}
\begin{equation}
    \partial_{g_L}R(u)
    =
    -i^L
    \frac{\sqrt{L}}{g\,x^L}
    \frac{1}{x-x^{-1}}\,.
\end{equation}

\paragraph{Final answer.}
The one-point function follows from
$-(1+(-1)^{L})\partial_{g_L}\log\mathcal Z_L/N$ in the planar limit, as in
\eqref{eq:1ptderformula}. There are two contributions,
\begin{align}
    -\frac{1+(-1)^{L}}{N}\partial_{g_L}\log\mathcal Z_L
    =
    2N\frac{dS[\rho]}{dg_L}
    +
    2\varepsilon_{\mathbb{K}}
    \hspace{5pt}\rightlunateb\hspace{-14pt}\int
    \frac{du}{2\pi i}\,
    \partial_{g_L}R(u)\,
    \log\left|\tanh(\pi u)\right|\,,
\end{align}
where
$\hspace{5pt}\rightlunate\hspace{-10pt}\int$
denotes a contour integral encircling only the right half of the branch cut.

As shown in \cite{Komatsu:2020sup}, the variation of the saddle-point action
vanishes identically. The one-point function therefore comes entirely from the
second term:
\begin{align}\label{eq:finalloc}
\begin{aligned}
\langle \mathcal O_{(L)}\rangle_{\mathbb{RP}^4_{\mathsf C}}
&=
-2i^L\varepsilon_{\mathbb{K}}\sqrt{L}
\hspace{5pt}\rightlunateb\hspace{-14pt}\int_{|x|=1}
\frac{dx}{2\pi i\,x^{L+1}}
\log\left|
\tanh\pi g\left(x+\frac{1}{x}\right)
\right|
\\
&=
-\frac{4i^L\pi g\,\varepsilon_{\mathbb{K}}}{\sqrt{L}}
\hspace{5pt}\rightlunateb\hspace{-14pt}\int_{|x|=1}
\frac{dx}{2\pi i\,x^{{L}}}
\left(1-\frac{1}{x^2}\right)
\frac{1}{
\sinh\left[
2\pi g\left(x+\frac{1}{x}\right)
\right]
}
\\
&=
-\frac{2i^L\pi\,\varepsilon_{\mathbb{K}}}{\sqrt{L}}
\oint
\frac{du}{2\pi i\,x^L}
\frac{1}{\sinh 2\pi u}\,.
\end{aligned}
\end{align}
In the second line, we integrated by parts. In the final line, the contour
encircles the cut $[-2g,2g]$, with the singularity at $u=0$ understood in the
principal-value sense.

\paragraph{Connection to integrability.}The last expression in \eqref{eq:finalloc} admits a useful residue expansion.
Closing the contour at infinity and summing the poles of $1/\sinh 2\pi u$
gives
\begin{align}\label{eq:1ptRP4C}
\langle \mathcal O_{(L)}\rangle_{\mathbb{RP}^4_{\mathsf C}}
&=
\frac{\varepsilon_{\mathbb{K}}}{\sqrt{L}}
\left[
    1
    +
    2i^L
    \sum_{m=1}^{\infty}
    \frac{(-1)^m}{x(im/2)^L}
\right]
\nonumber\\
&=
\frac{i^L\varepsilon_{\mathbb{K}}}{\sqrt{L}}
\sum_{m=-\infty}^{\infty}
\frac{(-1)^m}{x(im/2)^L}\,.
\end{align}
The first term in the first line arises from the principal-value prescription
at $u=0$, while the remaining terms come from the poles at
$u=\pm im/2$, with $m\in\mathbb Z_{>0}$. 

This can be further recast into a form suggestive of the underlying integrable structure \cite{Beisert:2010jr}. In the integrability approach to $\mathcal{N}=4$ SYM, we describe single-trace operators in terms of a collection of ``magnon'' excitations on a spin chain. The half-BPS operators that we have been studying correspond to the vacuum states in the spin chain. Hence there are no physical magnons. However, non-trivial corrections are known to arise from the production of virtual magnons, often referred to as mirror magnons. The mirror magnons obey a dispersion relation
\begin{align}
\tilde{E}(\widetilde p)
 =2{\rm arcsinh}\!\left(\frac{\sqrt{1+\widetilde p^{\,2}}}{4g}\right)\,,
\end{align}
where $\tilde{E}$ and $\tilde{p}$ are the energy and the momentum of the mirror magnon. There are also bound states of mirror magnons labelled by an integer $m$, whose dispersion relation is given by
\begin{align}
\tilde{E}_m(\widetilde p)
 =2{\rm arcsinh}\!\left(\frac{\sqrt{m^2+\widetilde p^{\,2}}}{4g}\right)\,.
\end{align}
Using these, we can rewrite \eqref{eq:1ptRP4C} as
\begin{align}
\langle \mathcal O_{(L)}\rangle_{\mathbb{RP}^4_{\mathsf C}}=\frac{\varepsilon_{\mathbb{K}}}{\sqrt{L}}\left[1+2\sum_{m=1}^{\infty}(-1)^m
 e^{-\frac L2\widetilde E_m(0)}\right]\,,
\end{align}
which suggests that the quantum corrections come from the contribution from virtual magnons---which, in the integrability literature, is called ``wrapping effects''. We leave the question of reproducing this expression directly from integrability for the future.
\subsection{Testing the localization result in various limits}
We now study \eqref{eq:finalloc} and \eqref{eq:1ptRP4C} in various limits and test them against gauge theory, supergravity and string theory.
\paragraph{Weak coupling.} The weak coupling expansion can be obtained straightforwardly by expanding \eqref{eq:1ptRP4C}. Expanding it to sufficiently high orders, we find that it has the following closed-form expression
\begin{align}
   \langle O_{(L)}\rangle_{\mathbb{RP}^4_{\mathsf C}}
=\frac{\varepsilon_{\mathbb{K}}}{\sqrt{L}}
\left[1+\sum_{r=\frac{L}{2}}^\infty\frac{(-1)^{L/2}\,L\left(2^{2r-1}-1\right)B_{2r}(2\pi g)^{2r}}
{r\left(r-\frac{L}{2}\right)!\left(r+\frac{L}{2}\right)!}\,,
 \right]
\end{align}
where $B_{2r}$ is the Bernoulli number. This correctly reproduces the tree-level answer \eqref{eq:vacVEV} and predicts the next correction to be
\begin{align}
   \langle O_{(L)}\rangle_{\mathbb{RP}^4_{\mathsf C}}
=\frac{\varepsilon_{\mathbb{K}}}{\sqrt{L}}
\left(1-2(2^{L}-2)\zeta_L
 g^{L}
 +O(g^{L+2})\right)\,,
\end{align}
where $\zeta_L$ is the zeta value.
\paragraph{Supergravity limit.}
We first consider the strong-coupling limit $g\to\infty$ with the operator
length $L$ held fixed. For this purpose, it is useful to rewrite
\eqref{eq:finalloc} as
\begin{align}\label{eq:strongcouplingint}
\begin{aligned}
    \langle \mathcal O_{(L)}\rangle_{\mathbb{RP}^4_{\mathsf C}}
    &=
    \frac{4i^L\pi\,\varepsilon_{\mathbb{K}}}{\sqrt{L}}
    \int_{0}^{2g}
    \frac{du}{2\pi i}
    \frac{x^{L}-x^{-L}}{\sinh 2\pi u}
    \\
    &=
    \frac{4\,\varepsilon_{\mathbb{K}}}{\sqrt{L}}
    \int_{0}^{2g}du\,
    \frac{
        \sin\!\left(
            L\arcsin\frac{u}{2g}
        \right)
    }{
        \sinh 2\pi u
    }\,.
\end{aligned}
\end{align}
At large $g$ and fixed $L$, we expand
\begin{equation}
 \sin(L\arcsin s)
 =
 Ls
 -\frac{L(L^2-1)}{6}s^3
 +\frac{L(L^2-1)(L^2-9)}{120}s^5
 +\cdots,
\end{equation}
and use
\begin{equation}
 \int_0^\infty\frac{u^q}{\sinh 2\pi u}\,du
 =
 \frac{2\Gamma(q+1)(1-2^{-q-1})\zeta_{q+1}}
 {(2\pi)^{q+1}}\,.
 \label{eq:sinhmoments}
\end{equation}
This gives
\begin{align}\label{eq:supergravity}
    \langle \mathcal O_{(L)}\rangle_{\mathbb{RP}^4_{\mathsf C}}
    =
    \varepsilon_{\mathbb{K}}
    \frac{\sqrt L}{8g}
    \left[
        1
        -
        \frac{L^2-1}{192g^2}
        +
        O(g^{-4})
    \right].
\end{align}
The leading term agrees with the supergravity prediction \cite{Zhou:2024ekb}, while the
subleading term gives a prediction for the first stringy correction.

\paragraph{Semiclassical string limit.}
A second natural strong-coupling regime is obtained by sending both $g$ and
$L$ to infinity while keeping
$\mathcal J=L/g$ fixed. This is the regime in which the operator is dual to a
macroscopic semiclassical string in $AdS_5\times S^5$. The same integral
\eqref{eq:strongcouplingint} applies, but its asymptotic expansion is now
different:
\begin{align}
    L\arcsin\frac{u}{2g}
    =
    \frac{\mathcal J u}{2}
    +
    \frac{\mathcal J u^3}{48g^2}
    +
    \frac{3\mathcal J u^5}{1280g^4}
    +
    \cdots.
\end{align}
Substituting this expansion into \eqref{eq:strongcouplingint} and extending
the upper limit of integration to infinity gives
\begin{align}\label{eq:strongsemi}
    \langle \mathcal O_{(L)}\rangle_{\mathbb{RP}^4_{\mathsf C}}
    =
    \frac{\varepsilon_{\mathbb{K}}}{\sqrt{\mathcal J g}}
    \left[
        \tanh\left(\frac{\mathcal J}{8}\right)
        +
        \frac{\mathcal J}{1536g^2}
        \frac{
            2-\cosh\frac{\mathcal J}{4}
        }{
            \cosh^4\frac{\mathcal J}{8}
        }
        +
        O(g^{-4})
    \right].
\end{align}

Unlike some observables in the same semiclassical regime
(cf.~\cite{Komatsu:2020sup}), the one-point function does not contain an
exponential factor of the form $e^{-g\,\#}$. This suggests that the relevant
classical worldsheet saddle has vanishing action, with the nontrivial
$\mathcal J$ dependence arising from one-loop and higher-order fluctuations.
It would be interesting to verify this interpretation directly from the
worldsheet theory.

As a consistency check, when $\mathcal J\ll1$, the semiclassical result
\eqref{eq:strongsemi} smoothly reduces to the supergravity expansion
\eqref{eq:supergravity}.

\paragraph{BMN limit.}
The same result can be compared directly with the pp-wave computation. Using
the standard BMN dictionary,
\begin{equation}
    m\ell
    =
    \frac{2\pi L}{\sqrt{\lambda}},
\end{equation}
the pp-wave crosscap overlap \eqref{eq:finalBMN} becomes
\begin{equation}
    c_{\ell}
    =
    c_0
    \tanh\left(\frac{L}{8g}\right).
\end{equation}
This agrees precisely with the leading term of \eqref{eq:strongsemi}, since
$\mathcal J=L/g$. The comparison also fixes the previously undetermined
normalization to be
\begin{align}
    c_0
    =
    \frac{\varepsilon_{\mathbb{K}}}{\sqrt{L}}\,.
\end{align}
Thus the localization result interpolates consistently between the
supergravity, semiclassical-string, and pp-wave descriptions.
\paragraph{Comments on $1/N$ corrections.}
The localization formula \eqref{eq:finiteNmatrix}, valid at finite $N$ and
$g_{\rm YM}$, in principle determines the full $1/N$ expansion of the
one-point functions. There are two distinct sources of corrections.

The first comes from fluctuations of the matrix integral around the planar
saddle. These are the usual matrix-model $1/N$ corrections and can, in
principle, be computed systematically using standard techniques such as
topological recursion. The second comes from the operator insertion itself.
The polynomial $f_L$ in \eqref{eq:GramSchmidtlargeN} was obtained by resolving
operator mixing only at leading order in large $N$. At subleading orders,
additional mixing must be taken into account, producing corresponding
$1/N$ corrections to $f_L$.

Combining these two effects should determine the $1/N$ expansion at finite
't~Hooft coupling. In Appendix~\ref{sec:AOLsubN}, we compute the first
subleading correction to the tree-level one-point functions directly in
perturbative $\mathcal N=4$ SYM. It would be interesting to reproduce this
result from the finite-$N$ localization formula, and thereby provide a
nontrivial check of our result beyond the planar limit. 

We leave this as an exercise for interested readers---or for an AI, provided it can be persuaded to perform the calculation rather than merely claim to have done so.\footnote{This sentence was written by an AI.}
\section{Conclusion}\label{sec:conclusion}
In this paper, we discussed planar $\mathcal{N}=4$ SYM on $\mathbb{RP}^{4}$ preserving 16 supercharges, whose study was initiated in\cite{Caetano:2022mus}. We provided evidence that the integrability is preserved once we combine the involution defining $\mathbb{RP}^{4}$ with the action of the charge conjugation of the SU$(N)$ gauge group. Concretely, we demonstrated that the computation of the one-point functions of single-trace operators can be mapped to the computation of overlaps between the energy eigenstates and the crosscap state in the integrable spin chain \cite{Caetano:2021dbh}. We calculated the overlaps in SU$(2)$, SO$(6)$ and SU$(1|2)$ sectors at tree level, deriving closed-form expressions in terms of a ratio of Gaudin determinants. Notably, unlike the former two bosonic sectors, the result for the SU$(1|2)$ sector includes a multiplicative factor in terms of a ratio of $Q$-functions, analogous to the overlap for the boundary states \cite{Piroli:2017sei,deLeeuw:2015hxa}. Requiring consistency with the results in subsectors and the fermionic duality, we conjectured a formula for the full $\PSU(2,2|4)$ sector. We also studied the holographic description. At strong coupling, we identified the relevant classical crosscap in the AdS$_5\times \text{S}^5$ string sigma model and constructed the corresponding quantum crosscap state in the pp-wave limit. Furthermore, using supersymmetric localization, we computed the one-point functions of half-BPS operators exactly at finite $N$ and $g_{\rm YM}$. In the planar limit, the result correctly interpolated between the weak-coupling prediction of $\mathcal N=4$ SYM, the strong-coupling supergravity result, and the BMN limit described by string theory on the pp-wave background.

The planar integrability of $\mathcal{N}=4$ SYM has been applied in the past to a variety of observables corresponding to different worldsheet topology; a cylinder \cite{Beisert:2010jr}, a pair of pants \cite{Basso:2015zoa}, spheres with multiple punctures \cite{Fleury:2016ykk,Eden:2016xvg}, disks \cite{Basso:2013vsa,Sever:2020jjx,deLeeuw:2015hxa,Jiang:2019xdz,Jiang:2019zig,Gombor:2020kgu,Gombor:2020auk,Komatsu:2020sup,Coronado:2025xwk} and higher genus worldsheets \cite{Bargheer:2017nne,Bargheer:2018jvq,Ben-Israel:2018ckc}. This paper initiates the study of the last missing corner;  worldsheets with crosscap insertions. It therefore naturally leads to many open questions.

First and foremost, we hightlight the importance of studying crosscap overlaps for super spin chains. Crosscap overlaps for bosonic spin chains are given purely by a ratio of Gaudin determinants \cite{Caetano:2021dbh,Gombor:2022deb}. By contrast, as observed in this paper, crosscap overlaps for super spin chains come with a multiplicative factor given by a ratio of $Q$-functions. In hindsight, this is not surprising since a ratio of Gaudin determinants does not transform nicely under the fermionic duality. 
Note that similar multiplicative factors are common in the boundary state overlaps \cite{deLeeuw:2015hxa,Jiang:2019xdz,Jiang:2019zig,Gombor:2020kgu,Gombor:2020auk,Komatsu:2020sup,Coronado:2025xwk}. In those cases, the multiplicative factors depend on the details of the boundary state, and they are constrained by the algebraic consistency conditions such as the boundary Yang-Baxter equation. By contrast, such algebraic understanding is missing for crosscap overlaps, and we currently do not know general rules for determining the ratio of $Q$-functions. We expect that there exists a simple and uniform description for the multiplicative factor, yet to be uncovered.

Understanding crosscap states for super spin chains is also crucial for promoting the result in this paper to finite coupling. As discussed in related contexts \cite{Jiang:2019xdz,Jiang:2019zig,Gombor:2020kgu,Gombor:2020auk,Komatsu:2020sup,Coronado:2025xwk}, it is straightforward to write a finite-coupling version of the ratio of Gaudin determinants. The stumbling block is again the multiplicative factor. On a positive side, once one understands the structure of the multiplicative factor at tree level, it is likely straightforward to generalize it to finite coupling.

Speaking about the finite-coupling generalization, another important direction to pursue is the reformulation of crosscap overlaps in terms of Quantum Spectral Curve \cite{Gromov:2013pga}. A simple starting point towards this goal would be to analyze the bosonic sector at tree level, for which the overlaps reduce to a ratio of Gaudin determinants.

Our localization analysis led us to two-dimensional Yang--Mills theory
(2d YM) on $\mathbb{RP}^2$ with charge conjugation. Although only its
zero-instanton sector was needed here, the full theory is interesting in its
own right, particularly in light of the recent revival of interest in the
worldsheet description of large-$N$ 2d YM \cite{Aharony:2023tam,Komatsu:2025sqo,Aharony:2025owv,Komatsu:2025dqv}. It would be worthwhile to
study this theory beyond the zero-instanton sector and to clarify its possible
string-theoretic interpretation.

Another interesting direction is to identify broader classes of integrable observables
associated with orientifolds in AdS. $\mathcal{N}=4$ SYM on $\mathbb{RP}^4$ studied in this
paper is just one example of a more general family. In forthcoming work \cite{inprogress}, we construct supersymmetric {\it crosscap defects}, a novel class of defects introduced in \cite{Drukker:2026nvv}, in
$\mathcal N=4$ SYM.
When combined with charge conjugation, they also preserve planar integrability and
their one-point functions can be described by crosscap states both in the
spin chain and on the string worldsheet. Hence many of the analyses developed
in this paper extend naturally to this broader class of defects.

\section*{Acknowledgement}
We thank Leonardo Rastelli and Yifan Wang for correspondence at the early stage of this project. 
\paragraph{Use of artificial intelligence tools.} The authors used OpenAI ChatGPT (Astra) to assist with technical analysis in Sections \ref{subsec:det} and \ref{sec:localization}, building on initial calculations performed independently by the authors. The last sentence of Section \ref{sec:localization} was written entirely by Astra. All AI-assisted work was reviewed, verified, and modified as necessary by the authors, who take full responsibility for the accuracy and content of the manuscript.
\appendix
\section{Gauge propagators}\label{sec:Gauge}
In this appendix, we explain the gauge choice that leads to the propagators \eqref{eq:propbefore} and \eqref{eq:propagators}. Since we will only be interested in the spacetime dependence of the propagators, we will strip off the color indices.
\subsection{Propagators on \texorpdfstring{$\mathbb{R}^4$}{R4}}
Our starting point is $d$-dimensional Yang-Mills theory
\begin{equation}
    S_{\text{YM}}=\frac{1}{4g_\text{YM}^2}\int\text{d}^d x\sqrt{g}\Tr F_{\mu\nu}F^{\mu\nu}\,,\label{eq:YangMills}
\end{equation}
We want to add an $R_\xi$ gauge-fixing term $S_{\text{gf}}(\xi)\sim (\nabla_\mu A^\mu)^2/\xi$. However, the gauge-fixing term is generically not conformally invariant, and we have to specify whether we are adding an $R_\xi$ term in flat space or on the sphere. This leads to two different gauge-fixings, given by
\begin{subequations}
\begin{align}
    S^{\text{flat}}_{\text{gf}}&=\frac{1}{2g^2_{\text{YM}}\xi}\int_{\mathbb{R}^d}\text{d}^dx\,\Tr(\partial_\mu A^\mu)^2\,,
    \\
    S^{\text{sphere}}_{\text{gf}}&=\frac{1}{2g^2_{\text{YM}}\xi}\int_{S^d}\text{d}^d\theta\,\Tr(\nabla_\mu A^\mu)^2
    \\
    \notag
    &=\frac{1}{2g^2_{\text{YM}}\xi}\int_{\mathbb{R}^d}\text{d}^dx\, \left(\frac{2}{1+x^2}\right)^{d-4}\Tr\left(\partial_\mu A^\mu-\frac{2(d-2) x_\mu A^\mu}{1+x^2}\right)^2\,.
\end{align}
\end{subequations}
The two choices are inequivalent and lead to very different propagators on $\mathbb{R}^4$. They are given by
\begin{subequations}
\label{eq:bothprop}
\begin{align}
    G_{\mu\nu}^{\text{flat}}(r)&=\frac{g^2_{\text{YM}}\Gamma(\Delta)}{8\pi^{d/2}r^{2\Delta}}\Big\{(1+\xi)\delta_{\mu\nu}+2\Delta(1-\xi)\frac{r_\mu r_\nu}{r^2}\Big\}\,,\label{eq:flatprop}
    \\
    G^{\text{sphere}}_{\mu\nu}(\eta)&=\frac{g_{\text{YM}}^2\Gamma(2\Delta)}{2^{d-1}\pi^{d/2}\Gamma(\Delta+1)}\Big\{[g_{\Delta,\xi}(\eta)-f_\Delta(\eta)](\partial_\mu^x\partial_\nu^y\eta)+g'(\eta)(\partial_\mu^x\eta)(\partial_\nu^y\eta)\Big\}\,,\label{eq:roundprop}
\end{align}
\end{subequations}
respectively,
where $\Delta=d/2-1$ is the scaling dimension of the gauge field, $r_\mu=x_\mu-y_\mu$, $\eta$ was defined in \eqref{eq:etadef} and the functions $f_\Delta$ and $g_{\Delta,\xi}$ are given by
\begin{equation}
\begin{split}
    f_\Delta(\eta)&={}_2F_1\left(2\Delta,1;\Delta+1;1-\eta\right)\comma 
    \\
    g_{\Delta,\xi}(\eta)&=-\frac{f(\eta)}{2\Delta-1}-\frac{\Delta}{2\Delta+1}\left(\xi-\frac{2\Delta+1}{2\Delta-1}\right)\sum_{n=1}^{\infty}\frac{(2\Delta+1)_n}{(\Delta+1)_n}\left[(\psi(n+d-1)-\psi(n)\right](1-\eta)^{n-1}\,,
\end{split}
\end{equation}
where $(a)_n$ is the Pochhammer symbol and $\psi$ is the digamma function.
In $d=4$, these functions simplify to
\begin{equation}
    f_1(\eta)=\frac{1}{\eta}\comma\quad g_{1,\xi}(\eta)=-\frac{\xi-1}{2\eta}+\frac{3-\xi}{6(1-\eta)^2}[\eta-1+(2\eta-3)\log\eta]
\end{equation}
There are a number of distinguished gauge-choices. One standard choice is conformal gauge $\xi_{\text{conf}}=\frac{\Delta+1}{\Delta-1}$. This choice satisfies
\begin{equation}
    G_{\mu\nu}^{\text{flat}}(r)\Big|_{\xi=\xi_{\text{conf}}}=\frac{2\Delta}{\Delta-1}\frac{g^2_{\text{YM}}\Gamma(\Delta)}{8\pi^{d/2}r^{2\Delta}}I_{\mu\nu}(r)\,,
\end{equation}
where $I_{\mu\nu}$ was defined in \eqref{eq:vetr}. This makes the propagator transform covariantly under conformal transformations. However, it is singular in $d=4$, since in this case the Yang-Mills action \ref{eq:YangMills} is conformally invariant on its own. There is another natural choice, called Fried-Yennie gauge $\xi_{\text{FY}}=\frac{2\Delta+1}{2\Delta-1}$. This gauge satisfies
\begin{equation}
    r^\mu G^{\text{flat}}_{\mu\nu}(r)=0\comma \quad g_{\Delta,\xi_{\text{FY}}}(\eta)=-\frac{f(\eta)}{2\Delta-1}\,.
\end{equation}
We will be particularly interested in the second property, which implies that the round propagator has no logarithmic singularities in this gauge. It corresponds to $\xi=3$ in $d=4$, in which the round propagator takes the form
\begin{equation}
    G_{\mu\nu}^{\text{sphere}}(\eta)\Big|_{d=4,\xi=3}=\frac{g^2_{\text{YM}}}{8\pi^2}\left\{\frac{(\partial_\mu^x\eta)(\partial_\nu^y\eta)}{\eta^2}-\frac{2(\partial_\mu^x\partial_\nu^y \eta)}{\eta}\right\}\,,
\end{equation}
which is the form of the propagator reported in \eqref{eq:propbefore}.
\subsection{Propagators on \texorpdfstring{$\mathbb{RP}^4$}{RP4}} We now consider what happens to these propagators once we add the images under $\iota$. Recall that the gauge-fixing term, and thus the propagators, are not conformally invariant. Which symmetries are preserved depends on the choice of propagator. 
\begin{itemize}
    \item Flat space $R_\xi$ gauge \eqref{eq:flatprop}: By construction, this transforms correctly under translations, rotations, dilatations and parity. But inversion $\mathsf{I}$ and therefore special conformal transformations on $\mathbb{R}^4$ are not symmetries away from the conformal point. Since this is singular in $d=4$, flat space $R_\xi$-gauge is not a good choice of gauge for computations on $\mathbb{RP}^4$.
    \item Spherical $R_\xi$ gauge \eqref{eq:roundprop}: This has the symmetry group of the sphere. This includes rotations, sphere translations generated by $P_\mu+K_\mu$, reflections, and inversion. Since inversions are included, round space $R_\xi$ gauge is a natural choice for computations on $\mathbb{RP}^4$. In particular, it allows us to use the method of images \eqref{eq:methodofim} to compute the propagator on $\mathbb{RP}^4$.
\end{itemize}
Since $\eta(x',y)=1-\eta(x,y)$, we can then write the propagator on $\mathbb{RP}^4$ in the form
\begin{equation}
    \Big<A_\mu(x),A_\nu(y)\Big>_{\mathbb{RP}^4}=G_{\mu\nu}^{\text{round}}(\eta)-G_{\mu\nu}^{\text{round}}(1-\eta)\,,
\end{equation}
where the minus sign comes from the sign-flip of $A$ under $\mathsf{I}$. Specifying Fried-Yennie gauge $\xi=3$ leads to the propagators in \eqref{eq:propagators}.
\section{Diagrammatics for one-point functions on \texorpdfstring{$\mathbb{RP}^4$}{RP4}}\label{sec:1ptF}
In this appendix, we provide details on the tree-level computation of one-point functions on $\mathbb{RP}^{4}$ with charge conjugation. In section \ref{sec:maxtwist}, we prove that the tree-level Wick contraction at large $N$ is dominated by the ``maximally twisted diagrams'' given in Figure \ref{fig:maxtwist}. We then compute the one-point functions of half-BPS operators both at large $N$ (section \ref{sec:AOLN}) and at all orders in $1/N$ (section \ref{sec:AOLsubN}).
\subsection{Dominance of the maximally twisted diagram}\label{sec:maxtwist}
\begin{figure}
\centering
\begin{tikzpicture}
  \def\R{8}          
  \def\t{22.5}         
  \def\toff{3.5}       
  \def\roff{6.8}       
  \def\innerangle{70}
  \def\outerangle{25}
  \def\looseness{1}  

  \coordinate (Li) at
    ({\R*sin(-\t/2+\toff)},{\R*cos(-\t/2+\toff)});
  \coordinate (Lo) at
    ({\R*sin(-\t/2-\toff)},{\R*cos(-\t/2-\toff)});
  \coordinate (Ri) at
    ({\R*sin(\t/2-\toff)},{\R*cos(\t/2-\toff)});
  \coordinate (Ro) at
    ({\R*sin(\t/2+\toff)},{\R*cos(\t/2+\toff)});
  \coordinate (C) at (0,\roff);


  \draw[thick,red] (Li)
    to[out={-\innerangle},in=180,looseness=\looseness] (C)
    to[out=0,in={180+\outerangle},looseness=\looseness] (Ro);
  \draw[thick,red,dashed] (Lo)
    to[out={-\outerangle},in=180,looseness=\looseness] (C)
    to[out=0,in={180+\innerangle},looseness=\looseness] (Ri);
  \draw[thick,red] (Li)
    arc[start angle={90+\t/2-\toff},
        end angle={90-\t/2+\toff},radius=\R];
  \draw[thick,red] (Lo)
    arc[start angle={90+\t/2+\toff},
        end angle={90+3*\t/2-\toff},radius=\R];
  \draw[thick,red] (Ro)
    arc[start angle={90-\t/2-\toff},
        end angle={90-3*\t/2+\toff},radius=\R];

\draw[black, dotted, thick] ({\R*sin(-3*\t/2-\toff)},{\R*cos(-3*\t/2-\toff)})
    arc[start angle={90+3*\t/2+\toff},
        end angle={90+5*\t/2-\toff},radius=\R];

\draw[black, dotted, thick] ({\R*sin(3*\t/2+\toff)},{\R*cos(3*\t/2+\toff)})
    arc[start angle={90-3*\t/2-\toff},
        end angle={90-5*\t/2+\toff},radius=\R];
        
    \node at ({\R*sin(-3/2*\t)},
              {\R*cos(-3/2*\t)}) {$Z_{L}$};
    \node at ({\R*sin(-1/2*\t)},
              {\R*cos(-1/2*\t)}) {$Z_{1}$};
    \node at ({\R*sin(1/2*\t)},
              {\R*cos(1/2*\t)}) {$Z_{2}$};
    \node at ({\R*sin(3/2*\t)},
              {\R*cos(3/2*\t)}) {$Z_{3}$};

\end{tikzpicture}
\caption{A cut-out of a diagram contributing to the one-point function of $\Tr(Z_{1}Z_{2}Z_{3}\dots Z_{L})$. This diagram contains a Wick-contraction between $Z_{1}$ and $Z_{2}$. This does not lead to a closed face with only two edges; indeed, even with 4 edges, the face is still open.}\label{fig:Neigh}
\end{figure}
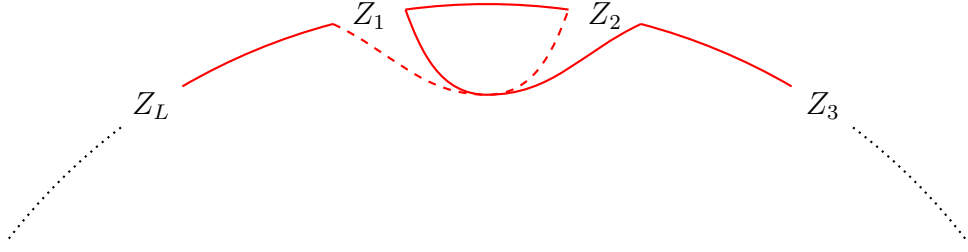

Consider the one-point function of the single-trace operator $\Tr(Z_1Z_2\dots Z_L)$, where $Z_i$ are any set of fields and $L$ is even. We will argue that at order $O(N^0)$ there is a unique diagram contributing to the one-point function, namely the maximally twisted diagram \ref{fig:maxtwist}. To see this, note that the maximally twisted diagram has exactly 4 edges for each face. Since each edge connects two fields and each field has 4 edges ending on it, this makes for a total of
\begin{equation}
    \text{\#faces}=\frac{\text{\#edges}}{4}=\frac{L}{2}\,.
\end{equation}
Now we will argue that for any other diagram, it must be true that:
\begin{enumerate}
    \item[$\mathbf{(1)}$] It can not contain a face with \textit{fewer} than 4 edges;
    \item[$\mathbf{(2)}$] It must contain at least one face with \textit{more} than 4 edges.
\end{enumerate}
We will consider the one-point function of $\Tr (Z_{1}Z_{2}\dots Z_{L})$, where $Z_i$ are any set of field. 
To argue for $\mathbf{(1)}$, consider how one might construct such a face. Since every other edge of a face will correspond to the trace, which connects neighbouring nodes, it must arise from Wick-contracting two neighbouring nodes. But this results in the diagram seen in Fig. \ref{fig:Neigh}, whose face does not close. Thus, such a construction is impossible using a twisted propagator. Contrast this with the non-twisted $\mathsf{P}$-$\mathbb{RP}^4$ propagator: there, the dominant diagram seen in Fig. \ref{fig:onepointN} arose precisely because faces could be closed with just two edges. 

To argue for $\mathbf{(2)}$, consider a diagram which Wick contracts the field $Z_{1}$ with $Z_{k+1}$, as seen in Fig. \ref{fig:ZZK}, assuming without loss of generality that $2k<L$. The only way the red face can then be closed using only four edges is to contract $Z_{2}$ with $Z_{k+2}$. But the argument then carries forward: the only way the red face can be closed with only four edges is to contract $Z_{3}$ with $Z_{k+3}$ and so on, contracting $Z_{n}$ with $Z_{k+n}$. Assume without loss of generality that $k<L/2$. Eventually, we will reach $n=k+1$, implying that we must contract $Z_{k+1}$ and $Z_{2k+1}$. But we started with the assumption that $Z_{k+1}$ was contracted with $Z_{1}$ and $2k\neq L$, and we reach a contradiction. 

\begin{figure}
\centering
\begin{tikzpicture}
\def\R{8}
\def\t{22.5}
\def\toff{3.5}
\def\looseness{1}
\pgfmathsetmacro{\Cy}{\R*cos(5*\t/2)}
\coordinate (C) at (0,\Cy);

\foreach \j in {1,2,3} {
  \pgfmathsetmacro{\thetaL}{(2*\j-9)*\t/2}
  \pgfmathsetmacro{\thetaR}{(2*\j+1)*\t/2}
  \pgfmathsetmacro{\startangle}{(2-\j)*\t}
  \pgfmathsetmacro{\endangle}{180+\startangle}
  \coordinate (L\j) at ({\R*sin(\thetaL)},{\R*cos(\thetaL)});
  \coordinate (R\j) at ({\R*sin(\thetaR)},{\R*cos(\thetaR)});
  \foreach \s/\ls/\rs in {1/i/o,-1/o/i} {
    \coordinate (L\j\ls) at
      ({\R*sin(\thetaL+\s*\toff)},{\R*cos(\thetaL+\s*\toff)});
    \coordinate (R\j\rs) at
      ({\R*sin(\thetaR+\s*\toff)},{\R*cos(\thetaR+\s*\toff)});
    \pgfmathsetmacro{\angleIn}{atan2(\R*cos(\thetaL+\s*\toff)-\Cy,\R*sin(\thetaL+\s*\toff))}
    \pgfmathsetmacro{\angleOut}{atan2(\R*cos(\thetaR+\s*\toff)-\Cy,\R*sin(\thetaR+\s*\toff))}
    \expandafter\xdef\csname curve\j\ls\endcsname{
      (L\j\ls) to[out=\startangle,in=\angleIn,looseness=\looseness] (C)
      to[out=\angleOut,in=\endangle,looseness=\looseness] (R\j\rs)
    }
    \expandafter\xdef\csname reverse\j\ls\endcsname{
      to[out=\endangle,in=\angleOut,looseness=\looseness] (C)
      to[out=\angleIn,in=\startangle,looseness=\looseness] (L\j\ls)
    }
  }
}

\foreach \j/\col in {1/red,2/CornflowerBlue} {
  \pgfmathtruncatemacro{\next}{\j+1}
  \pgfmathsetmacro{\thetaL}{(2*\j-9)*\t/2}
  \pgfmathsetmacro{\thetaR}{(2*\j+1)*\t/2}
  \fill[\col,opacity=0.2]
    \csname curve\j i\endcsname
    arc[start angle={90-\thetaR-\toff},end angle={90-\thetaR-\t+\toff},radius=\R]
    \csname reverse\next o\endcsname
    arc[start angle={90-\thetaL-\t+\toff},end angle={90-\thetaL-\toff},radius=\R]
    -- cycle;
}

\foreach \j/\solidcol/\dashedcol in {1/red/Dandelion,2/CornflowerBlue/red,3/ForestGreen/CornflowerBlue} {
  \draw[thick,\solidcol] \csname curve\j i\endcsname;
  \draw[thick,\dashedcol,dashed] \csname curve\j o\endcsname;
  \node at (L\j) {$Z_{\j}$};
  \node at (R\j) {$Z_{k+\j}$};
}

\foreach \j/\col in {1/red,2/CornflowerBlue} {
  \pgfmathsetmacro{\thetaL}{(2*\j-9)*\t/2}
  \pgfmathsetmacro{\thetaR}{(2*\j+1)*\t/2}
  \draw[thick,\col] (L\j i)
    arc[start angle={90-\thetaL-\toff},end angle={90-\thetaL-\t+\toff},radius=\R];
  \draw[thick,\col] (R\j o)
    arc[start angle={90-\thetaR-\toff},end angle={90-\thetaR-\t+\toff},radius=\R];
}

\draw[black,dotted,thick] (L3i)
  arc[start angle={90+3*\t/2-\toff},end angle={90-3*\t/2+\toff},radius=\R];
\draw[black,dotted,thick] (L1o)
  arc[start angle={90+7*\t/2+\toff},end angle={90+9*\t/2-\toff},radius=\R];
\draw[black,dotted,thick] (R3o)
  arc[start angle={90-7*\t/2-\toff},end angle={90-9*\t/2+\toff},radius=\R];
\end{tikzpicture}
\caption{A cut-out of a diagram contributing to the one-point function of $\Tr(Z_{1}Z_{2}Z_{3}\dots Z_{L})$. This diagram contains a Wick-contraction between $Z_{1}$ and $Z_{k+1}$. This only leads to a closed face with 4 edges if the diagram also contracts $Z_{2}$ with $Z_{k+2}$, which then requires contracting $Z_{3}$ with $Z_{k+3}$ and so on.}\label{fig:ZZK}
\end{figure}
\subsection{Computing \texorpdfstring{$A_{\mathcal{O}_{(L)}}$}{AOL} at leading order in \texorpdfstring{$N$}{N}}\label{sec:AOLN}
We will now focus on the computation of the 1-point function of the operator $\mathcal{O}_{(L)}$ \eqref{eq:OL}. We will write the operator in the form
\begin{equation}
    \mathcal{O}_{(L)}=\frac{\Tr \hat{Z}^L}{\sqrt{L}}\,,\quad \hat{Z}=\sqrt{\frac{4\pi^2}{\lambda}}Z\,,
\end{equation}
where the field $\hat{Z}$ is normalised such that at leading order in $N$, the Wick-contraction takes the form
\begin{equation}
\left<:[\hat{Z}]_{pq}[\hat{Z}]_{rs}:(x)\right>=\frac{1}{(1+x^2)^2}\frac{\mathbb{K}_{pr}\mathbb{K}_{sq}^{-1}}{N}\,.\label{eq:ZhatWick}
\end{equation}
 As a warm-up, let us compute the one-point function of $\mathcal{O}_{(2)}$. This simply amounts to performing a single Wick contraction, and we get
 \begin{equation}
  \begin{split}
     A_{\mathcal{O}_{(2)}}=(1+x^2)^2\left<\mathcal{O}_{(2)}(x)\right>&=\frac{(1+x^2)^2}{\sqrt{2}}\sum_{p,q,r,s}\delta_{qr}\delta_{ps}\left<:[\hat{Z}]_{pq}[\hat{Z}]_{rs}:(x)\right>=\frac{1}{\sqrt{2}}\sum_{p,q}\frac{\mathbb{K}_{pq}\mathbb{K}_{pq}^{-1}}{N}
     \\
     &=\frac{\varepsilon_{\mathbb{K}}}{\sqrt{2}}\sum_{p,q}\frac{\mathbb{K}_{pq}\mathbb{K}^{-1}_{qp}}{N}=\frac{\varepsilon_{\mathbb{K}}}{\sqrt{2}}\sum_{p}\frac{\delta_{pp}}{N}=\frac{\varepsilon_{\mathbb{K}}}{\sqrt{2}}\,.
      \end{split}\label{eq:A2}
 \end{equation}
 We can then proceed with the computation of $A_{\mathcal{O}_{(L)}}$ for higher $L$. This is where the previous section becomes important, as we only need to compute the contribution of the maximally twisted diagram, since every other diagram is subleading. Keeping track of the color indices, we find that the maximally twisted diagram contribution takes the form
\begin{equation}
    \frac{(1+x^2)^L}{\sqrt{L}}\sum_{p_1,p_2,\dots p_{2L}}\prod_{n=1}^{L/2}\delta_{p_{2n+1},p_{2n}}\delta_{p_{L+2n+1},p_{L+2n}}\left<:[\hat{Z}_{n}]_{p_{2n-1},p_{2n}},[\hat{Z}_{n+L/2}]_{p_{L+2n-1},p_{L+2n}}:(x)\right>\,,
\end{equation}
where the trace enforces $p_{2L+n}=p_{n}$.
Using the Wick contraction rules \eqref{eq:ZhatWick}, we then obtain
\begin{equation}
\begin{split}
   &\frac{1}{\sqrt{L}}\sum_{p_1,p_2,\dots p_{2L}}\prod_{n=1}^{L/2}\delta_{p_{2n+1},p_{2n}}\delta_{p_{L+2n+1},p_{L+2n}}\frac{\mathbb{K}_{p_{2n-1},p_{L+2n-1}}\mathbb{K}_{p_{L+2n},p_{2n}}^{-1}}{N}
    \\
    =&\frac{1}{\sqrt{L}}\sum_{p_2,p_4,\dots p_{2L}}\prod_{n=1}^{L/2}\frac{\mathbb{K}_{p_{2n-2},p_{L+2n-2}}\mathbb{K}^{-1}_{p_{L+2n},p_{2n}}}{N}
    \\
    =&\frac{1}{\sqrt{L}}\sum_{p_2,p_4,\dots p_{2L}}\frac{1}{N}\mathbb{K}_{p_{2L},p_L}\prod_{n=1}^{L/2-1}\left(\frac{\mathbb{K}^{-1}_{p_{L+2n},p_{2n}}\mathbb{K}_{p_{2n},p_{L+2n}}}{N}\right)\mathbb{K}^{-1}_{p_{2L},p_L}\,,
\end{split}
\end{equation}
where we used the telescoping nature of the product of the $\mathbb{K}$-matrices to regroup the terms. Each term in the product are equal to $\Tr(\mathbb{K}^{-1}\mathbb{K})/N=1$, which leaves us with the first and last $\mathbb{K}$-matrices, which equal $\Tr(\mathbb{K}^\mathsf{T}\mathbb{K}^{-1})/N=\varepsilon_{\mathbb{K}}$. We thus conclude as claimed
\begin{equation}
    A_{\mathcal{O}_{(L)}}=\frac{\varepsilon_{\mathbb{K}}}{\sqrt{L}}+O(N^{-1})\label{eq:tree}
\end{equation}
\subsection{Sub-leading orders}\label{sec:AOLsubN}
For the operator $\mathcal{O}_{(L)}$, it is also possible to compute all the sub-leading contributions. To ensure that the operator is unit normalised, it should take the form\cite{kristjansen_new_2002}\footnote{Note that $\mathcal{O}_{(L)}$ is orthogonal to all single-trace operators at finite $N$, but not multi-trace operators.} 
\begin{equation}
    \mathcal{O}_{(L)}(N)=\frac{\Tr\hat{Z}^L}{\sqrt{L U_L(1/N)}}\,,\quad U_L(t)=(L-1)!t^{2(L-1)}+\sum_{n=0}^{L-1}(-1)^nn!\binom{L}{n}\binom{L-1}{n} t^{2n} \hat{U}_{L-n}(t)\,,
\end{equation}
where the latter polynomials $\hat{U}_n(t)$ are given by
\begin{equation}
    \hat{U}_n(t)=\frac{2}{n(n+1)}\sum_{m=0}^{\lfloor(n-1)/2\rfloor}\stir{n+1}{n-2m}t^{2m}\,,
\end{equation}
where $\stir{n}{m}$ are the unsigned Stirling numbers of the first kind. Note that this is a polynomial in even powers of $1/N$.
On the other hand, as we shall see shortly, the 1-point function of $\Tr \hat{Z}^L$ takes the form
\begin{equation}
\begin{split}
    \left<:\!\Tr\hat{Z}^L\!:(x)\right>&=\frac{\varepsilon_{\mathbb{K}}}{(1+x^2)^L}V^{\varepsilon_{\mathbb{K}}}_L(1/N)\,,\quad V^{\varepsilon_{\mathbb{K}}}_L(t)=\sum_{n=0}^{L/2}\frac{L!(-1)^nt^{2n}}{2^n n!(L-2n)!}\hat{V}^{\varepsilon_{\mathbb{K}}}_{L-2n}(t)\,,
    \\
    \hat{V}^{\varepsilon_\mathbb{K}}_n(t)&=\sum_{m=0}^{n/2-1}2^m\stir{n/2}{n/2-m}(\varepsilon_{\mathbb{K}}t)^m\,,\quad \hat{V}_0^{\varepsilon_{\mathbb{K}}}(t)=\frac{\varepsilon_{\mathbb{K}}}{t}\,.
\end{split}
\end{equation}
In both the two- and one-point functions, the hatted polynomials correspond to the result in $\U(N)$ gauge theory, which differs from $\SU(N)$ by the sub-leading term in the propagator \eqref{eq:propbefore}. Indeed, this is one source for the sub-leading terms in the 1-point function, which only gives contributions in even powers of $1/N$, plus a contribution at order $1/N^{L-1}$. Another is the non-maximally twisted diagrams, which contribute at all orders in $1/N$.
As hinted by the formulae, the one- and two-point functions are related via
\begin{equation}
    \hat{U}_n(t)=\frac{\hat{V}^+_{2(n+1)}(t/2)-\hat{V}^{-}_{2(n+1)}(t/2)}{n(n+1)t}\,.
\end{equation}
\\
To establish these formulae, we will first introduce some terminology. We will first do the computation for $\U(N)$ gauge theory: the computation for $\SU(N)$ will follow at the end.
\\

Let $S_L$ denote the group of permutations of $L$ objects. Let $P_L$ and $C_L$ denote the conjugacy classes consisting of pairings and full cycles, respectively, with $L$ even. Some distinguished members of these are 
\begin{equation}\label{eq:pi0}
    \pi_0=(12)(34)\dots (L-1,L)\in P_L\,,\quad \kappa_0=(123\dots L)\in C_L\,.
\end{equation}
Notice that the possible choice of Wick contractions among the $L$ fields $\hat{Z}$ is in one-to-one correspondence with $P_L$, such that
\begin{equation}
    \hat{A}_L(N)=\sum_{\pi\in P_L}\hat{A}^{(\pi)}_L(N)\,,
\end{equation}
where $\hat{A}^{(\pi)}_L(N)$ denotes the contribution from the diagram corresponding to $\pi$ and the hats again denote that we are working in the $\U(N)$ theory. Similarly, the trace identifies neighbouring sites and is mapped to $\kappa_0$. This allows us to write
\begin{equation}
    \hat{A}^{(\pi)}_L(N)=\frac{1}{\sqrt{L\hat{U}_L(1/N)}N^{L/2}}\sum_{p_1,p_2,\dots p_{L}}\prod_{\overset{n=1}{n<\pi(n)}}^L\mathbb{K}_{p_n,p_{\pi(n)}}\mathbb{K}^{-1}_{p_{\kappa_0\pi(n)},p_{\kappa_0(n)}}\,.
\end{equation}
We are then asking how many closed cycles this generates. Each cycle is decomposed into alternating $\mathbb{K}$ and $\mathbb{K}^{-1}$ edges. Having gone through one set of edges, we have mapped $n$ to $\kappa_0\pi\kappa_0^{-1}\pi(n)$. The answer is then given by $\frac{1}{2}\ell(\kappa_0\pi \kappa_0^{-1}\pi)$, where $\ell(\sigma)$ counts the number of closed loops in $\sigma$ and the factor of $1/2$ takes into account that each edge in $\kappa_0\pi \kappa_0^{-1}\pi$ is two edges in the diagram. Taking the appropriate factors of $\varepsilon_{\mathbb{K}}$ into account,\footnote{
Since $\mathbb{K}^{\mathsf{T}}=\varepsilon_\mathbb{K}\mathbb{K}$, each transposition of a matrix's
index order contributes one factor of $\varepsilon_\mathbb{K}$. The relative number of transpositions modulo two is given by $\sgn(\kappa_0\pi)=(-1)^{1+L/2}$.
After orienting the factors along the loops, there is one additional factor coming from a cycle shift of even length per loop. Thus the number of required reversals is
$1+(L-\ell(\kappa_0\pi\kappa^{-1}_0\pi))/2$ modulo two.} we end up with the formula
\begin{equation}
    \hat{A}_L^{(\pi)}(N)=\frac{\varepsilon_{\mathbb{K}}}{\sqrt{L\hat{U}_L(1/N)}}\left(\frac{\varepsilon_{\mathbb{K}}}{N}\right)^{\frac{L-\ell\left(\kappa_0\pi \kappa_0^{-1}\pi\right)}{2}}\quad  \Rightarrow\hat{V}^{\varepsilon_\mathbb{K}}_L(t)=(\varepsilon_\mathbb{K}t)^{L/2}f_L(\varepsilon_\mathbb{K}t^{-1})\,,\label{eq:Ahatf}
\end{equation}
where $f_L$ is given by $f_L(x)=\sum_{\pi\in P_L}x^{\frac{1}{2}\ell(\kappa_0\pi\kappa_0^{-1}\pi)}$.

The trick to computing $f_L$ is to notice that nothing distinguishes $\kappa_0$ in the space of cycles, and we can replace it by the average over all of $C_L$ instead
\begin{equation}
    \sum_{\pi\in P_L}x^{\frac{1}{2}\ell(\kappa_0\pi\kappa_0^{-1}\pi)}=\frac{1}{|C_L|}\sum_{\pi\in P_L,\kappa\in C_L}x^{\frac{1}{2}\ell(\kappa\pi\kappa^{-1}\pi)}\,.
\end{equation}
Using that $\kappa\pi\kappa^{-1}\in P_L$ distributes uniformly, we can then trade the average over $C_L$ by the average over $P_L$
\begin{equation}
    \frac{1}{|C_L|}\sum_{\pi\in P_L,\kappa\in C_L}x^{\frac{1}{2}\ell(\kappa\pi\kappa^{-1}\pi)}=\frac{1}{|P_L|}\sum_{\pi,\rho\in P_L}x^{\frac{1}{2}\ell(\pi\rho)}\,,
\end{equation}
Finally, we can trade this average for a fixed contribution from any given member. We choose $\pi_0$ as defined in \eqref{eq:pi0}, such that we have established
\begin{equation}
   f_L(x)=\sum_{\rho\in P_L}x^{\frac{1}{2}\ell(\pi_0 \rho)}
\end{equation}
We can compute this sum by recursion. Let $\pi_0'=(12)(34)\dots(L-3,L-2)$. For fixed $\rho\in P_L$, we will map it to an element $\rho'\in P_{L-2}$. If $\rho$ contains the pairing $(L-1,L)$, map it to the element obtained by removing this pairing. In this case, $\ell(\rho\pi_0)=2+\ell(\rho'\pi_0')$, since both $L$ and $L-1$ would be fixed points of $\rho\pi_0$. Otherwise, $\rho$ contains the pairing $(n,L-1)$ and $(m,L)$ with $n,m+1\neq L$. We can then generate an element $\rho'=\in P_{L-2}$ by replacing $(n,L-1)(m,L)$ by $(n,m)$. This mapping $\rho\mapsto\rho'$ preserves the number of closed loops but is evidently not one-to-one: each such element $\rho'$ has $L-2$ pre-images in $P_L$. Since each pairing $(n,L)$ appear equally often, this establishes
\begin{equation}
    f_L(x)=(x+L-2)f_{L-2}(x)\,.
\end{equation}
But this continues recursively. Alongside $f_2(x)=x$, as shown in \eqref{eq:A2}, this allows us to establish
\begin{equation}
    f_L(x)=x\prod_{n=1}^{L/2-1}(x+2n)\,.
\end{equation}
Using the definition of the Stirling numbers, this allows us to establish\footnote{By matching this with the number of permutations of $L/2$ elements with $k$ loops, this also allows us to establish $\sum_{\rho\in P_L} x^{\frac{1}{2}\ell(\pi_0\rho)}=\sum_{\sigma\in S_{L/2}}2^{L/2-\ell(\sigma)}x^{\ell(\sigma)}$.}
\begin{equation}
    \hat{V}_n^{\varepsilon_{\mathbb{K}}}(t)=\sum_{m=0}^{n/2-1}2^m\stir{n/2}{n/2-m}(\varepsilon_{\mathbb{K}}t)^m\,.
\end{equation}
Finally, to pass to the $\SU(N)$ theory, we have to take the finite $N$ term in the propagator \eqref{eq:propbefore} into account. This replaces the Wick contraction by \eqref{eq:ZhatWick} by
\begin{equation}
\left<:[\hat{Z}]_{pq}[\hat{Z}]_{rs}:(x)\right>=\frac{1}{(1+x^2)^2}\left\{\frac{\mathbb{K}_{pr}\mathbb{K}_{sq}^{-1}}{N}-\frac{\delta_{pq}\delta_{rs}}{N^2}\right\}\,.
\end{equation}
Notice that $\mathbb{K}$ and $\mathbb{K}^{-1}$ cancelled in the second term. This implies that using this second part of the Wick contraction amounts to multiplying by $-1/N^2$ and removing two fields from the trace. Since there are $\binom{L}{2n}(2n-1)!!=\frac{L!}{2^n n!(L-2n)!}$ ways to perform $n$ such contractions, we have established
\begin{equation}
    V_{L}^{\varepsilon_{\mathbb{K}}}(t)=\sum_{n=0}^{L/2}\frac{L!(-1)^nt^{2n}}{2^n n!(L-2n)!} \hat{V}_{L-2n}^{\varepsilon_{\mathbb{K}}}(t)\,.
\end{equation}
Similar arguments allows one to compute $U_L(t)$ and $\hat{U}_L(t)$, by replacing sums over $P_L$ by $C_L$. Thus, we have established
\begin{equation}
    A_L(N)=\frac{\varepsilon_{\mathbb{K}}}{\sqrt{L}}\frac{V_L(1/N)}{\sqrt{U_L(1/N)}}=\frac{\varepsilon_{\mathbb{K}}}{\sqrt{L}}\left\{1+\frac{\varepsilon_{\mathbb{K}}L(L-2)}{4N}+\frac{L(L-2)(L^2-14L+10)}{96N^2}+O(N^{-3})\right\}\,.
\end{equation}
The fact that this contains contributions with odd powers of $1/N$ is a distinguishing feature of orientifolds.
\section{Coordinate Bethe ansatz and Gaudin determinant for the \texorpdfstring{$\SU(1|2)$}{SU(1|2)} spin chain}\label{sec:CBA}
Since the Dynkin diagrams and Bethe states for the SU(1$|$2) spin chain in Sections \ref{sec:SU12} and \ref{sec:PSU224} might be unfamiliar to some of the readers, in this appendix, we briefly summarize the construction of the coordinate Bethe states. We also provide explicit expressions for the Gaudin matrices that appear in the computation of one-point functions in the main text.
\subsection{Dynkin diagrams for superspin chains}\label{sec:Dynkin}
We consider an $\SU(n|m)$ superspin chain of length $L$, where each site transforms in the same representation $R$. We will use $K=n+m-1$ to denote the number of nodes in the corresponding Dynkin diagram. Like the $\SU(K+1)$ Dynkin diagram, the $\SU(n|m)$ diagram is created by having $K+1$ nodes, such that each line connects at most two nodes and every node has two lines ending on it, forming a chain. There should be $n$ bosons (represented by solid lines) and $m$ fermions (represented by dashed lines). 

We can read off the Cartan matrix by looking at the Dynkin diagram. A solid line between node $a$ and node $b$ corresponds to $C_{ab}=-1$, while a dashed line corresponds to $C_{ab}=1$. The diagonal elements $C_{aa}$ are minus the sum of the lines ending at the node. On top of the Dynkin diagram, we also need to specify the Dynkin labels $q_a$ at each node, which tells us which representation $R$ we are considering. Nodes for which $q_a\neq 0$ are called momentum-carrying. We restrict our attention to representations for which only one of the nodes is momentum-carrying. 

Let us see this in action by considering the $\SU(1|2)$ superspin chain, whose Dynkin diagram is
\begin{align}
\begin{tikzpicture}[cross/.style={path picture={ 
			\draw[black,thick]
			(path picture bounding box.south east) -- (path picture bounding box.north west) (path picture bounding box.south west) -- (path picture bounding box.north east);
	}}]
	\draw [thick](0,0)--(1,0);
	\draw[thick,dashed] (1,0)--(2,0)--(3,0);
	\draw (1,0)node[circle,draw=black,fill=white,cross]{};
	\draw(2,0) node[circle,draw=black,fill=white]{};
\end{tikzpicture}\label{eq:SU12Dynkinapp}
\end{align}
and whose Cartan matrix and Dynkin labels are given by
\begin{equation}
    C=\begin{pmatrix}
        0&1\\1&-2
    \end{pmatrix}\,,\quad q=\begin{pmatrix}
        1\\0
    \end{pmatrix}
\end{equation}
By reading \eqref{eq:SU12Dynkinapp} from left to right, we can then see how a state is created in the CBA. Recalling the Hilbert space defined in \eqref{eq:0updown}, the lines tell us that the nested Bethe ansatz works by exciting in steps $\ket{0}\mapsto\ket{\!\!\uparrow}\mapsto \ket{\!\!\downarrow}$. Thus, we are considering as a vacuum the state $\otimes \ket 0_j$. Next, we are creating $M_1$ momentum-carrying fermionic magnons $\{u^1_1,\dots u^1_{M_1}\}$ on top of this state, which work by exciting $\ket{0}\mapsto \ket{\!\!\uparrow}$, which indeed corresponds to acting with a fermionic element of the algebra. This creates a new auxiliary spin chain of length $M_1$, for which the momentum-carrying magnons act as inhomogeneities.
Lastly, we are furthermore creating $M_2$ auxiliary bosonic magnons $\{u^2_1,\dots u^2_{M_2}\}$ on this emergent spin chain, which flip $\ket{\!\!\uparrow}\mapsto\ket{\!\!\downarrow}$: since we cannot have more magnons than sites, this immediately implies that $M_1\geq M_2$. In the end, we have a state with $L-M_1$ $\ket{0}$'s, $M_1-M_2$ $\ket{\!\!\uparrow}$'s and $M_2$ $\ket{\!\!\downarrow}$'s.
\paragraph{Baxter polynomials and the Bethe equations.}
Like in the SU(2) case, the Bethe equations for an arbitrary Dynkin diagram and labels can be written compactly in terms of Baxter polynomials
\begin{equation}
    Q_0(x)=x^L\,,\qquad Q_a(x)=\prod_{j=1}^{M_a}(x-u_j^a)\,.\label{eq:GenQ}
\end{equation}
For a diagram with $K$ nodes, Bethe equations can be written in the form
\begin{equation}
    R_a(u^a_j|\bm{u})=1\,,\qquad R_a(x|\bm{u})=(-1)^{1+\epsilon_a}\frac{Q_0(x-iq_a/2)}{Q_0(x+iq_a/2)}\prod_{b=1}^K\frac{Q_b(x+iC_{ab}/2)}{Q_b(x-iC_{ab}/2)}\,,\label{eq:GenBethe}
\end{equation}
where $\epsilon_a=0$ for a bosonic node and $\epsilon_a=1$ for a fermionic.
\subsection{The Bethe state}\label{sec:BetheState}
What is the corresponding Bethe state? We focus for simplicity on $\SU(1|2)$, though a similar construction works for higher-rank superspin chains. We again consider the case where only one node is momentum carrying (which we can conveniently label as the first node); the $\SU(1|2)$ wave function of the corresponding Bethe state is then relatively straightforward to write down. It will be useful to introduce the partial Baxter polynomials
\begin{equation}
    Q^{\sigma_a}_{a,m}(x)=\prod_{j=1}^m(x-u^a_{\sigma^a_j})\,,
\end{equation}
which depends on a permutation $\sigma^a\in S_{M_a}$. For rapidities that solve the Bethe equations, the corresponding Bethe wave-function $\Psi(\bm{u}^1,\bm{u}^2|\bm{n}^1,\bm{n}^2)$, where $\bm{n}^a=(n^a_1,\dots n^a_{M_a})$ denotes the location of the excitation on the spin-chain at level $a$, takes the form
\begin{equation}
    \Psi(\bm{u}^1,\bm{u}^2|\bm{n}^1,\bm{n}^2)=\sum_{\sigma^1\in S_{M_1}}\sum_{\sigma^2\in S_{M_2}}k_1\left(\bm{u}^1;\sigma^1\right)\,k_2\left(\bm{u}^1,\bm{u}^2;\sigma^1,\sigma^2;\bm{n}^2\right)\,\left(\prod_{j=1}^{M_1}e^{ip_{\sigma^1_j}n_j}\right)\,,
\end{equation}
where
\begin{equation}
\begin{split}
    &e^{ip_j}=\frac{u^1_j+iq_1/2}{u^1_j-iq_1/2}\,,\quad k_1\left(\bm{u}^1;\sigma^1\right)=\sgn(\sigma^1)^{\epsilon_1}\left(\prod_{j=1}^{M_1}\frac{Q^{\sigma^1}_{1,j-1}\left(u^1_{\sigma^1_j}-iC_{11}/2\right)}{Q^{\sigma^1}_{1,j-1}\left(u^1_{\sigma^1_j}\right)}\right)\,,\\
    &k_2\left(\bm{u}^1,\bm{u}^2;\sigma^1,\sigma^2;\bm{n}^2\right)=\sgn(\sigma^2)^{\epsilon_2}\left(\prod_{j=1}^{M_2}\frac{Q^{\sigma^1}_{1,n^2_j-1}\left(u^2_{\sigma^2_j}-iC_{12}/2\right)}{Q^{\sigma^1}_{1,n^2_j}\left(u^2_{\sigma^2_j}+iC_{12}/2\right)}\frac{Q^{\sigma^2}_{2,j-1}\left(u^2_{\sigma^2_j}-iC_{22}/2\right)}{Q^{\sigma^2}_{2,j-1}\left(u^2_{\sigma^2_j}\right)}\right)\,.\label{eq:SU12WaveFunc}
    \end{split}
\end{equation}
While the above formula looks rather complicated, the physical picture is quite clear. $k_1$ accounts for the phase factor being picked up by the momentum-carrying magnons as they move past one another. $k_2$ subsequently accounts for the phase factor picked up by the auxiliary magnons, which arise both from moving magnons past each other (the ratio of $Q_2$'s) and from moving past the auxiliary spin chain inhomogeneities, namely the momentum-carrying magnons (the ratio of $Q_1$'s). These formulae hold for either of the two Dynkin diagrams \eqref{eq:SU12Dynkin} and \eqref{eq:SU12DynkinAlt}.

Fortunately, we will typically not need to know the state explicitly in order to compute various observables. Indeed, for any (super)spin chain, knowing the Bethe roots is enough to write down the corresponding Gaudin matrix which appears in the norm of the state. It takes the form
\begin{equation}
    G_{jk}^{ab}=-i\frac{\partial}{\partial u^a_j}\log R_b(u_k^b|\bm{u})\,.\label{eq:genG}
\end{equation}
It naturally organises into $K^2$ blocks, labelled by $a,b$. Note that if $a\neq b$ and $C_{ab}=0$, $G^{ab}=0$.
\subsection{Parity-symmetric states}\label{sec:parstate}
The crosscap state is an integrable state for any superspin chain. Thus, it has non-vanishing overlap only with parity-symmetric states. For the crosscap state relevant for describing $\mathcal{N}=4$ SYM on $\mathbb{RP}^4$, this corresponds to the Bethe roots on each node being invariant under $\bm{u}\mapsto-\bm{u}$.\footnote{In the notation of \cite{Gombor:2022deb}, integrable states with the property that the overlap vanishes unless $\bm{u}=-\bm{u}$ are called \textit{twisted}.}. However, a new feature of the superspin chain is that the ``spin'' of the crosscap state is not necessarily integer. This implies that each set of roots can be written in one of two forms, depending on whether the number of roots ($M_a=2m_a+\delta_a$) on the node is even or odd ($\delta_a=0,1$)
\begin{equation}
    \{u^a_j\}_{j=1}^{2m_a}=\{u_1^a,\dots,u_{m_a}^a,-u_1^a,\dots, -u^a_{m_a}\}\,,\quad \text{or}\quad \{u^a_j\}_{j=0}^{2m_a}=\{0,u_1^a,\dots, u_{m_a}^a,-u_1^a,\dots,- u_{m_a}\}\,.
\end{equation}
When the total number of roots is odd, the root of the $Q$-function $Q_a(x)$ at $x=0$ plays a distinguished role.

Like in the $\SU(2)$ case, the Gaudin matrix \eqref{eq:genG} of parity-symmetric states factorises. Define the matrices 
\begin{equation}
    \begin{split}
        &G^{\pm,ab}_{jk}=G_{j,k}^{ab}\pm G^{ab}_{j+m_a,k}\quad \text{for}\quad j,k\neq 0\,,
        \\
        &G^{+,ab}_{jk}=\sqrt{2}G_{jk}^{ab}\quad \text{for}\quad j=0\,,\,k\neq 0\,,\\
         &G^{+,ab}_{jk}=G_{jk}^{ab}\quad \text{for}\quad j=k=0\,.
    \end{split}\label{eq:genGpGm}
\end{equation}
Note that in the presence of zeroes, the two block matrices $G^{+,ab}$ and $G^{-,ab}$ have different dimensions, being $(m_a+\delta_a)\times(m_b+\delta_b)$ and $m_a\times m_b$ respectively. The factorisation then works as for $\SU(2)$, namely $\det G=\det G^+\,\det G^-$, and only their ratio $\det G^+/\det G^-$ appears in the formula for the crosscap overlap.
\bibliographystyle{utphys2}
\bibliography{refs}
\end{document}